\documentclass[12pt]{iopart}
\usepackage[utf8]{inputenc}
\usepackage[english]{babel}
\usepackage{amsmath}
\usepackage{amsfonts}
\usepackage{amssymb}
\usepackage{graphicx}
\usepackage[backend=biber,style=numeric-comp,sorting=none]{biblatex}

\usepackage{tabu}
\usepackage[font=small]{caption}
\usepackage{hyperref}
\usepackage[dvipsnames]{xcolor}

\begin{document}
\title[Atomic-magnetometer tutorial]{How to build a magnetometer with thermal atomic vapor: A tutorial}

\author{Anne Fabricant$^{1,2}$, Irina Novikova$^3$, and Georg Bison$^4$}

\address{$^1$Helmholtz Institute Mainz, GSI Helmholtzzentrum für Schwerionenforschung, Darmstadt, Germany}
\address{$^2$Johannes Gutenberg University of Mainz, Mainz, Germany}
\address{$^3$Department of Physics, William \& Mary, Williamsburg, Virginia 23187, USA}%https://www.overleaf.com/project/60e5a8e59c005029afe64bed
\address{$^4$Paul Scherrer Institute, Villigen 5232, Switzerland}
%\author[1,2]{Anne Fabricant}
%\author[3]{Irina Novikova}
%\author[4]{Georg Bison}
%\affil[1]{Helmholtz Institute Mainz, GSI Helmholtzzentrum für Schwerionenforschung, Darmstadt, Germany}
%\affil[2]{Johannes Gutenberg University of Mainz, Mainz, Germany}
%\affil[3]{Department of Physics, William and Mary, Williamsburg, Virginia 23187, USA}%https://www.overleaf.com/project/60e5a8e59c005029afe64bed
%\affil[4]{Paul Scherrer Institute, Villigen 5232, Switzerland}
% \vspace{10pt}
% \begin{indented}
% \date{\today}
% \end{indented}

\begin{abstract}
    This article is designed as a step-by-step guide to optically pumped magnetometers based on alkali atomic vapor cells. We begin with a general introduction to atomic magneto-optical response, as well as expected magnetometer performance merits and how they are affected by main sources of noise. This is followed by a brief comparison of different magnetometer realizations and an overview of current research, with the aim of helping readers to identify the most suitable magnetometer type for specific applications. Next, we discuss some practical considerations for experimental implementations, using the case of an $M_z$ magnetometer as an example of the design process. Finally, an interactive workbook with real magnetometer data is provided to illustrate magnetometer-performance analysis.
    % \todo{A 10-15 page article ready by Spring 2022 ;)}
    % \\ \todo{General to-do: fix bibtex issues}
\end{abstract}

% \tableofcontents

%\maketitle
%\pagebreak
\section{Why should you build an atomic magnetometer?}
\label{sec:1-Intro} 
% \noindent Suggested outline:
% \begin{itemize}
%     \item Brief overview of available magnetometer technologies
%     \item Is atomic magnetometry right for my application?
%     \item Compare performance figures to other magnetometer technologies---sensitivity, bandwidth, standoff distance, spatial resolution, accuracy v. stability---strengths and challenges---GRAPHIC
%     \item Summarize working principle of atomic magnetometer
%     \item Basics of orientation and polarization
%     \item Discussion of issues like heading error
%     \item Focus on specific magnetometer throughout paper
%     \item Free induction decay
%     \item Not covering rare gases
% \end{itemize}
% \bigskip

The purpose of this review is to introduce students and newcomers to the field of atomic magnetometry in a friendly and didactic manner. We aim to bridge the gap between the various excellent reviews already existing in literature \cite{Budker2007,Aleksandrov2009,Budker2013,Weis2017,Tierney2019,Fu2020,Romalis2022} and the nitty-gritty of actually setting up and operating an atomic magnetometer in the lab, for either academic or industrial purposes.
%and using it to detect real signals.

The origins of magnetometry --- the measurement of magnetic fields --- most likely date back to the time of the dinosaurs 150 million years ago, when the first birds evolved. It is known that birds, not only those of the migratory variety, possess an innate magnetic compass for orientation along the geomagnetic field lines, which is activated by exposure to light \cite{Wiltschko2021,Xu2021}. In fact, a wide range of animal species are blessed with built-in geomagnetic-field sensors, based on a variety of proposed biochemical mechanisms \cite{Gould2008}. Humans, being less gifted by nature, had to develop their own navigational compasses --- the earliest type of magnetometer --- through arduous scientific work from the 12th century onward, enabling the formation of empires in Europe and Asia. After Gauss  developed, in 1832, the first known absolute magnetometer able to directly measure the geomagnetic field, the 20th century saw an explosion of new magnetometer technologies based on both classical- and quantum-physics phenonema. This enabled the sensitive measurement of hypogeomagnetic fields (fields smaller than Earth field) from a diversity of sources (Table~\ref{tab:commercial}). Chances are that the reader is routinely carrying around a state-of-the-art magnetometer: most modern smartphones are outfitted with a tiny Hall-effect or magnetoresistive (MR) based compass, which can be integrated onto a chip with an accelerometer (to sense linear acceleration) and usually a gyroscope (to sense angular acceleration).%\footnote{The difference between accelerometers and gyroscopes was acutely felt by some smartphone users during the Pokemon Go mobile-app craze of 2016, as a gyroscope was required for proper functioning of the game.}
% Typical handheld compass: precision +/- 2deg

Atomic magnetometers are quantum devices making use of the fact that the electron spin of each atom in a gas reacts to its own magnetic-field environment. During magnetometer operation, atoms are first spin-polarized via optical pumping with laser light along a given direction (Sec.~\ref{subsec:2.1-OpticalPumping}), then allowed to evolve under the influence of an external magnetic field (Sec.~\ref{subsec:2.3-BlochEq}). Magnetic-field measurement is enabled by subsequent optical readout of the atomic spin state, which exploits magneto-optical effects --- for example, shining linearly polarized off-resonant ``probe'' light on the atoms and measuring its polarization rotation via the Faraday effect, or measuring the change in absorption of resonant probe light (Sec.~\ref{subsec:2.5-OpticalDetection}). Such atomic magnetometers may be called by different names, including optical(-atomic) magnetometers or optically pumped magnetometers (OPMs), although the magnetic-sensing volume itself consists of atoms.

\begin{figure}[h]
  \centering
  \includegraphics[width=0.9\columnwidth]{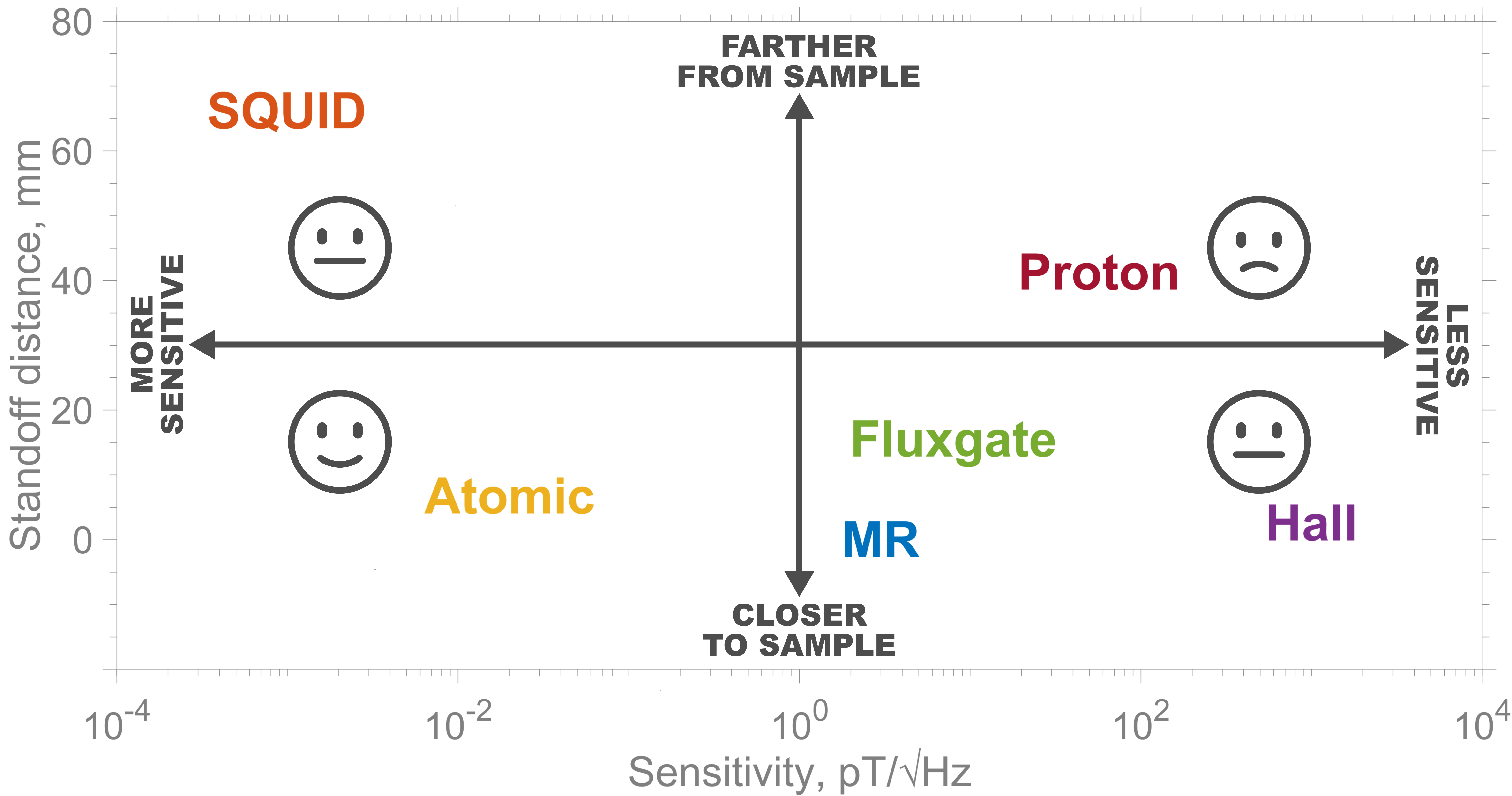}
  \caption{A silly pictorial representation of the information in Table~\ref{tab:commercial}.}
\label{fig:compass}
\end{figure}

Before designing and building a sensor in the lab, one should first ask an important question:
\emph{Is atomic magnetometry right for my application?} 
There exists a plethora of magnetometer technologies, many of which are commercially available, with each technology having its own pros and cons. In Table~\ref{tab:commercial} and Fig.~\ref{fig:compass}, we compare the performance figures of some common commercial magnetometers operating in the magnetic-field range up to Earth field.
%at the Earth-like range of magnetic fields. 

\renewcommand{\arraystretch}{1.5}
\begin{table}[h]
\tiny
\begin{center}
\begin{tabular}{ |p{0.11\textwidth}|>{\raggedright}p{0.08\textwidth}|>{\raggedright}p{0.08\textwidth}|>{\raggedright}p{0.08\textwidth}|>{\raggedright}p{0.08\textwidth}|>{\raggedright}p{0.08\textwidth}|>{\raggedright}p{0.08\textwidth}|>{\raggedright}p{0.08\textwidth}|>{\raggedright\arraybackslash}p{0.08\textwidth}|}
 \hline
 \textbf{Sensor type} & \textbf{Birth year} & \textbf{Example(s)} & \textbf{Sensitivity} ($\text{pT}/\sqrt{\text{Hz}}$) & \textbf{Standoff distance (mm)} & \textbf{Smallest linear dimension} (mm) & \textbf{Bandwidth} (kHz) & \textbf{Approx. price range} (USD) & \textbf{Notes} \\ 
 \hline\hline
 Fluxgate & 1936 \cite{Aschenbrenner1936} & SENSYS, Bartington & 6--10 & 14.5--16.5 & 23--26 & 3--4 & 3--5\,000 & better lab sensitivity achieved \cite{Janosek2017,Koshev2021} \\
 \hline
 Proton (incl. Overhauser) & 1958 \cite{Waters1958} & GEM, Geometrics & 15--100 & 37.5--45 & 75--90 & 15-30 & 100\,--\,1\,000 & geophysical applications \\
 \hline
 Atomic & 1962 \cite{Bloom1962} & QuSpin, Twinleaf & 0.01--0.03 (0.09--0.13) & 6.5 & 12.4 & 0.10--0.15 (0.5--1.1) & 9--10\,000 & biomedical applications \\ 
 \hline
 SQUID & 1964 \cite{Jaklevic1964} & Quantum Design, Magnicon & 0.001 & 66 & 840 & 1 & up to 1\,000\,000 & cheaper partial systems available \\ 
 \hline
 Hall-effect & 1967 \cite{Owston1967} & Metrolab & 1000 & 2.5 & 1.2 & 1 & 4--10\,000 & useful for field mapping \\
 \hline
 Magnetoresistive & 1982 \cite{Hoffmann1982} & TE Connectivity, Sensitec & 3 & 0 & N/A & 10 & 2\,000 & only bare sensor sold commercially \\
 \hline
%  NV diamond & 2008 & Qnami, QSabre & - & - & - & - & - & microscopy applications \\
%  \hline
\end{tabular}
\caption{Performance comparison of commonly available commercial magnetometers. The choice of examples is intended to be representative and not exhaustive; quoted specs are approximate. Non-commercial laboratory magnetometers may surpass the quoted typical performance figures \cite{Grosz2017}. See Sec.~\ref{sec:3-Requirements} for definitions and discussion of the performance figures. Note that optically pumped magnetometers based on nitrogen-vacancy (NV) centers in diamond \cite{Jensen2017} are not included in this table, since commercial NV sensors at the time of writing are sold only as scanning-microscopy imaging systems. Atomic magnetometers based on parametric resonance in He-4 are currently undergoing commercialization by the company MAG4Health.
%\todo{TDK has asked us not to include their sensor in the article - look for alternative commercial MR devices.}
%\todo{Double-check the specs.}
} 
\end{center}
\normalsize
\end{table}
\label{tab:commercial}
% Other sensor types: cavity optomechanical, induction-coil, giant-magnetoimpedance (GMI), magnetoelectric (ME), mechanical/force
% I remember something about oil exploration possibly using MR sensors---check for references
\renewcommand{\arraystretch}{1}

Fig.~\ref{fig:compass} illustrates that atomic magnetometers can offer the best overall performance for applications requiring both relatively high magnetic-field sensitivity and relatively small stand-off distance from the sample to be measured. Unlike superconducting-quantum-interference-device (SQUID) magnetometers, they are non-cryogenic and can operate at ambient temperatures, while also offering greater portability. Although the leading commercial models are designed specifically for biomedical applications such as magnetoencephalography (MEG) \cite{Sander2020} and fetal magnetocardiography (fMCG) \cite{Batie2018}, there is great flexibility in the design of laboratory atomic magnetometers. For example, the atomic sensing volume can be miniaturized or otherwise  adapted to the particular sample geometry \cite{Nardelli2020,Jensen2018}.

This review is structured as follows. In Sec.~\ref{sec:2-BlochEq}, we briefly introduce the technique of optical pumping, through which atomic-spin polarization can be created using resonant laser light. Then we present what we consider to be the most important equation of magnetometry, the Bloch equation, which describes the interaction of atomic spins with an external magnetic field --- the heart of an atomic magnetometer. To illustrate the process of magnetometer operation, we consider a specific example of an $M_z$ magnetometer. Sec.~\ref{sec:3-Requirements} provides an overview of the relevant figures of merit used to characterize magnetometer performance, some of which were already introduced in Table~\ref{tab:commercial}. Sec.~\ref{sec:4-OperatingModes} compares the different types of atomic magnetometers and guides the reader in selecting an appropriate operating mode for the desired application (e.g., biological/non-biological, applied/fundamental) and performance requirements. From Sec.~\ref{sec:5-Assembly}, the tutorial becomes more hands-on. Here we describe how to choose proper components for the magnetometer, including a suitable atomic vapor cell for the sensing volume, as well as laser and magnetic-shielding systems. Next, we outline the step-by-step operational principles of the $M_z$ magnetometer and describe how to obtain 
%and analyze 
a free-induction-decay (FID) signal --- steps which can be applied by the reader to other types of atomic magnetometers. Sec.~\ref{sec:6-Characterization} focuses on how to characterize magnetometer performance according to the figures of merit discussed in Sec.~\ref{sec:3-Requirements}.
%, as well as technical challenges such as noise, broadening, and heading error. 
% Finally, in Sec.~\ref{sec:7-Measurement} we provide some tips and tricks about working with real-life biological and nonbiological samples, in the context of both applied and fundamental physics.

As the title of this review suggests, we will be covering magnetometers based on thermal atomic vapors. In cold-atom magnetometry, laser cooling is employed to trap and control the sensing atoms, and to prevent spin-destructive atomic collisions which worsen magnetometer performance (Sec.~\ref{sec:3-Requirements}). Although cold atoms offer advantages in terms of spatial resolution and stand-off distance, the experimental requirements are more stringent and require the use of a vacuum system, and the number of atoms which can be trapped is limited. Despite progress in recent years \cite{Gawlik2013,Cohen2019}, the sensitivity of cold-atom sensors is not yet competitive with thermal sensors utilizing buffer-gas or anti-relaxation-coated atomic vapor cells (Sec.~\ref{subsec:5.1-VaporCell}). Therefore, the latter are preferred for most atomic-magnetometry applications (Sec.~\ref{sec:7-Applications}). 

Furthermore, we focus specifically on alkali-vapor sensing. References are provided for readers interested in learning more about magnetic sensing with nuclear or electron spins in helium \cite{Heil2017,Fourcault2021,Bertrand2021,LeGal2021,Leger2015,Maul2018,Nikiel2014}, as well as comagnetometers incorporating both noble-gas and alkali vapors in tandem, whereby the nuclear-spin signal is enhanced via spin-exchange optical pumping \cite{Limes2018,Jiang2021}. Helium magnetometers have traditionally been used for military and aerospace applications, due to their instrinsic robustness and stability, and optically pumped He-4 sensors have recently begun to approach the sensitivity levels of their alkali cousins \cite{Fourcault2021,Palacios-Laloy2022}. Alkali/noble-gas comagnetometers are primarily used in fundamental-physics and gyroscope experiments \cite{Limes2018,Park2019,Terrano2022,Padniuk2022,Su2021}.

%\pagebreak
% \section{Getting to know the Bloch equation}
\section{The Bloch equation and all that jazz}
\label{sec:2-BlochEq}

%We choose the $M_z$ method because it is the simplest to set up in practice.

\begin{figure} [h]
  \centering
  \includegraphics[width=12cm]{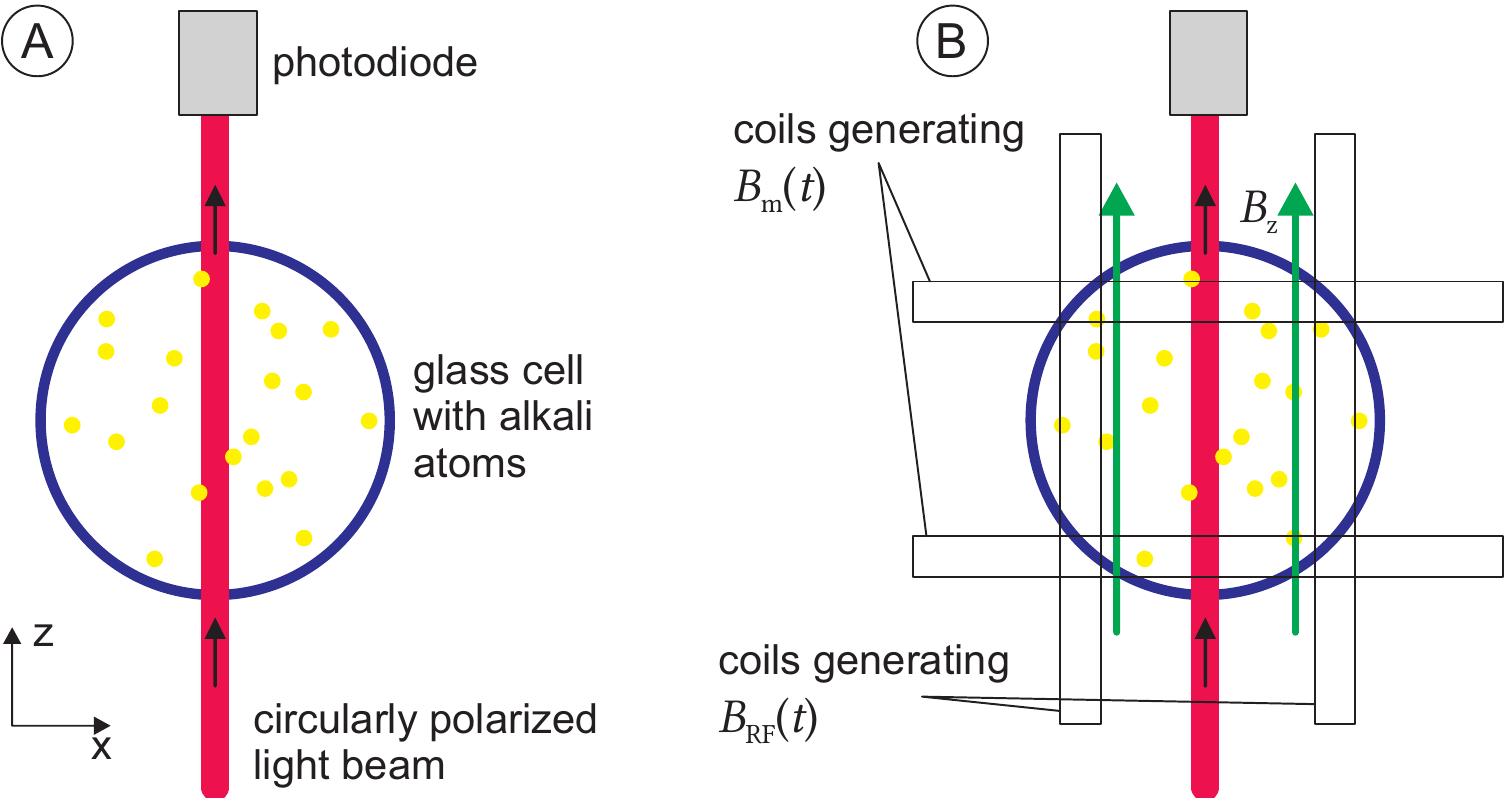}
  \caption{A) Basic setup of an optically pumped magnetometer.  B) Setup for the $M_z$ mode of operation. 
    }
\label{fig:setup}
\end{figure}

% This section \AMF{is the title still broad enough, since we're now only talking about Bloch equation?} describes the experimental components of an optical-atomic magnetometer and how they interact. \IBN{I suggest an alternative: This section introduce the methodology for theoretical modeling of the magnetometer performance.}
This section introduces the methodology for theoretical modeling of an atomic magnetometer.
As a starting point for the concrete description of such a magnetometer, we choose the experimental setup shown in Fig.~\ref{fig:setup}A. 
% \AMF{This section uses the terms ``optical magnetometer'' and ``OPM'' throughout. Although these are introduced in Section 1, there we prioritize the term ``atomic magnetometer''. Let's discuss how to make this more consistent.}
%
The core of the magnetometer is the alkali atoms which form a vapor in a glass cell.
A laser beam traverses the cell, and the power of the transmitted light is measured by a photodiode.
Since atoms interact with both the ambient magnetic field $\vec{B}$ and the laser light, they mediate magneto-optical coupling, such that information about $\vec{B}$ is transferred to the measurable optical signal.
At the level of individual atoms, this interaction is characterized by the total atomic angular momentum $\vec{F}$, often referred to as the atomic spin.
However, since the laser beam interacts with a large number of atoms at the same time, the ensemble average of %the atomic spin 
$\vec{F}$ is the relevant quantity here. 
This ensemble average, which we denote as $\vec{M}$, is a macroscopic quantity describing the magnetization state of the atomic medium.
Because $\vec{M}$ is derived from atomic spins, it is a macroscopic sum of atomic angular momenta. In the atomic-magnetometry literature, $\vec{M}$ is commonly referred to as the \textbf{bulk magnetization} or just \textbf{magnetization} of the atomic medium. (We note that in electromagnetism, $\vec{M}$ is defined as bulk magnetization per unit volume.)
%it is both a macroscopic magnetization and a macroscopic sum of spin angular momentum --- in other words, as is often the case in physics, both a classical and a quantum description are possible.

Although in this section we primarily rely on the classical description of spin dynamics, we'll first briefly introduce some basics of the quantum description which are relevant for understanding. Our chosen theoretical treatment of the magnetometer can be described as semiclassical, where quantum mechanics is only explicitly invoked in the discussion of optical pumping (Sec.~\ref{subsec:2.1-OpticalPumping}). We find that this approach, based on classical Bloch formalism, is typically the most intuitive and accessible for new experimentalists. However, we wish to emphasize that physically equivalent quantum treatments are also possible. These include the density-matrix formalism --- whereby the atomic spin state is encoded in multipole moments of a density matrix which evolves, e.g. in the presence of any electromagnetic fields, according to the quantum Liouville equation \cite{Budker2008,Auzinsh2010} --- as well as the Floquet description \cite{Florez2021} and representations within the framework of quantum optics \cite{Julsgaard2003}.

In spectroscopic notation, the electronic energy levels of an atom are denoted $n L_J$. Here $n$ is the principal quantum number and $\vec{L}$ is the orbital angular momentum with quantum number $L$ (S=0, P=1, D=2); $\vec{J} = \vec{L} + \vec{S}$ is the total electronic angular momentum, with $\vec{S}$ being the electron spin ($S = 1/2$).
%\footnote{The more complete spectroscopic notation is $n^{2\mathit{S}+1}L_J$, where $2\mathit{S}+1$ is the multiplicity.}
The total angular momentum of the atom is then $\vec{F} = \vec{I} + \vec{J}$, where $\vec{I}$ is the nuclear spin. %
Let us take as an example the alkali atom Cs-133, the only stable isotope of cesium, with $I = 7/2$: the ground state $6\text{S}_{1/2}$ ($\vec{L} = 0$) splits into two hyperfine states with $F=3$ and $F=4$.

Atomic magnetometers make use of this \textbf{hyperfine structure} of electronic energy levels, which arises from the magnetic interaction between the nucleus and the electron. The magnetic structure becomes experimentally discernible as a result of the \textbf{Zeeman effect}, whereby the otherwise-degenerate hyperfine energy levels split in the presence of a static magnetic field (linearly, in the low-field regime). Summing over all $N$ atomic spins in the ensemble gives us the bulk magnetization: 
\begin{equation}
    \vec{M} = \sum_{n=1}^N \vec{F_n} \,.
\end{equation}
Both quantities, $\vec{F_n}$ and $\vec{M}$, can be measured experimentally from outside the medium \cite{Cohen-Tannoudji1969,Scott1962}.

The magnetic quantum number $m_F = F, F-1, ..., -F$ measures the atomic-spin component along a given quantization axis. In the absence of optical pumping with laser light (Sec.~\ref{subsec:2.1-OpticalPumping}), each atom's $m_F$-substates will be thermally populated --- corresponding to an unpolarized medium in which the atomic spins are randomly oriented, with an average magnetization given by $\vec{M}=0$. This is the state which the medium always tends to approach, due to various effects that disturb the atomic spins and thus randomize them. If we are able to put most of the atoms into one of the extreme hyperfine substates $m_F = \pm F$, we have achieved macroscopic spin polarization along the direction of the quantization axis --- % also known as a ``coherent spin state'' (Sec.~\ref{subsec:2.2-SpinRelaxation}) or, classically, bulk magnetization 
which manifests classically as bulk magnetization
--- and we say that there is \textbf{orientation} of the atomic-spin ensemble. The spin state of an individual %\AMF{I think ?} 
fully polarized atom, in $m_F = \pm F$, is known as a ``stretched state''.

Other interesting situations are possible: 
% by starting with an oriented spin ensemble and then applying an electric field \IBN{There are different ways to create alighnent, including pumping with a linearly polarized optical field. Should we just state that for some magnetometers alignemnt is used, and leave for interested readers to see how?}, one can obtain
some magnetometers utilize atomic-spin
\textbf{alignment} --- similar to orientation in the sense that there is a preferred polarization \textit{axis}, but different in that there is no preferred polarization \textit{direction}. Alignment is important for atomic magnetometers based on the principle of nonlinear magneto-optical rotation (NMOR), described in Sec.~\ref{sec:4-OperatingModes}. The concept of orientation-to-alignment conversion is treated in a didactic and visual way in \cite{Rochester2012}.

% In an unpolarized medium, %namely when 
% where the atomic spins are randomly oriented, the average magnetization is given by $\vec{M}=0$.
% %
% This is the state the medium always approaches, due to different effects that disturb the atomic spins and thus randomize them.

% \begin{figure}
%   \centering
%   \includegraphics[width=10cm]{Section2-BlochEq/setup.pdf}
%   \caption{A) Basic setup of an optically pumped magnetometer. \AMF{I believe ``photodiode'' is one word, as in the main text.} B) Setup for the $M_z$ mode of operation. \AMF{Will the field coils and applied magnetic field be labeled in the figure?}
%     }
% \label{fig:setup}
% \end{figure}

Note that throughout this tutorial and in other atomic-physics literature, the word ``polarization" is commonly used to refer to both atomic-spin polarization and light polarization (direction of the light electric field). We hope that the distinction will be clear from the context, such that this overlapping terminology does not cause confusion.

\subsection{Optical pumping}
\label{subsec:2.1-OpticalPumping}

The first step toward making the medium useful for magnetometry purposes is to create spin polarization, equivalent to magnetizing the medium: orienting a sizeable fraction of the atomic spins in the same direction. 
In nuclear-magnetic-resonance (NMR) experiments this is typically achieved using a strong (few T) magnet, but alkali atoms can be much more efficiently magnetized using optical pumping.
%
%Interact with near-resonant optical fields, transitions between atomic energy levels occur via absorption and stimulated emission, a phenomenon exploited by the experimental technique of optical pumping. This IBN: it is an awkward sentense, and I am not sure it is 100% accurate.
Optical pumping is a field of atomic physics in itself, and a detailed treatment may be found in the canonical reference \cite{Happer1972}. For our purposes it suffices to say that using the Hamiltonian for a light-atom dipole interaction, one can derive the following angular-momentum \textbf{selection rules} for the driven atomic transitions:%IBN: I think we can skip those if we need to cut things down.
\begin{align}
    \Delta L &= \pm 1 \,, \\
    \Delta S &= 0 \,, \\
    \Delta J &= \pm 1, 0 \,, \\
    \Delta F &= \pm 1, 0 \,.
\end{align}
We see that a photon can transfer angular momentum, but it does not interact with the electron spin. The selection rules for the magnetic quantum numbers depend on the polarization of the light. For circularly polarized light, we have
\begin{align}
    \sigma_+\,\text{photon} & \Rightarrow \Delta m_F = +1 \,, \\
    \sigma_-\,\text{photon} & \Rightarrow \Delta m_F = -1 \,.
\end{align}
Here $\sigma_+$ and $\sigma_-$ denote, respectively, right- and left-circular polarization. 
%The light is propagating along the direction of the static magnetic field. 
When we talk about circular polarization, we should be clear about the convention we are using for handedness. If we visualize the light as a classical wave, $\sigma_+$ ($\sigma_-$) polarization means that the polarization vector is rotating in the clockwise (counterclockwise) direction \textit{as seen from the source}. An atom may also decay by spontaneous emission with $\Delta m_F = \pm 1, 0$. Examples of the pumping process in alkali atoms are illustrated in \ref{sec:Sup-OpticalPumping} and in \cite{Auzinsh2008,Pouliot2018}.

Thus, the first interaction we need in our model is the absorption of circularly polarized photons by the atoms, in order to create orientation in the atomic ground state.
%IBN: technically it is crusial that spontaneous emission is isotropic, so we add AM at absorption, but don't change (on average) as atoms decay back down. But that's probably too fine a point :)
%
Each photon carries a quantum of angular momentum which is transferred to the atoms. 
As a consequence, $\vec{M}$ builds up in the direction of the laser beam.
This build-up can be modeled classically by the following differential equation:
\begin{equation}
 \frac{d}{dt}
 \left(\begin{array}{c}
M_x(t)\\
M_y(t)\\
M_z(t)\\
\end{array}\right) =
 \Gamma_p \left(\begin{array}{c}
0-M_x(t)\\
0-M_y(t)\\
M_p-M_z(t)\\
\end{array}\right)
 - \Gamma_r \left(\begin{array}{c}
M_x(t)\\
M_y(t)\\
M_z(t)
\label{eq:bloch1}
\end{array}\right) \,.
\end{equation}
Here, $\Gamma_p$ is the pumping rate, $\Gamma_r$ is the relaxation rate due to as-yet-unspecified relaxation mechanisms (for now assumed to be the same for all magnetization components), and $M_p$ is the maximum magnetization which can be achieved in the absence of relaxation according to this equation, such that eventually $M_p/M_z = 1$.
%the pumping process strives to create.
%
This expression explicitly assumes that the laser-light direction is parallel to the $z$-axis, as shown in Fig.~\ref{fig:setup}.
The pumping rate quantifies how fast the optical pumping works and is proportional to the power of the laser beam.
We see that in the absence of relaxation mechanisms, any magnetization in the $x$- or $y$-directions tends toward 0 with a rate $\Gamma_p$, while the magnetization in the $z$-direction approaches the value $M_p$.

Eq.~\eqref{eq:bloch1} can be simplified by defining an effective relaxation rate $\Gamma = \Gamma_p + \Gamma_r$ and an effective asymptotic (steady-state) magnetization  % $M_a = \Gamma_p/(\Gamma+\Gamma_p) M_p$ 
% $M_a = M_p\,\Gamma_p/(\Gamma+\Gamma_p)$
$M_a = M_p\,\Gamma_p/\Gamma$, which yields
\begin{equation}
 \frac{d}{dt}
 \left(\begin{array}{c}
M_x(t)\\
M_y(t)\\
M_z(t)\\
\end{array}\right) =
 \Gamma \left(\begin{array}{c}
-M_x(t)\\
-M_y(t)\\
M_a-M_z(t)\\
\end{array}\right) \,.
\label{eq:bloch2}
\end{equation}
This means that in the general case of optical pumping along the $z$-direction, any magnetization in the $x$- or $y$-directions tends toward 0 with a rate $\Gamma$, while the magnetization in the $z$-direction approaches the value $M_a$.

Traditionally, the electromagnetic field that creates the magnetization is referred to as \textbf{pump} light, and the one that is used to detect it as \textbf{probe} light. However, in many magnetometer arrangements, light from the same laser plays both roles.
If all atomic spins in the medium are parallel or antiparallel to the direction of the laser beam --- i.e., each atom is in a stretched state, having
%This case is called a stretched state, in which all atoms have
the maximum amount of angular momentum with respect to the light direction --- 
 no atom can absorb a photon, since it would be physically impossible to transfer even more angular momentum to the atom.
Thus, the medium has become transparent to the laser light.
In general, one can detect the direction of atomic magnetization optically, since the amount of light transmitted through the medium is proportional to the projection of $\vec{M}$ onto the light wavenumber $\vec{k}$.
In most systems, transmission is high (absorption is low) for $\vec{M}$ parallel to $\vec{k}$, while transmission is low (absorption is high) for $\vec{M}$ antiparallel to $\vec{k}$. Further discussion of optical detection is provided in Sec.~\ref{subsec:2.5-OpticalDetection}.

% \begin{figure}
%   \centering
%   \includegraphics[width=10cm]{Section2-BlochEq/pumping.pdf}
%   \caption[pumping]{Atomic spin magnetization as it can be measured by monitoring transmitted light power as a function of time. The different curves show the magnetization behavior for different laser powers and thus different pumping rates $\Gamma_p$, defined with respect to the spin relaxation rate $\Gamma_r$. \AMF{Can the figure labels be made larger in general, for readability?}
%     }
% \label{fig:pumping}
% \end{figure}

\begin{figure} [h]
  \centering
  \includegraphics[width=12cm]{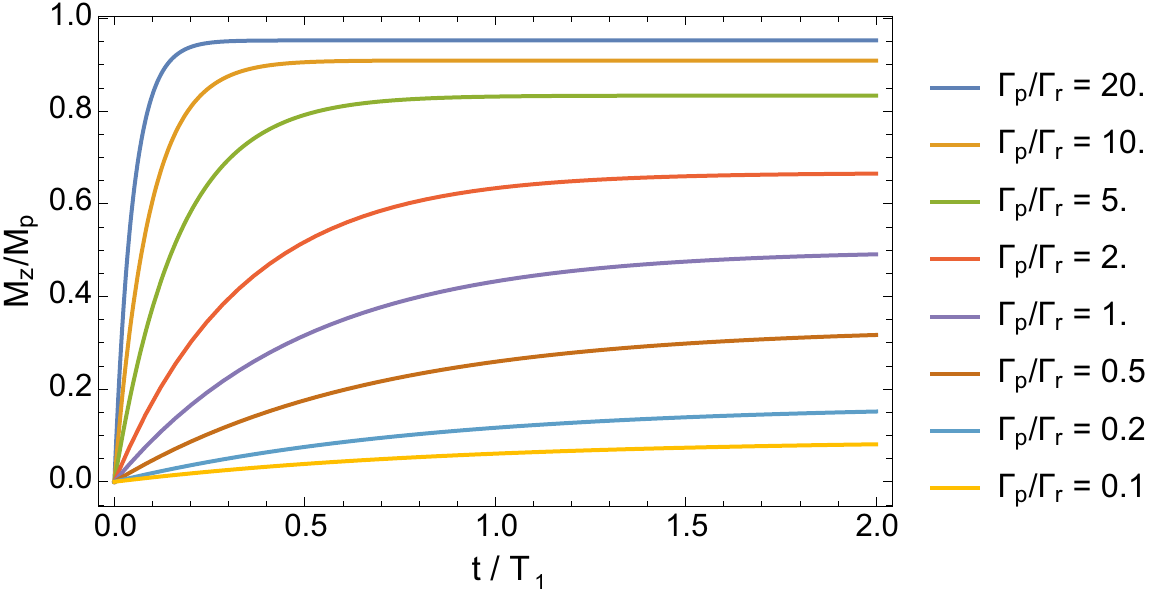}
  \caption[pumping]{Simulation of atomic magnetization as it can be measured by monitoring transmitted light power as a function of time. The different curves show the magnetization behavior for different laser powers and thus different pumping rates $\Gamma_p$, defined with respect to the spin relaxation rate $\Gamma_r$. For $\Gamma_p/\Gamma_r=20$, a spin-ensemble polarization of approximately 95\% is achieved.
    }
\label{fig:pumping}
\end{figure}

Figure \ref{fig:pumping} shows how the magnetization accumulates after optical pumping is started at time $t=0$.
Time is measured in units of the characteristic relaxation time $T_1 \equiv 1/\Gamma_r$,
% $T_2=1/\Gamma_r$ 
and amplitude is measured in units of $M_p$.
The different curves show how the magnetization behavior changes as a function of laser power, proportional to the pumping rate $\Gamma_p$.
All quantities in the graph are dimensionless, and thus the curves are universally applicable when the scaling coefficients $\Gamma_p$ and $M_p$ are known.
Since the transmitted light power is proportional to $\vec{M}\left(t\right)$, the displayed curves can be easily measured.

Solving Eq.~\eqref{eq:bloch2} yields
\begin{equation}
    M_z(t) = M_a \left( 1-e^{-\Gamma\,t} \right) = M_p\, \frac{\Gamma_p}{\Gamma_r+\Gamma_p} \, \left[1-e^{-\left(\Gamma_r+\Gamma_p\right)\,t} \right] \,.
    \label{eq:magnetization}
\end{equation}
With increasing laser power, the asymptotic magnetization $M_a$ comes closer to its maximum value, given by $M_p$.
It is also evident that the time constant of the exponential approach to steady state is shortened by increasing the laser power.
Solving for the steady state, we find that optical pumping alone gives
\begin{equation}
 \frac{d}{dt}
 \left(\begin{array}{c}
M_x(t)\\
M_y(t)\\
M_z(t)
\end{array}\right) = 0 \quad \Rightarrow \quad
\left(\begin{array}{c}
M_x(t)\\
M_y(t)\\
M_z(t)
\end{array}\right) =
\left(\begin{array}{c}
0\\
0\\
M_a
\end{array}\right) \,.
\label{eq:stpumping}
\end{equation}

\subsection{Spin-relaxation times}
\label{subsec:2.2-SpinRelaxation}

Let us assume that we are able to create a high degree of magnetization $M_z/M_p \approx 1$ along the $z$-direction via optical pumping. Looking at Eq.~\eqref{eq:bloch1}, we see that if we then turn off the pump beam at time $t=0$ such that $\Gamma_p = 0$, the longitudinal magnetization decays as
\begin{equation}
    M_z \left( t \right) = M_z \left( 0 \right) \, e^{-\Gamma_r\,t} = M_p \, e^{-t/T_1} \,,
\end{equation}
%where here the relaxation is occurring ``in the dark'' and 
Here we have explicitly introduced the traditional \textbf{spin-depopulation time} $T_1 = 1/\Gamma_r$, which describes how fast the atomic magnetization decays to the unpolarized thermal state --- or, in in the presence of a longitudinal magnetic field, to a nonzero equilibrium magnetization. %\AMF{I'm not sure if this definition is kosher according to Georg's model, but I'm trying to connect somehow to the previous discussion. Irina: I this it is good, I meddled with it a bit more. GB: This is fine with me! For relaxation in the dark T1 is 1/Gamma} 

What happens to the magnetization components transverse to the pumping direction, $M_x$ and $M_y$, in this case?
Until now, we have assumed that the transverse and longitudinal relaxation rates are the same. Although it is possible to add by hand different relaxation rates for each magnetization component in Eq.~\eqref{eq:bloch1}, here we begin to reach the limits of the classical description. According to our modeling thus far, in the absence of any additional electromagnetic fields (Sec.~\ref{subsec:2.3-BlochEq}), after optical pumping we should effectively have $M_x = 0 = M_y$. This is not the full picture, however, since atomic magnetometers are quantum devices which make use of \textbf{spin coherence}. Although the expectation value of each atomic spin along either transverse direction is zero, there are nonzero quantum fluctuations in accordance with the Heisenberg uncertainty principle, as discussed further in Sec.~\ref{subsec:3.1-Sensitivity}. In the density-matrix description of the atomic spin state \cite{Auzinsh2010,Budker2008}, the diagonal elements of the density matrix give the $m_F$-sublevel population probabilities, while the off-diagonal elements represent the coherences (superpositions) among sublevels. These coherences give rise to the transverse spin components, which also decay exponentially with a time constant called the \textbf{spin-(de)coherence time} $T_2$. As explained in Sec.~\ref{subsec:3.1-Sensitivity}, the $T_2$ time is a crucial parameter which defines the sensitivity of an atomic magnetometer. In general, the longer the $T_1$ time in an atomic ensemble, the longer will be the $T_2$ time as well. Although typically $T_1 \geq T_2$, for purposes of calculation it is often convenient to set $T_1 \approx T_2$. In thermal atomic vapor cells, a dominant spin-relaxation mechanism is collision of atoms with the cell walls (Sec.~\ref{subsec:5.1-VaporCell}).

\subsection{Interaction of atoms with a magnetic field}
\label{subsec:2.3-BlochEq}

Apart from optical pumping, the second interaction that drives an atomic magnetometer is the effect of external magnetic fields on the magnetization $\vec{M}$ of the medium.
A magnetic field acting on polarized atoms
induces a torque that aims to align the magnetization with the field lines.
Since the magnetization results from atomic spins with angular momentum, this torque induces a precession motion of the spins, known as Larmor precession.
The precession frequency or \textbf{Larmor frequency} $\Omega$ is proportional to the magnetic field, and it is convenient to express all magnetic-field components in terms of precession frequencies.
For example, we define for the $x$-component $\Omega_x = \gamma B_x$, with the gyromagnetic ratio $\gamma$ (see %\todo{Table XXX} and 
\cite{Steck2021} for the $\gamma$ values of different alkali atoms).
This allows us to model the evolution of $M$ in the following way:
\begin{equation}
 \frac{d}{dt}
 \left(\begin{array}{c}
M_x(t)\\
M_y(t)\\
M_z(t)\\
\end{array}\right) =
\left(\begin{array}{c}
M_x(t)\\
M_y(t)\\
M_z(t)\\
\end{array}\right) \times \left(\begin{array}{c}
\Omega_x\\
\Omega_y\\
\Omega_z\\
\end{array}\right)
 + \Gamma \left(\begin{array}{c}
-M_x(t)\\
-M_y(t)\\
M_a-M_z(t)\\
\end{array}\right).
\label{eq:bloch3}
\end{equation}
The first part of this expression is the classical Bloch equation, which has been extended to model optical pumping along the $z$-direction as well as relaxation of the magnetization.
The equation is useful for a basic description of virtually all processes in optically pumped atomic magnetometers. 
%and will therefore be used throughout this paper.
%
For example, it defines how the optical absorptive and dispersive properties of the atomic medium are modified, which enables optical detection of the change in spin state as discussed in Sec.~\ref{subsec:2.5-OpticalDetection}. 

More detailed descriptions of the atom-field interaction can be derived from quantum density-matrix calculations \cite{Auzinsh2010,Budker2008,Blum2012}. %\todo{cite Blume book, Mathematica library}.

% How do we characterize the spin lifetime of our atoms after they have been polarized via optical pumping? Assuming simple exponential decay, we have for the longitudinal magnetization component
% \begin{equation}
%  \frac{\text{d} \vec{M}_z}{\text{d}t} = - \beta\,\vec{M}_z \,,
% \end{equation}
% where $\beta$ is a decay constant with units of $s^{-1}$. Solving the differential equation and defining the decay time $T_1 = 1/\beta$, we find the behavior $\vec{M}_z\left(t\right) = \vec{M}_z\left(0\right) e^{-t/T_1}$.
% %
% We refer to $T_1$ as the \textbf{spin-depopulation time}. The decay time $T_2$ of the transverse magnetization components $\vec{M}_x$ and $\vec{M}_y$ (\textbf{spin-decoherence time}) is defined analogously. 
% % To measure $T_1$, we use a setup like the one shown in \todo{FIGURE}. Measurement of $T_2$ can be performed using the magnetometer setup, since the $T_2$ time is proportional to the width of the magnetic resonance. A typical $M_z$ magnetometer scheme is depicted in Table~\ref{Table:magn_overview}.}
% \todo{Connect to the relaxation rates already defined.}

\subsection{A specific example: Measuring the transverse magnetic field}
%the longitudinal and transverse magnetization compoments %magnetic fields %($M_z$ magnetometer)}

Let us have a look at some special cases.
First, we assume that we have a magnetic field parallel to the laser beam.
In this situation, the pumping process creates magnetization only along the magnetic-field direction.
The field induces precession around an axis given by the field direction, which in this case does not change the magnetization at all.
Due to the cross-product in Eq.~\eqref{eq:bloch3}, the $z$-component of the field mixes the $x$- and $y$-components of the magnetization, which are zero in this case.
Thus, the situation is stable, and we find the same result as in Eq.~\eqref{eq:stpumping}.

The situation changes when we apply a magnetic field along a direction perpendicular to $z$ --- for example, the $x$-direction.
When we solve for the steady state (by setting the time derivatives equal to zero) with $\Omega_y=0$ and $\Omega_z=0$, Eq.~\eqref{eq:bloch3} simplifies to
\begin{equation}
%   \left(\begin{array}{c}
% 0\\
% 0\\
% 0
% \end{array}\right) =
 \frac{d}{dt}
 \left(\begin{array}{c}
M_x(t)\\
M_y(t)\\
M_z(t)\\
\end{array}\right) =
\left(\begin{array}{rcl}
&-&\Gamma\,M_x \\
M_z \,\Omega_x &-&  \Gamma\, M_y\\
- M_y\, \Omega_x &+& \Gamma\, (M_a-M_z) \\
\end{array}\right) =
  \left(\begin{array}{c}
0\\
0\\
0
\end{array}\right) \,.
\label{eq:bloch4}
\end{equation}
Apart from the trivial result $M_x=0$, this equation contains the essence of many atomic magnetometers.
In the term for $M_y$, two contributions have to balance each other in the steady state.
The first contribution is a source term that increases the modulus of $M_y$ proportionally to $\Omega_x$, by rotating magnetization from the $z$-direction to the $y$-direction.
The second contribution is the relaxation, which decreases the modulus of $M_y$.
%
%Both terms describe a rate of change for $M_y$, and they have to cancel each other in the steady state.

\begin{figure} [h]
  \centering
  \includegraphics[width=10cm]{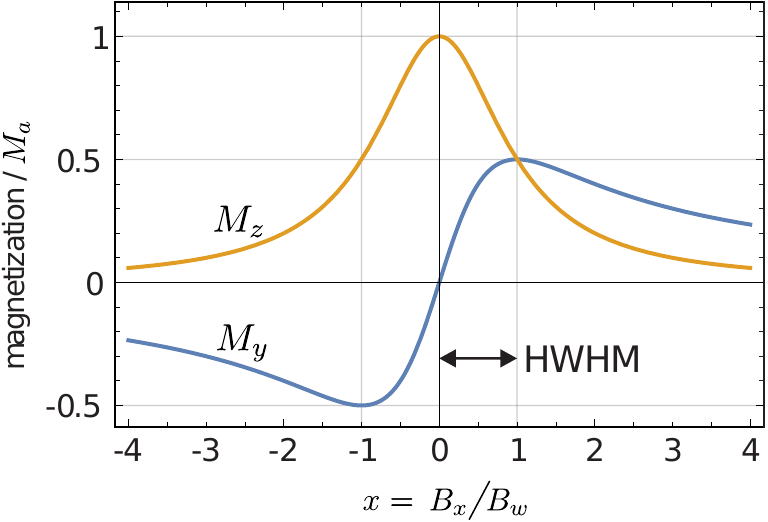}
  \caption{Magnetic-resonance spectrum for the case $B_y=B_z=0$. The $M_y$ and $M_z$ components of the magnetization show resonant behavior as a function of $B_x$. The scale for the $x$-axis is given by the magnetic linewidth $B_w$.
  $M_y$ displays a dispersive and $M_z$ an absorptive Lorentzian resonance. 
    }
\label{fig:lorentzian}
\end{figure}

The term for $M_z$ has similar contributions, which here have reversed roles.
The first one models the loss of $M_z$ as it is rotated to $M_y$, and the second one is a relaxation towards the asymptotic magnetization $M_a$.
Solving the system of equations yields
\begin{equation}
M_y= M_a\,\frac{\Gamma \, \Omega_x }{\Gamma^2 + \Omega_x^2} = M_a\,\frac{x }{1 + x^2} \,, \quad \quad
M_z= M_a\,\frac{\Gamma^2}{\Gamma^2 + \Omega_x^2} = M_a\,\frac{1 }{1 + x^2} \,,
\label{eq:lorentzian}
\end{equation}
where $x = \Omega_x / \Gamma$ is a dimensionless variable that expresses the magnetic-field component scaled to the corresponding precession frequency and divided by the relaxation rate.
Both $M_y$ and $M_z$ show resonant behavior, and the dimensionless parameterization allows us to generate the universally applicable plots shown in Fig.~\ref{fig:lorentzian}.

The amplitude of both resonances is given by the asymptotic magnetization $M_a$.
When $B_x = 0$ and consequently $x=0$, $M_z$ reaches $M_a$ and $M_y$ vanishes.
This is the same situation as in Eq.~\eqref{eq:stpumping} for optical pumping alone.
The magnetization $M_z$ shows an absorptive Lorentzian peak and thus gets smaller with an increasing magnitude of $B_x$.
At $x=\pm 1$, we find $M_z = M_a/2$; this means that the half width at half maximum (HWHM) gives the points where the modulus of the precession frequency $\Omega_x$ is equal to the relaxation rate $\Gamma$.
We can define the magnetic linewidth $B_w$ by rearranging the terms that define $x$:
$$
x = \frac{\gamma B_x}{ \Gamma} = \frac{B_x}{B_w} \,, \quad \text{with} \quad B_w = \frac{\Gamma}{\gamma} \,.
$$
This concept is not limited to the $x$-component of $B$, as we shall see in Sec.~\ref{subsec:5.4-Operation}.

The $M_y$ component follows a dispersive Lorentzian curve with a linear zero crossing.
Close to $x=0$, the $M_y$ component is thus proportional to the applied magnetic field. The larger the change in magnetization signal (relative to noise) produced by a small change in the magnetic field, the more sensitive the magnetometer. 
This exact method is used in magnetometers based on the ground-state Hanle effect, also known as zero-field resonance, which often operate in the spin-exchange-relaxation-free (SERF) regime (see Sec.~\ref{sec:4-OperatingModes}).
%
%In sensitive magnetometers, such signals proportional to the magnetic field show a large signal change, relative to the noise in the signal, for a small change in magnetic field.

\subsection{Optical detection}
\label{subsec:2.5-OpticalDetection}

The propagation of light through an atomic medium can be modeled semiclassically, whereby the electronic energy levels are quantized but the light is treated as a classical electromagnetic field in which the electrons are oscillating dipoles \cite{Milonni2010}. Based on its electric polarizability, the atomic medium has a complex index of refraction for light:
\begin{equation}
    n\left(\omega\right) = n_{\text{Re}}\left(\omega\right) + i\,n_{\text{Im}}\left(\omega\right) \,,
\end{equation}
expressed as a function of the angular frequency $\omega$ of the incident light. Here $n_{\text{Re}}$ defines the real index of refraction, characterized by dispersive behavior as a function of $\omega$; $n_{\text{Im}}$ defines the absorption, characterized by resonant behavior at the resonance frequency $\omega=\omega_0$. At resonance, $n_{\text{Re}}=0$ and $n_{\text{Im}}=1$, in analogy to the behavior of the magnetization curves in Fig.~\ref{fig:lorentzian}. This duality enables two different modes of optical detection in atomic magnetometry, either absorptive (based on light transmission) or dispersive (based on light polarization rotation).

An external magnetic field creates birefringence in the atomic medium, such that the $\sigma_+$ and $\sigma_-$ polarization components of incident light along the magnetic-field direction experience different values of $n$. Depending on the detection mode, this manifests as a change in absorption or polarization rotation of the transmitted probe light. In both cases, the change is proportional to the projection of the atomic magnetization in the direction of light propagation, from which the Larmor frequency can be extracted.

%\clearpage
%\pagebreak
\section{Define your requirements: Figures of merit}
\label{sec:3-Requirements} 
% \textbf{Anne's Sec.}
% \bigskip

% \noindent ADD REFERENCES
% \bigskip

When designing a magnetometer, the following questions should first be considered in order to select an appropriate operating mode (Sec.~\ref{sec:4-OperatingModes}). They relate the application requirements to associated magnetometer performance figures, some of which were already introduced in Sec.~\ref{sec:1-Intro}.
\begin{enumerate}
    \item What are the amplitude and frequency ranges of the signals I want to measure? $\rightarrow$ \textbf{sensitivity, bandwidth}
    \item How close must the sensing volume be to the sample? How well do I need to localize the measured signals? $\rightarrow$ \textbf{standoff distance, spatial resolution}
    \item Am I interested in measuring absolute or relative field values? $\rightarrow$ \textbf{accuracy}
    \item Over what time scale do I plan to measure? $\rightarrow$ \textbf{stability}
\end{enumerate}

\subsection{Sensitivity and noise}
\label{subsec:3.1-Sensitivity}

Sensitivity to magnetic fields is arguably the most important benchmark used to characterize magnetometer performance. In atomic magnetometry, sensitivity is typically reported in units of $\text{T}/\sqrt{\text{Hz}}$ (SI) or $\text{G}/\sqrt{\text{Hz}}$ (CGS). Although these units may appear mysterious to newcomers, they can be understood by noting that magnetic-field noise fluctuates over the frequency bandwidth of a magnetometer. In order to measure sensitivity, one records the magnetometer response for some time duration. To convert from the time domain to the frequency domain, a fast Fourier transform is performed to produce a magnetic-field-noise power spectrum (see Sec.~\ref{sec:6-Characterization}). Calculating root-mean-square (RMS) amplitudes of the noise in 1\,Hz bins leads to a sensitivity expressed per $\sqrt{\text{Hz}}$. Note that although both sensitivity and \textbf{resolution} are essentially a measure of the smallest change in magnetic field detectable by the sensor (at a given frequency), the units are different --- resolution is reported in magnetic-field units alone.

For measurement of relatively small magnetic fields, we would like the sensitivity of our magnetometer to be as good as possible --- in the ideal case, limited only by fundamental quantum mechanics. It is important to identify the various contributions to noise in our measurements, and to understand how they behave.

The basic sensitivity limits for an atomic magnetometer are set by the Heisenberg uncertainty principle, which manifests in the \textbf{spin-projection noise} $\delta B_{\text{PN}}$ of the atoms and the \textbf{shot noise} $\delta B_{\text{SN}}$ of the probe light. These contributions to the magnetometer sensitivity scale as \cite{Budker2007}
\begin{align}
    \delta B_{\text{PN}} & \propto \frac{1}{\sqrt{N\,T_2\,t}} \,,
    \label{eq:sensitivityPN}
    \\
    \delta B_{\text{SN}} & \propto \frac{1}{\sqrt{\Phi\,t}} \,,
    \label{eq:sensitivitySN}
\end{align}
where $N$ is the number of atoms in the ensemble, $T_2$ is the spin-coherence time of the ensemble introduced in Sec.~\ref{subsec:2.2-SpinRelaxation}, $\Phi$ is the photon flux of the probe light, and $t$ is the total measurement time. We see that noise decreases the longer we measure, corresponding to a better sensitivity and underscoring the value of averaging over many measurement cycles. In order to express Eqs.~(\ref{eq:sensitivityPN}--\ref{eq:sensitivitySN}) as noise densities (per $\sqrt{\text{Hz}}$), we can multiply by $\sqrt{t}$.

Spin-projection noise is a type of \textbf{atomic noise} that ultimately limits the sensitivity of any atomic magnetometer regardless of the particular modality or detection method. In order to understand this limit, it is instructive to consider a single atom. Imagine that we prepare the atom in a stretched state along the $z$-direction, as described in Sec.~\ref{sec:2-BlochEq}, and then measure the atomic-spin projection $m_F$ along some axis. If this experiment is repeated a number of times, we find the expectation value $\left<F_z\right>=F$. Along the $x$- and $y$-directions we measure zero on average, but each individual measurement yields a random result according to the uncertainty principle:
\begin{equation}
    \Delta F_x \, \Delta F_y \geq \frac{\left<F_z\right>}{2} \,,
\end{equation}
which in the case of the minimum-uncertainty stretched state becomes
\begin{equation}
    \Delta F_x \, \Delta F_y = \frac{F}{2} \,.
    \label{eq:MinUncertainty}
\end{equation}
%(Recall that for any two observables $\hat{A}$ and $\hat{B}$, we must have ${\Delta \hat{A} \Delta \hat{B} \geq | \langle [ \hat{A}, \hat{B} ] \rangle | /2}$, where ${\Delta A = [ \langle \hat{A}^2 \rangle - \langle \hat{A} \rangle^2 ]^{1/2} = [ \text{Var} ( \hat{A} ) ]^{1/2}}$.) Irina: I really feel like that's too much of introductory information! 
A derivation of Eq.~\eqref{eq:sensitivityPN} may be found in ~\ref{sec:Sup-Sensitivity}. As explained in quantum-optics textbooks, it is possible to beat the limit set by Eq.~\eqref{eq:MinUncertainty} for $\Delta F_x$ or $\Delta F_y$ (but not both simultaneously) via a technique known as \textbf{spin squeezing}, such that the magnetometer sensitivity may be improved~\cite{Kominis2008}. Squeezing of the probe light may also be employed to beat sensitivity limits set by photon shot noise. Such ``quantum-enhanced'' atomic magnetometry is an active area of research \cite{Kuzmich1997,Auzinsh2004,Wolfgramm2010,Sewell2012,Horrom2012,Troullinou2021,Li2022}.

Another quantum-mechanical source of atomic noise is \textbf{back-action noise}. This stems from the fundamental problem of not being able to measure a quantum system without disturbing it. In the case of atomic magnetometry, we use light as a detection tool since we cannot read out the atomic spin states directly. This interference by the experimenters results in ``back action'' of the light onto the atoms --- essentially an AC Stark shift, also known as light shift, caused by the probe light which shifts the hyperfine energy levels and changes the Larmor precession frequency. In effect, a ``fictitious magnetic field'' is generated on top of the real one that we actually want to measure. Various evasion techniques have been developed for reducing or eliminating back-action noise \cite{Wasilewski2010,Vasilakis2011}. Since, in contrast to shot noise (Eq.~\eqref{eq:sensitivitySN}), the contribution of back-action noise to sensitivity scales as ${\sqrt{\Phi}}$, one does not generally gain by increasing probe power --- which can also introduce additional atomic noise due to undesirable depumping effects.

In practice, there may be additional systematic sources of noise that worsen the sensitivity of a magnetometer, including \textbf{electronic noise} and \textbf{technical (classical) noise}.
%, although these are not fundamental limits.
% For detection schemes based on polarization rotation of probe light, Using a balanced-polarimetry detection scheme \todo{[REFS]} helps to suppress classical noise in the probe light, such that the optical readout is hopefully limited primarily by shot noise.
% \IBN{I think back-action is often a part of the fundamental quantum limit, so I am not sure I agree with the last statement. Also - not too be too pedantic - some people consider projection noise more fundamental -- since it limits the sensitivity regardless of the detection method, while the photon shot noise is more application-dependent.}
% \AMF{I agree and am in the process of overhauling the noise discussion.}
These categories are, of course, somewhat artificial, as some electronic noise may be classical. Furthermore, some classical noise may be atomic --- according to our definition, atomic noise is anything that causes \textbf{broadening} of the magnetic resonance. Since the resonance linewidth is inversely proportional to the spin-coherence time $T_2$, sensitivity worsens in the presence of any such broadening. One common source of classical atomic noise is magnetic-field gradients. Systematics are discussed further in Sec.~\ref{subsec:4.2-MagPerformance}.

Whether or not it is necessary to suppress systematic noise sources enough to achieve so-called quantum-noise-limited operation depends on the application requirements. In any case, it is considered best practice to remove electronic noise from the photodetector signal by measuring and subtracting dark counts. For detection modalities based on light polarization rotation (Sec.~\ref{sec:4-OperatingModes}), many classical noise sources can be removed via homodyne-detection schemes, a common technique in quantum optics \cite{Fox2006}. Absorption-based detection modalities, while generally simpler to implement, are more prone to systematics due to, e.g., laser-intensity noise.

Going back to Eq.~\eqref{eq:sensitivityPN}, we also see that in order to optimize sensitivity --- i.e., to make it as small as possible --- one should increase both the number of atoms and the coherence time inside the vapor cell. The latter can be accomplished in two ways: (1) by using \textbf{antirelaxation wall coating} that has low absorption energy for the atoms, so that they spend less time at the cell walls, or (2) by using a \textbf{buffer gas} to prevent atoms from diffusing to the walls. Both these techniques, which are discussed further in Sec.~\ref{subsec:5.1-VaporCell}, are effective because spin relaxation at the cell walls is a key relaxation mechanism \cite{Bouchiat1966,Scholtes2014}. A popular way to increase the density of atoms is by heating the vapor cell --- SERF magnetometers work at temperatures over 100$^\circ$C (Sec.~\ref{sec:4-OperatingModes}).

\subsection{Bandwidth and spatial performance}
\label{subsec:3.2-Bandwidth_SpatialRes}

One often hears the statement that in magnetometry, there exists a trade-off between sensitivity and bandwidth, i.e.\ that improving the sensitivity comes at the cost of reduced bandwidth and vice versa. This is because these two parameters tend to have an opposite dependence on the  the width of the atomic resonance. Narrow resonances may provide better sensitivity (which is good), but smaller bandwidth (which is usually bad). %Keeping in mind that, in general, a smaller sensitivity is \textit{better} while a smaller bandwidth is \textit{worse}, these two performance figures can be thought of as having opposite dependence on the coherence time. 
%\GB{Would it help to make the connection between sensitivity and statistical uncertainty?} \AMF{Probably, but I'm not sure what this means.}
Another way to think about this intuitively is to recall that bandwidth gives us information about how quickly a magnetometer responds to sudden changes in magnetic field: the smaller the change to which the magnetometer is sensitive, the longer it takes for the magnetometer to react to this change. A directly related concept is \textbf{dynamic range}, or the range of field magnitudes that can be measured within the magnetometer response bandwidth.

The sensitivity-bandwidth trade-off is not the only one arising in the context of magnetometer performance figures. 
% GB: I prefer tradeoff over paradox
We have already seen that we want to maximize the number of atoms in the sensing volume, as per Eq.~\eqref{eq:sensitivityPN}. Naively, this suggests that we should make the sensing volume as large as possible. Regardless of the cell volume, for atoms in thermal motion, the probe light effectively interacts with all atoms equally, such that the motion of individual atoms averages out and we can think of the light as addressing a homogeneous atomic ensemble, a principle known as motional averaging. In the case of buffer-gas cells, where the atoms are prevented from diffusing to the cell walls, the size of the probe beam is typically chosen to interact with a majority of atoms in the cell. However, a larger volume tends to increase the sensor standoff distance and decrease the spatial resolution of measurements. 
% For atoms in thermal motion, the probe light effectively interacts with all atoms equally --- such that the motion of individual atoms averages out and we can think of the light as addressing a single \IBN{uniform? homogeneous?} atomic ensemble, a principle known as motional averaging. In the case of buffer-gas cells, the size of the probe beam is typically chosen to interact with as many atoms as possible. \IBN{I think the previous two sentences can be removed as they are tangential to the discussion, imho.} \AMF{Perhaps we can move them elsewhere.}

More precisely, the standoff distance of a sensor is generally defined as the distance between the sample of interest and the geometric center of the sensing volume; the spatial resolution is the smallest spatial variation in the measured signal that can be resolved. Thus, we may run into trouble if we want to localize signals to within distance scales smaller than the diameter of the sensing volume. One way around this is to decrease the size of the sensing volume but increase the atomic density via heating, ideally in the SERF regime. Arrays of miniaturized vapor cells can be used to obtain excellent spatial localization and mapping of magnetic signals (Secs.~\ref{subsec:4.2-MagPerformance} and \ref{subsec:5.1-VaporCell}). An alternative way to effectively increase the atomic density, by actually increasing the light-atom interaction, is to place the vapor cell in an optical cavity or to use a multipass cell \cite{Sheng2013,Vasilakis2014,Cai2020,Lucivero2021}.

\subsection{Stability and accuracy}

Stability essentially refers to how long a magnetometer can operate continuously, or continually in pulsed mode, before it needs to be recalibrated. For example, some popular commercially available sensors operate at zero field via integrated field coils which null the local magnetic field. However, any drifts in the local magnetic-field environment over time eventually push the sensor out of its sensitive range, limiting the duration of continuous magnetometer operation. For longer-term sensor usage demanding greater stability with limited power consumption, e.g. in non-lab settings, a self-oscillating mode of operation is preferred which renders the magnetometer insensitive to such drifts; the NMOR magnetometers discussed in Sec.~\ref{sec:4-OperatingModes} are ideal in this case. Feedback control of various other time-dependent sensor parameters, such as laser frequency and cell temperature, is typically also required for practical operation. A standard way to quantify the stability of a magnetometer, or other frequency-based measurement device, is to calculate the Allan variance over time \cite{Allan2012}. %\AMF{Irina, what was the website of Allen that you suggested?} 

Finally, the accuracy of a sensor has to do with whether we are measuring absolute or relative magnetic-field values --- in the latter case, this performance figure is less critical. In essence, magnetometers measure frequencies which must be converted to corresponding magnetic-field values. For some applications, such as precise mapping of the geomagnetic field and other coarse magnetic structures, or precise measurement of physical constants, accuracy is paramount and may outrank sensitivity as the dominant figure of merit \cite{Leger2015,Mora2019}. Since atoms form the heart of the magnetometers under consideration here, these can be used as an absolute reference during magnetometer calibration, in order to improve accuracy. In practice, however, various systematics may worsen measurement accuracy during magnetometer operation (Sec.~\ref{subsec:4.2-MagPerformance}). 

% \bigskip

% \noindent Original suggested outline:
% \begin{itemize}
%     \item Figures of merit---requirements, characterization
%     \item Choice of alkali vapor
%     \item Vapor cells, laser systems, magnetic shielding
%     \item Buffer gas v. antirelaxation coating; microfabrication
% \end{itemize}

%\pagebreak
\section{Choose your operating mode}
\label{sec:4-OperatingModes} 
% \textbf{Irina's section}
%\bigskip

%\noindent Suggested outline:
%\begin{itemize}
%    \item Decode jargon---$M_z$/$M_x$, ground-state Hanle, Bell-Bloom, RF, NMOR, SERF (close to historical order)
%    \item Scalar v. vector
%    \item Zero-field v. finite-field
%    \item Gradiometry
%    \item Examples of applications
%    \item Miniaturization
%\end{itemize}
%\bigskip

As discussed in Sec.~\ref{sec:2-BlochEq}, most atomic magnetometers have a similar underlying physical principle of operation, relying on a resonant change in either light absorption or rotation of the light polarization to determine the Larmor precession frequency and extract the magnetic-field value. Yet, this operational principle allows for many different configurations, each having its own strengths and weaknesses. The choice of a particular approach is usually dictated by the desired magnetometer performance and other application-specific restrictions. In this section, we provide a brief comparative analysis of the main approaches for magnetic-field measurements and their basic performance characteristics, as summarized in Table~\ref{tab:magn_overview}.

\renewcommand{\arraystretch}{2}
\begin{table}[h] 
\tiny
\centering
\begin{tabular}{|p{0.13\textwidth}|>{\raggedright}p{0.11\textwidth}|>{\raggedright}p{0.11\textwidth}|>{\raggedright}p{0.11\textwidth}|>{\raggedright}p{0.11\textwidth}|>{\raggedright}p{0.11\textwidth}|>{\raggedright\arraybackslash}p{0.11\textwidth}|}
\hline
\textbf{Magnetometer type} & \textbf{Mz} & \textbf{Mx} & \textbf{Bell-Bloom} & \textbf{Hanle/SERF} & \textbf{NMOR}  & \textbf{CPT/EIT} \\ 
\hline\hline
\textbf{Typical geometry} &                 \begin{minipage}{.12\textwidth}
      \includegraphics[width=\linewidth]{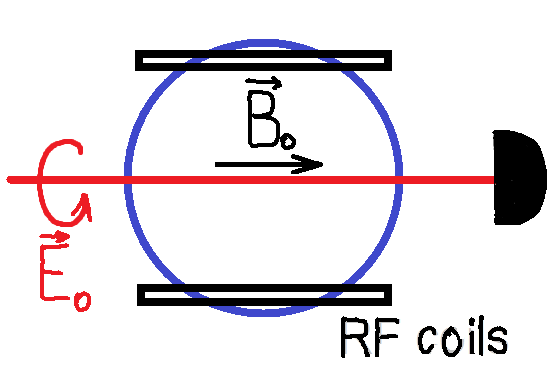}
    \end{minipage} &  \begin{minipage}{.12\textwidth}
      \includegraphics[width=\linewidth]{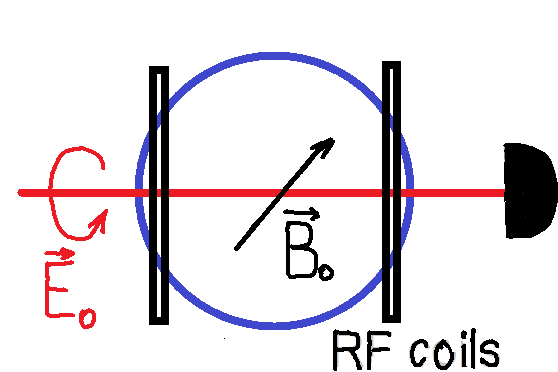}
    \end{minipage} & \begin{minipage}{.12\textwidth}
      \includegraphics[width=\linewidth]{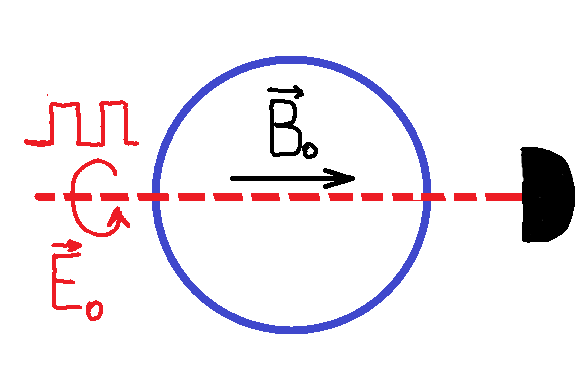}
    \end{minipage} & \begin{minipage}{.12\textwidth}
      \includegraphics[width=\linewidth]{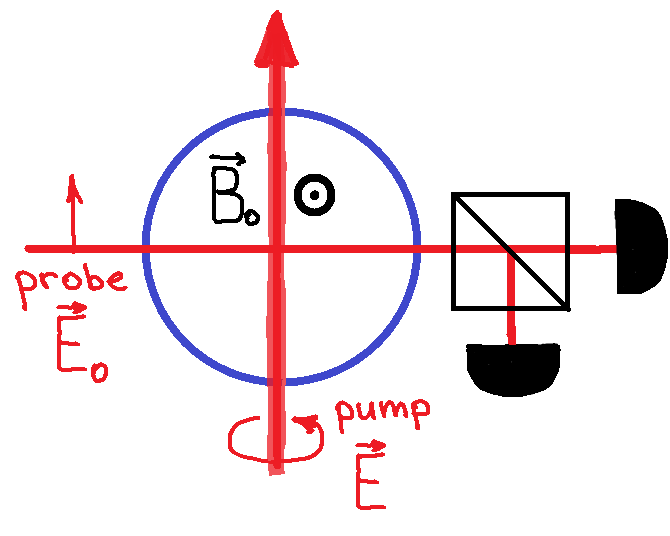}
    \end{minipage} &  \begin{minipage}{.12\textwidth}
      \includegraphics[width=\linewidth]{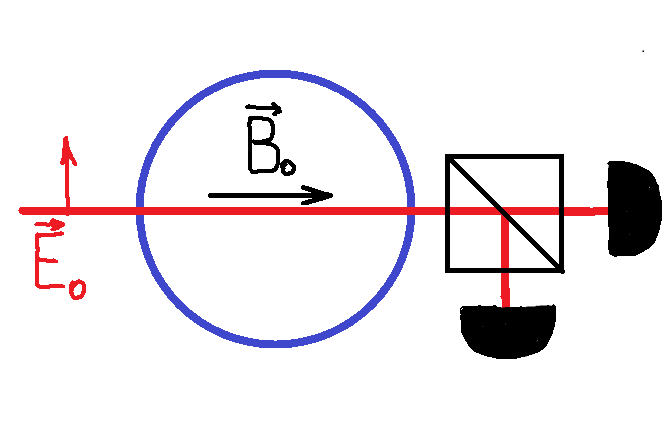}
    \end{minipage} & \begin{minipage}{.12\textwidth}
      \includegraphics[width=\linewidth]{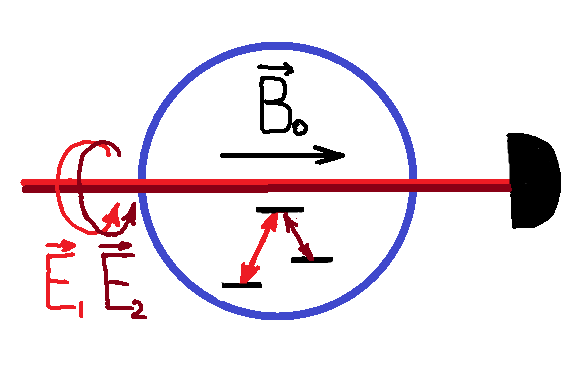}
    \end{minipage} \\ 
\hline
\textbf{Operational principle} & Circularly polarized light $\vec{E}_0$, parallel to the bias magnetic field $\vec{B}_0$,  magnetizes atoms along the $z$-axis. An orthogonal resonant RF field at the Larmor frequency produces a narrow absorption resonance~\cite{Bloom1962,Alexandrov1992}.
& Circularly polarized light $\vec{E}_0$ propagates at an angle to the bias magnetic field $\vec{B}_0$ and along the oscillating RF field. Light absorption is defined by the transverse magnetization component~\cite{Bloom1962,Groeger2006EPJD,Aleksandrov1995,Tiporlini2013}.  
& Intensity modulation of the laser light $\vec{E}_0$ produces stroboscopic optical pumping when modulated at the Larmor frequency~\cite{Allred2002,Bell1961}.
& Circularly polarized pump light $\vec{E}$ sets the atomic magnetization, whose orientation imprints on the polarization angle of a weak probe $\vec{E}_0$. Often operates in the spin-exchange relaxation-free (SERF) regime~\cite{Sheng2013,Li2018}.   
& Linearly polarized light $\vec{E}_0$ aligns the atoms, whose precession in the magnetic field $\vec{B}_0$ imprints back onto the light polarization angle~\cite{Budker2000,Budker2002RMP,Rosner2022}. Use of the modulated light enables operation at nonzero magnetic field~\cite{Budker2002PRA,Gawlik2006}. 
& A bichromatic optical field $\vec{E}_{1,2}$ prepares atoms in a superposition of magnetosensitive hyperfine sublevels with increased transmission (``dark state'')~\cite{Nagel1998,Staehler2001}. \\ 
\hline
\textbf{Sensitivity}, $\mathrm{fT}/\sqrt{\mathrm{Hz}}$
& $10$~\cite{Schultze2017} 
& $1$~\cite{Groeger2006EPJD} 
& $20$~\cite{Bevilacqua2019PRA,Zhang2020,Yang2021} 
& $0.16$~\cite{Dang2010}
& $70$~\cite{Lucivero2014}
& 10,000~\cite{Staehler2001,Matsko2005} \\ 
\hline
\textbf{Bandwidth}, kHz
& $0.6$~\cite{Schultze2017}
& $1$~\cite{Groeger2006SAA,Shah2010}
& $1$~\cite{Yang2021}
& $1$~\cite{Shah2010,Shah2007}
& $>100$~\cite{Li2020}
& $>0.1$~\cite{Belfi2007,Ishikawa2021} \\ 
\hline
\textbf{Can operate at Earth field?}  &  yes  &  yes  &  yes  &  no  &  yes  &  yes \\ 
\hline
\textbf{Has been modified for vector measurement? }  &  no  &  \cite{Yang2020}  &  \cite{Huang2015,Ding2018,Cai2020}  &  \cite{Seltzer2004,Papoyan2016}  &  \cite{Patton2014,Pyragius2019}  &  \cite{Yudin2010} \\ 
\hline
\textbf{Miniaturization}: cell volume $|$ sensor volume $|$ sensitivity 
& 1\,mm$^3$ $|$ 22\,cm$^3$ $|$ 15$\mathrm{pT}/\sqrt{\mathrm{Hz}}$ \cite{Korth2016,Oelsner2022}
& 2\,mm$^3$ $|$ 25\,mm$^3$ $|$ 5$\mathrm{pT}/\sqrt{\mathrm{Hz}}$ \cite{Schwindt2007,Korth2016,Jimenez-Martinez2010}
& 16\,mm$^3$ $|$ NA $|$ 0.07$\mathrm{pT}/\sqrt{\mathrm{Hz}}$ \cite{Gerginov2017,Jimenez-Martinez2010,Gerginov2020}
& 6\,mm$^3$ $|$ NA $|$ 0.07$\mathrm{pT}/\sqrt{\mathrm{Hz}}$ \cite{Shah2007,Zhang2021IEEE}
& $7\,\cdot\,10^{-4}$\,mm$^3$ $|$ NA $|$ 700$\mathrm{pT}/\sqrt{\mathrm{Hz}}$ \cite{Sebbag2021}
& 1\,mm$^3$ $|$ 12\,mm$^3$ $|$ 50$\mathrm{pT}/\sqrt{\mathrm{Hz}}$ \cite{Schwindt2004,Hong2021} \\ 
\hline
\end{tabular}
\caption{Brief summary of the main atomic-magnetometer technologies. } \label{tab:magn_overview}
\end{table}
\renewcommand{\arraystretch}{1}
 
\subsection{Common types of atomic magnetometers}
\label{subsec:4.1-OperatingModes}

We saw in Sec.~\ref{subsec:2.1-OpticalPumping} that many atomic magnetometers rely on circularly polarized light, tuned near one of the atomic optical transitions, to optically pump atoms and create the desired magnetization of the atomic ensemble. Among these devices, summarized in the first four columns of Table~\ref{tab:magn_overview}, the most common are probably $M_z$ and $M_x$ magnetometers, so named for the magnetization component to which the device is sensitive. Both of them utilize a resonant radiofrequency (RF) field to induce an optical absorption resonance. The other two magnetometers of this type are all-optical: in Bell-Bloom magnetometers, amplitude modulation of the optical field plays the role of the effective RF interrogation, and, unlike others, this type of magnetometer can operate as an active device if the output signal of a photodetector is fed back into the modulation circuit. Finally, Hanle magnetometers use a pump-probe geometry, in which atomic spins are oriented with a circularly polarized strong pump optical field, and then their precession in the bias magnetic field is probed through the polarization rotation of a weaker probe optical field. The most successful realization of the Hanle magnetometer operates in the SERF regime, which has now become a well-recognized magnetometer type in its own right. The magnetometer operating principles in the first four columns of Table~\ref{tab:magn_overview} may be loosely categorized under the umbrella of ``magneto-optical double resonance'' --- a general term which refers to the dual combination of optical and magnetic resonances, manifested physically when the frequency of an applied or effective RF field matches the Larmor frequency. This concept and its experimental demonstration in atomic spectroscopy predate magnetometry applications \cite{Brossel1952,Corney2006}.

By contrast, the devices in the last two columns of Table~\ref{tab:magn_overview} rely on evolution of a quantum superposition of atomic spin states, rather than on magnetization. Magnetometers based on nonlinear magneto-optical rotation (NMOR) generally employ linearly polarized light to create atomic alignment (Sec.~\ref{sec:2-BlochEq}) --- a quadrupole magnetic moment, corresponding to a coherent superposition of Zeeman sublevels --- and then detect the optical polarization rotation caused by alignment evolution in the magnetic field to be measured. Typically, such measurements are done using amplitude- or phase-modulated optical fields (similar to the Bell-Bloom approach) to enable detection of nonzero magnetic fields. Finally, if atoms are prepared in a non-interacting coherent superposition of the hyperfine atomic states (usually referred to as a ``dark state'') via coherent population trapping (CPT), the magnetometer uses an associated transmission resonance (electromagnetically induced transparency, or EIT) to measure the magnetic-field value.

    It is important to emphasize that this list is not exhaustive and  that there are no clear boundaries between different methods. Indeed, many researchers have successfully combined the characteristic features of two or more of these techniques: examples include the $M_z$-$M_x$ tandem~\cite{Allen1972,Pulz1999}, RF atomic magnetometers sensitive to RF magnetic fields~\cite{Chalupczak2012,Wasilewski2010}, magnetometers based on free spin precession~\cite{Grujic2015}, active self-oscillating magnetometers~\cite{Belfi2009,Matsko2005}, and many others. A commercial example is the QuSpin zero-field magnetometer mentioned in Table~\ref{tab:commercial}, which is essentially a Hanle/SERF device using a single-beam $M_z$-type geometry to measure transverse magnetic fields \cite{Shah2018}.
    Not included in Table~\ref{tab:magn_overview} are non-alkali atomic-magnetometer types --- such as those employing the technique of parametric resonance, an established subfield of fundamental and applied magnetometry research \cite{Fourcault2021,Bertrand2021,LeGal2021}.

    \subsection{Typical performance characteristics of different magnetometer types}
    \label{subsec:4.2-MagPerformance}
    
\begin{figure} [h]
  \centering
  \includegraphics[width=\columnwidth]{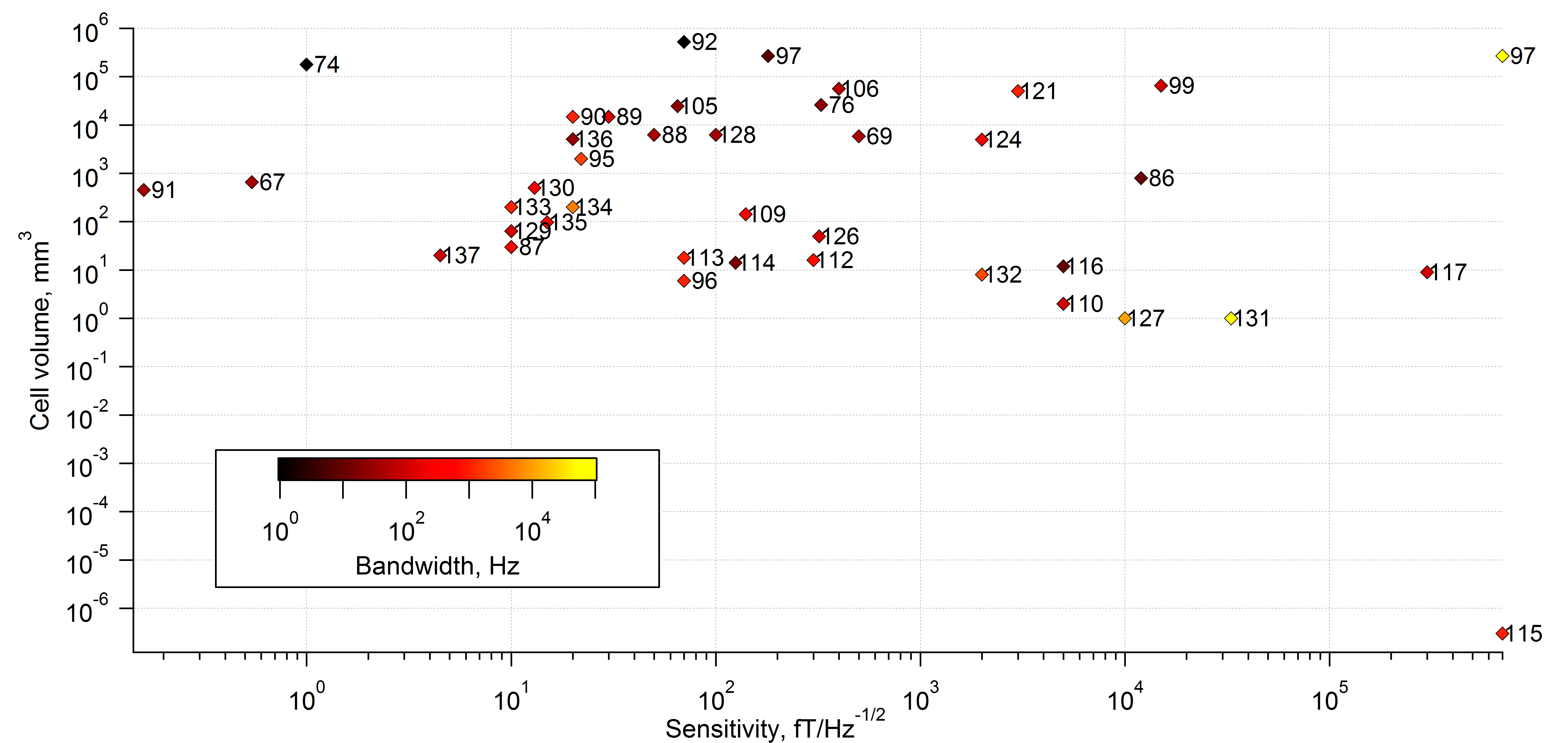}
  \caption[pumping]{Characteristics of various atomic magnetometers \cite{Sebbag2021, Lucivero2014, Tiporlini2013, Sheng2013, Schultze2017, Li2020, Dang2010, Groeger2006EPJD, Schwindt2007, Hong2021, Patton2014, Schwindt2004, Schultze2012, Shah2010, Gerginov2017, Jimenez-Martinez2012, Staehler2001, Pyragius2019, Pulz1999, Belfi2009, Shah2007, Bison2003, Zhang2021IEEE, Gerginov2020,Ishikawa2021,Sheng2017,Petrenko2021,Lebedev2020,Colombo2016,Tayler2017,Bodenstedt2021,Cai2020,Pratt2021,Bevilacqua2019PRA,Yang2021,Zhang2020,Kowalczyk2021,Oelsner2022,Borna2017} utilizing atomic vapor cells: magnetic-field sensitivity, cell volume, and measurement bandwidth (indicated by color).
    }
\label{fig:AM_performance}
\end{figure}

    The choice of an optimal method for magnetic-field measurements depends on the application and the expected performance. Altogether, existing atomic magnetometers span a very wide range of achievable characteristics, as shown in Fig.~\ref{fig:AM_performance}. It is easy to trace some general trends, discussed in Sec.~\ref{sec:3-Requirements}: for example, more sensitive devices tend to use larger vapor cells and operate at lower bandwidth. %At the same time, each type of magnetometer has its own strengths and weaknesses, and as a result may or may not be suitable for a particular application. Below we discussed some specific requirements that may be important while choosing a particular approach.
    Since a given type of magnetometer may or may not be suitable for a particular application, here we discuss some specific requirements that can be important when making the choice.
    % %%%
 % \cite{Sebbag2021}\cite{Lucivero2014}\cite{Tiporlini2013}\cite{Sheng2013}\cite{Schultze2017}\cite{Li2020}\cite{Dang2010}\cite{Groeger2006EPJD}\cite{Schwindt2007}\cite{Hong2021}\cite{Patton2014}\cite{Schwindt2004}\cite{Schultze2012}\cite{Shah2010}\cite{Gerginov2017}\cite{Jimenez-Martinez2012}\\
 % \cite{Staehler2001}\cite{Pyragius2019}\cite{Pulz1999}\cite{Belfi2009}\cite{Shah2007}\cite{Bison2003}\cite{Zhang2021IEEE}\cite{Gerginov2020}\cite{Ishikawa2021}\cite{Sheng2017}\cite{Petrenko2021}\cite{Lebedev2020}\cite{Colombo2016}\cite{Tayler2017}\cite{Bodenstedt2021}\cite{Cai2020}\cite{Pratt2021}\cite{Bevilacqua2019PRA}\cite{Yang2021}\cite{Zhang2020}\cite{Kowalczyk2021}\cite{Oelsner2022}\cite{Borna2017}
    % %%%%
    
    \emph{What sensitivity does my application require?} Hanle magnetometers operating in the SERF regime \cite{Allred2002} (usually just referred to as ``SERF magnetometers'') are currently the most sensitive among all atom-based devices by a wide margin, reaching down to the sub-fT range --- although their sensitivity may vary widely depending on the experimental parameters, such as cell size, light-source characteristics, etc. As such, they can easily compete with SQUIDs in performance without the need for bulky cryogenics. It is important to note, however, that such magnetometers are most sensitive at zero magnetic field, and thus optimal for measuring the faintest magnetic fields in a shielded environment, such as MEG signals. In fact, the commercially available atomic magnetometers (Table~\ref{tab:commercial}) are primarily designed with MEG applications in mind. As discussed in Sec.~\ref{sec:3-Requirements}, there is also an expected trade-off in performance, as SERF magnetometers tend to have limited dynamic range and a potentially larger volume. While these challenges can be overcome with clever engineering, for a number of applications with less stringent sensitivity requirements it may be simpler to use alternative approaches that can provide sensitivity at the level of tens of fT while operating in a fairly large background optical field (Sec.~\ref{sec:5-Assembly}). 
    
    \emph{What is the expected magnetic-field magnitude?} One important characteristic to consider is the base level of the measured magnetic fields. Some magnetometers ($M_z$/$M_x$, Bell-Bloom, CPT/EIT) are intrinsically designed to operate at nonzero bias magnetic field, since the measured Larmor frequency value directly appears as a frequency of the RF field or modulation. These methods are well-suited for applications in Earth's magnetic field --- for example, geomagnetometry or medical diagnostics in a magnetically unshielded environment \cite{Ingleby2022}. Notably, the basic NMOR effect was also initially restricted to near-zero-field measurements, but later modifications using amplitude- or phase-modulated lasers rendered it possible to extend the operational range of NMOR-based magnetometers up to Earth field.  
    
    \emph{Which magnetic-field parameters need to be measured?}
       Most atomic magnetometers are intrinsically \textbf{scalar}, as the measured spin-precession frequency depends only on the magnitude of the magnetic field. However, the expected direction of the magnetic field must be taken into account even when constructing a scalar magnetometer, due to the possibility of \textbf{heading error} (the dependence of the readout on the orientation of the sensor in space)~\cite{Oelsner2019,Lee2021,Bao2018} or blind spots (the inability of the sensor to measure magnetic field in certain orientations). If information about the direction of the magnetic-field vector is required by the application, various strategies have been explored to enable operation in the \textbf{vector} modality. Most magnetometers based on the Hanle effect are intrinsically sensitive to one vector component only. Apart from that, the most common approach is to use three pairs of mutually orthogonal external magnetic-field coils to zero the magnetic field at the sensor location. In this case, the electric current in each coil can be translated into a corresponding magnetic-field component. Alternatively, the field orientation, together with the light polarization, determines the selection rules and relative strength of the involved optical transitions, and thus the vector field information may be deduced from the optical-resonance position or amplitude (or both).
       
        \emph{How fast is the magnetic field changing?}
        In general, speed of operation is not one of the strengths of atomic magnetometers. Their exceptional magnetic sensitivity  originates from long lifetimes of atomic magnetization (Sec.~\ref{subsec:2.2-SpinRelaxation}), which may range from tens of milliseconds to tens of seconds depending on the size and quality of the vapor cell. However, that renders them unable to respond quickly to rapid changes in the magnetic field, limiting their operational bandwidth. Some strategies have been proposed to break this limit. For example, the operational bandwidth was successfully extended by several orders of magnitude by using active feedback to adjust the pump modulation frequency in NMOR and SERF magnetometers~\cite{Li2020,Bodenstedt2021}, albeit with proportional deterioration of sensitivity. Alternatively, a causal waveform estimation using Kalman filtering was demonstrated to accurately predict the magnetic-field variations at time scales shorter than the sensor’s intrinsic time resolution, without paying the sensitivity penalty~\cite{Jimenez-Martinez2018}.
        
        \emph{Can the sensor operate in a gradiometric configuration?} Accurate measurements of the magnetic field using optical-atomic magnetometers may also be challenging due to a number of potential systematics, such as heading errors caused by orientation-dependent light shifts and the nonlinear Zeeman effect~\cite{Oelsner2019,Lee2021,Bao2018}. Moreover, as mentioned in Sec.~\ref{subsec:3.1-Sensitivity}, the sensitivity of measurements may be limited by technical noise from the laser or electromagnetic background coupling to the atomic spins. Some of these issues can be resolved by operating the magnetometer in gradiometer mode, in which two or more nominally identical probe regions are introduced inside the sensor to conduct differential measurements. Of course, this approach is only effective if the expected source of the magnetic field is well-localized and has a distinctly different value for each probe region. In this case, any common-mode noise will be suppressed in the differential signal, while maintaining the same useful signal. On the other hand, when it is necessary to reconstruct a detailed spatial map of the magnetic field, one must usually employ arrays of sensors~\cite{Nardelli2020,Tierney2019,Deans2018,Pratt2021}. In these cases, additional system characteristics must then be considered, such as the common-mode rejection ratio, relative gain and phase stability, low crosstalk, etc., across multiple sensors.
        
        \emph{Are there any restrictions on the physical size of the sensor?}
        Eq.~\eqref{eq:sensitivityPN} clearly illustrates the advantages of larger sensor size: with a larger cell, one can increase the number of interacting atoms and reduce the magnetization decay due to collisions with the cell walls. Indeed, the best sensitivity for all types of atomic magnetometers has been demonstrated in cells with a few-cm$^{3}$ volume. However, as discussed in Sec.~\ref{subsec:3.2-Bandwidth_SpatialRes}, a large sensor size severely limits the spatial resolution as well as the achievable standoff distance. Many applications place restrictions on the sensor dimensions, unavoidably compromising their sensitivity, while also increasing the risk of magnetic-field gradients. Indeed, while most types of magnetometers have been successfully shrunk to few-mm$^3$ cell volumes, the average sensitivity of such miniature versions dropped to a few pT/$\sqrt{\text{Hz}}$. Notably, the magnetometers with purely optical interrogation (NMOR, EIT, Bell-Bloom) seem better suited for extreme miniaturization to chip-scale sensor sizes, whereas the need for rf coils around the cell poses additional geometrical limitations for $M_z$ and $M_x$ magnetometers. An alternative approach to enhancing spatial resolution in larger buffer-gas cells with low spin diffusion is to use multiple probe beams and a photodiode array, thereby subdividing the cell into multiple effective miniature sensing volumes \cite{Kominis2008}. Finally, to obtain true sub-wavelength magnetic-field resolution, nanocells can be utilized~\cite{Sargsyan2017}.

\section{Choose your components and assemble your magnetometer}
\label{sec:5-Assembly}

In Sec.~\ref{sec:2-BlochEq} we derived the basic equations (\ref{eq:lorentzian}) for the zero-field resonances of a Hanle/SERF magnetometer, which was further discussed in Sec.~\ref{sec:4-OperatingModes}. 
In the derivation we assumed that $\Omega_y=0$ and $\Omega_z=0$, which means that this kind of magnetometer works only in conditions where all components of the magnetic field are close to zero.
We assumed further that the optical pumping is along the $z$-direction (Fig.~\ref{fig:setup}).
Consequently, the measurement of the $M_y$ component must not contribute to the optical pumping and thus must rely on a measurement principle that does not involve resonant circularly polarized light.
Both conditions can be achieved experimentally but typically require good magnetic shields, magnetic compensation coils in all directions, and a second laser to produce probe light.
%\IBN{ should we specify the role of the second laser, or it is self-obvious?}.
%
For the rest of this section, however, we mainly focus on magnetometer schemes that are easier to implement, in particular the $M_z$ mode of operation (Table~\ref{tab:magn_overview}).

\begin{figure} [h]
  \centering
  \includegraphics[width=0.7\columnwidth]{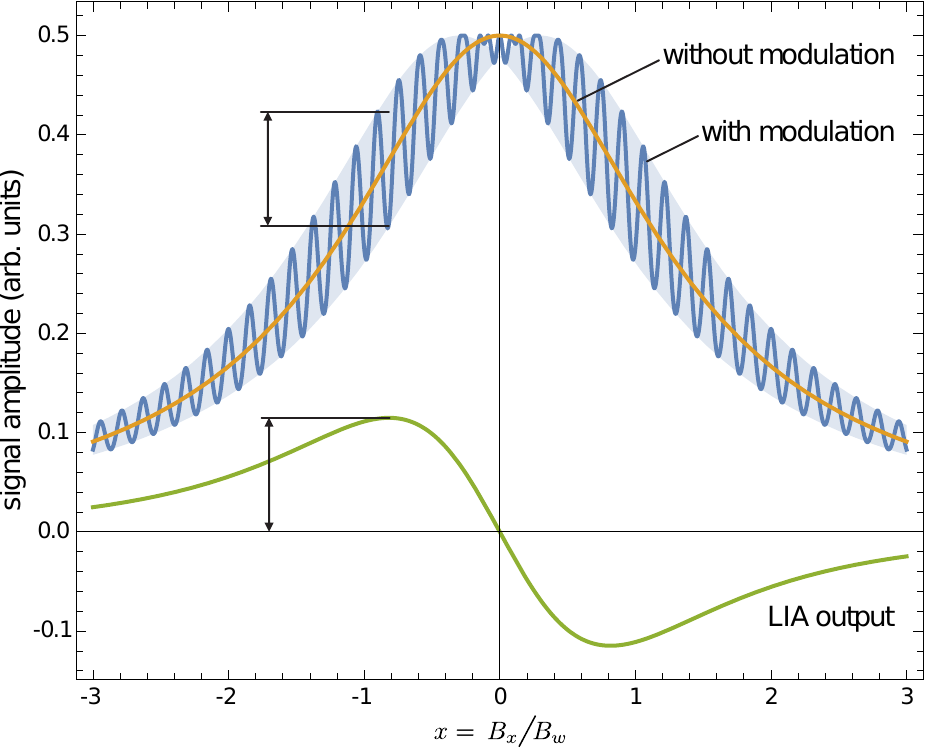}
  \caption{Generation of a dispersive signal from an absorptive signal, using a lock-in amplifier.} %\AMF{It looks like some labels are missing on the figure?}}.
\label{fig:modulation}
\end{figure}

%The existing laser in our setup
Monitoring the transmission of the laser, used for optical pumping, also allows us to measure the $M_z$ component.
%
%However, the corresponding resonance (see Fig.~\ref{fig:lorentzian}) is less desirable for magnetometry, since it is symmetric with respect to $B_x=0$ and does not have a linear zero crossing.
%
While the absorptive Lorentzian line shape is perfectly suited for the optimization of the magnetometer (Sec.~\ref{subsec:5.3-CellPerformance}--\ref{subsec:5.4-Operation}), it does not give us a signal proportional to the magnetic field, as was derived in Eq.~\eqref{eq:lorentzian}.
To obtain such a signal requires generating the derivative of the absorptive Lorentzian.
%
%This is typically done by modulating the magnetic field to be measured using an additional field generated by a coil. 
%\IBN{Alternative wording: this is typically achieved by applying an additional oscillating magnetic field using a coil. I presume we don't have control over the magnetic field we are measuring.}
This is typically achieved by applying an additional oscillating magnetic field using a coil.
Figure~\ref{fig:setup}B shows the setup in which a function generator drives the modulation coil with a sinusoidal current $B_m(t)=A_m \, \sin(\omega_m \,t)$.
The frequency $\omega_m$ of the modulation has to be slow enough that the magnetization does not deviate significantly from its steady state, which means that ${\omega_m < \Gamma}$.
The derivative of the Lorentzian is generated when the signal from the photodiode is demodulated using a lock-in amplifier \cite{DeVore2016}.
Internally, the lock-in determines the amplitude of the signal component at a given reference frequency, which should be the same as $\omega_m$.
%
%It is best practice to use the same function generator for both the modulation and the reference frequency.
%
The output of the lock-in (Fig.~\ref{fig:modulation}) is phase-sensitive: positive when the signal and the reference are in-phase and negative when they are out-of-phase.
Thus, on the rising slope of the Lorentzian the output is positive and on the falling slope it is negative.

This is still not the full picture for an $M_z$ magnetometer, since in general we do not measure only around $B_x=0$, and the magnetometer can operate in an unshielded environment at Earth field (Table~\ref{tab:magn_overview}). So how can a useful absorption signal be generated? %\AMF{Please definitely fact-check the following discussion.}
As usual, we return to the modified Bloch equation, Eq.~\eqref{eq:bloch3}, and write out the time derivatives of the magnetization components for our case:
\begin{equation}
 \frac{d}{dt}
 \left(\begin{array}{c}
  M_x(t)\\
  M_y(t)\\
  M_z(t)\\
 \end{array}\right) =
 \left(\begin{array}{rcl}
  M_y\,\Omega_z &-\,M_z\,\Omega_y &-\,\Gamma\,M_x \\
  M_z\,\Omega_x &-\,M_x\,\Omega_z &-\,\Gamma\,M_y \\
  M_x\,\Omega_y &-\,M_y\,\Omega_x &+\,\Gamma\,(M_a-M_z)
 \end{array}\right) %=
%  \left(\begin{array}{c}
%   0\\
%   0\\
%   0
%  \end{array}\right) 
 \,.
\label{eq:bloch5}
\end{equation}
For simplicity we continue to assume that the longitudinal and transverse atomic-spin components exhibit the same relaxation behavior, i.e. $T_1 \approx T_2$ (Sec.~\ref{subsec:2.2-SpinRelaxation}). Solving the system of equations for the steady state, we find the longitudinal term
\begin{equation}
    M_z = M_a \, \frac{\Gamma^2 + \Omega_z^2}{\Gamma^2 + \Omega_x^2 + \Omega_y^2 + \Omega_z^2} \,.
\end{equation}
Thus, we see that a small amplitude change in either of the transverse fields will modulate $M_z$, but there is also a messy dependence on changes in the longitudinal field. 

Luckily, clever tuning of the modulation coil to the Larmor frequency $\Omega_z$ generates a DC absorption dip which allows us to measure and track changes in $B_z$. This is best understood in the reference frame rotating around the $z$-axis at the Larmor frequency $\Omega_z$. An atomic spin at rest in this rotating frame effectively experiences $B_z=0$. If the additional field is applied along the $x$-direction, $B_{\text{RF}}=\sin(\omega_{\text{RF}}\,t)$ in the laboratory frame, it appears in the rotating frame as a static field $B_{\text{RF}}\,\hat{x}$ when the resonance condition $\omega_{\text{RF}} = \Omega_z$ is met. Thus, the spin will tend to rotate around $\hat{x}$ in the $y$-$z$ plane. This tipping of the spin away from $\hat{z}$ produces a resonant dip in the absorption signal measured by the photodetector . In order to generate a dispersive signal for locking to the resonance, the oscillation frequency of the applied field can be modulated around $\Omega_z$.

We shall now present some important aspects of magnetometer assembly, which will prepare us for discussion of actual operation in Sec.~\ref{subsec:5.4-Operation}.

\subsection{Choice of atomic system and vapor cell}
\label{subsec:5.1-VaporCell}

%Here we collect various questions that students and newcomers to the field may have regarding the atomic sensing volume, and try to answer them in a straightforward way.

\paragraph{Why do we use alkali atoms in atomic magnetometers?} An alkali atom has a single valence electron, which makes it an ideal candidate for optical pumping and control of the electronic angular momentum (Sec.~\ref{subsec:2.1-OpticalPumping}).

\paragraph{Does it matter which alkali atom I choose?} 
% \todo{Consider availability of lasers, vapor pressure / operation temperature, necessity of isotope and hyperfine resolution.}
In many cases, the choice of atom may be informed by strictly technical considerations such as availability of lasers at the appropriate wavelengths. Beyond that, requirements of the particular application should be considered. For example, different atomic species have different vapor pressures, with the vapor pressure of cesium being higher at any given temperature than that of rubidium, potassium, sodium, or lithium \cite{Steck2021}. This makes cesium a popular choice for operation near room temperature \cite{Wasilewski2010,Jensen2018}. The frequency resolution of different isotopic and hyperfine lines in the atomic spectra also varies by atomic species, and is an important consideration for magnetometry modalities requiring such resolution. For applications requiring robustness in non-laboratory conditions, potassium is a favorite due to its narrowband spectral lines.

\paragraph{Should I heat my vapor cell?}

As discussed in Sec.~\ref{subsec:3.1-Sensitivity}, since the sensitivity of a magnetometer is ultimately limited by the number of atoms  in the interaction volume (Eq.~\eqref{eq:sensitivityPN}), it is beneficial to increase this number as much as possible.
%In atomic magnetometry and related fields, we are always interested in enhancing the interaction between light and atoms. This can be accomplished either by using a large number of atoms or by using strong light --- but not so strong as to cause atomic decoherence. 
In order to increase the atomic density without increasing the overall cell size, one typically heats the cell above room temperature. Different techniques are possible, including ovens and coil-based heating systems using high-resistance wire \cite{Li2017}. Care must be taken to avoid heating gradients which can damage the cells, particularly if antirelaxation coating is used --- another reason why buffer-gas cells are preferred for higher-temperature operation. Coated cells are intended to be heated to temperatures up to around 40$^{\circ}$C, depending on the exact chemical composition of the coating used. Radiation trapping of spontaneous emission may also play a negative role at higher atomic densities, unless quenching buffer gas is introduced~\cite{Matsko2001,Han2017}. 

\paragraph{I need a microfabricated vapor cell --- what are my options?} Although we saw in Sec.~\ref{sec:3-Requirements} that reducing the number of atoms worsens the sensitivity of a magnetometer, there may be strong motivations for miniaturization --- including reduced standoff distance and improved spatial resolution, or integration of the cell into small quantum systems~\cite{Shah2007,Nardelli2020,Ledbetter2008,Jimenez-Martinez2010,Zhang2021IEEE,Gerginov2020,Sebbag2021,Brazhnikov2022,Campbell2021} with low size, weight, and power (SWaP). 
% \IBN{Anne, I've added references to all the magnetometers using microfabricated cells, hopefully it is helpful.} 
Production of MEMS (``micro-electromechanical systems'') vapor cells for magnetometers and atomic clocks is a rapidly growing field, typically incorporating anodic-bonding techniques \cite{Woetzel2011,Karlen2017,Vicarini2018}. Most MEMS vapor cells are filled with buffer gas, although antirelaxation-coated cells may offer advantages for biological applications where elevated-temperature operation is problematic \cite{Jensen2016,Jensen2018}. 

\subsection{Laser systems and magnetic shielding}

% \todo{Very short subsection, mainly references:
% \begin{itemize}
% \item Laser options for alkali wavelengths
% \item Power requirements probe v. pump
% \item Laser locking---AJP article
% \end{itemize}}

Nowadays there are a number of commercial laser systems suitable for atomic magnetometry available at various price ranges, including edge-emitting and distributed-feedback (DFB) diode lasers and vertical-cavity surface-emitting lasers (VCSELs). An inexpensive but more time-consuming option is to construct a homemade device using a laser-diode element. The main requirement in any case is single-mode operation around the alkali optical transition of interest, so frequency stabilization of the probe light is critical. (An introductory treatment of the theory of laser locking, also relevant to frequency stabilization generally, can be found in \cite{Black2001}.) Laser power requirements may vary according to the specific magnetometer implementation; in devices utilizing two separate probe and pump beams, higher power is required for optical pumping than for probe readout.

When conducting sensitive magnetometry measurements at zero field, for example in the SERF regime, one needs to reliably shield atoms from unwanted ambient fields. Therefore the cell is contained in a nested magnetic-shielding structure, which may be small enough to fit on a laser table or large enough for experimenters to enter (magnetically shielded room \cite{Knappe-Grueneberg2008}). The innermost layers of the shield typically contain a system of coils to produce the desired DC and/or RF magnetic fields for magnetometer operation, as well as to compensate for field gradients in the cell volume. The surrounding layers usually consist of mu-metal and iron, while strategically placed holes in the shield enable access for laser beams and cables. Mu-metal is a nickel-iron alloy with high magnetic permeability. The idea behind magnetic shielding is not actually to block external fields, but rather to provide a literal ``path of least resistance'' for the magnetic field lines. Since the permeability, and therefore the effectiveness, of mu-metal varies with field strength, several layers can be used to successively reduce the unwanted fields: an outermost iron layer takes care of stronger fields, since iron saturates at a higher field strength than does mu-metal. Such tabletop shielding systems are commercially available, offering shielding factors on the order of $10^6$ in the central region, or may be machined according to individual requirements. To further improve shielding performance, active compensation coils may be used, with current state-of-the-art solutions even enabling MEG measurements from human subjects in motion within shielded rooms \cite{Holmes2021}.

Of course, many atomic magnetometers operate unshielded within Earth's field (typically at the cost of sensitivity performance, see Table~\ref{tab:magn_overview}), or unshielded within field-canceling coil systems \cite{Belfi2007,Seltzer2004,Rushton2022}. For Earth-field operation, it may be especially desirable to operate in a gradiometric configuration \cite{Campbell2021}.

\subsection{Cell performance}
\label{subsec:5.3-CellPerformance}

Before being used for magnetometry or any other experiment, an atomic vapor cell must undergo a testing process in order to verify that several requirements are fulfilled:
\begin{itemize}
 \item There are in fact atoms in the cell.
 \item The density of atoms is sufficiently high.
 \item It is possible to create a high-quality spin-polarization state (magnetization).
 \item The buffer-gas or coating quality (see below) is such that the decay time $T_1$ of the magnetization is adequately long.
\end{itemize}
% In ~\ref{sec:Sup-CellCharacterization} we briefly discuss how to implement these tests in practice.
Further information about how to conduct such tests may be found in various introductory references and theses covering absorption spectroscopy, optical pumping, and spin-relaxation measurements \cite{Preston1998, Li2017, Krauter2011}.

\paragraph{What is the difference between buffer-gas and antirelaxation-coated vapor cells?}
The smaller the magnetization relaxation rate, the longer the atomic spins can precess in the magnetic field and the more accurately the magnetic field can be measured (Sec.~\ref{subsec:3.1-Sensitivity}). While ground-state spin coherence of alkali atoms isolated from the environment can live almost indefinitely long, there are many experimental factors that limit its lifetime. Random motion of thermal atoms inside a glass cell inevitably leads to their collisions with the cell walls, 
%Because of the unavoidable magnetic impurities inside the glass, such collisions 
which effectively destroy any prior magnetization and lead to thermalization of the atomic spin state. Thus, in an evacuated cell the effective spin relaxation rate is determined by the average time it takes for atoms to cross the laser beam (known as transient time). For a typical few-mm-wide laser beam, this time is limited to a few microseconds.

To extend the spin-relaxation time $T_2$, two common strategies are employed. In one, a buffer gas is added to the cell, so that alkali atoms undergo diffusive, rather than ballistic motion. This way, the $T_2$ time can be significantly lengthened. The choice and pressure of the buffer gas is typically customized for each cell geometry: while higher pressure increases atomic transient time, it also increases resonance collisional broadening and shifts~\cite{Happer1972}. Most common buffer gases are inert gases with low collisional cross-sections (Ne, He, Ar, etc.), although nitrogen (N$_2$) is commonly used to quench spontaneous emission. The main advantages of using cells with buffer gas are their scalability (the shorter transient time due to reduction in the cell volume can be at least partially compensated by an increase in buffer-gas pressure) and relative manufacturing simplicity. Notably, collisions with buffer gas are not always bad: a high collision rate plays a crucial role in the excellent performance of SERF magnetometers (Sec.~\ref{sec:4-OperatingModes}), as it allows for suppression of the spin-exchange decoherence mechanism in a dense vapor \cite{Allred2002}.
At the same time, there are some potentially serious drawbacks: buffer gas effectively ``freezes'' atoms in their locations, making the system more susceptible to local magnetic-field gradients and inhomogeneous optical pumping. Also, collisional dephasing of the excited state is typically much more significant than for the ground state, leading to additional homogeneous broadening of optical transitions. In conjunction with velocity-changing collisions, this can result in undesirable modifications of the atomic-spin response, especially in the presence of multiple excited states~\cite{Novikova2005}.

The second approach is to use a hydrocarbon antirelaxation coating for the inner cell walls, which is applied during cell production. The science of antirelaxation coatings for atomic vapor cells has actually been described in the literature as ``black magic''. This is because the details of how the coating works are not fully understood, and the coating process does not necessarily yield reproducible results in cell performance. Hence, for mass production of commercial systems, buffer-gas cells are typically preferred (Table~\ref{tab:commercial}) At the time of writing, there are is no commercially available anti-relaxation coated vapor cells, although such cells are used in some commercially available magnetometers \cite{GEM2022}. 

The process of alkali-atom relaxation on paraffin surfaces was modeled quantitatively, based on experimental work, already in the 1960s \cite{Bouchiat1966}. Some relevant qualitative points are the following. Atoms do not scatter elastically off the coated surface, but rather spent a finite time in the coating. For temperatures in the range $20 \text{--} 60^{\circ} \text{C}$, this is an adsorption process. (Adsorption is the process whereby molecules adhere to a surface; in absorption, they are taken up by a volume.) Spin relaxation takes place primarily at the walls and not in the cell body, as supported by the fact that the relaxation times scale with the cell dimensions \cite{Graf2005}. 

Antirelaxation coatings are conventionally based on paraffin, also known as alkane ($\text{C}_n \text{H}_{2n+2}$). In recent years, coatings based on alkene ($\text{C}_n \text{H}_{2n}$) have also shown promise. An important parameter is the number of atom-wall collisions (bounces) that the coating can sustain before the atomic spins depolarize: on order $10^4$ polarization-preserving bounces for paraffin and $10^6$ for alkene \cite{Balabas2010}. Although the interaction of alkali atoms with alkene surfaces has not been studied in detail, it appears that the performance enhancement is related to carbon double bonds which are present in alkene but absent in paraffin \cite{Budker2013}. However, paraffin coatings tend to offer better performance operation above room temperature \cite{Li2017}. In the SERF regime, buffer-gas cells are the only viable choice. Further details regarding production of various coated-cell types may be found in \cite{Seltzer2010,Castagna2009}, and specific considerations of optical pumping in coated cells are treated in \cite{Han2017}.

\subsection{Magnetometer operation}
\label{subsec:5.4-Operation}

% \todo{Here we can include the theoretical treatment of the $M_z$ magnetometer.}

% \bigskip

% \todo{
% \noindent Original suggested outline:
% \begin{itemize}
%     \item Step-by-step process---scan resonance, locking, optical pumping, FID
%     \item \textbf{Concrete example based on Rb/Cs}---$M_z$, $M_x$, or NMOR (probably $M_z$)
%     \item Cross-reference to other articles
%     \item Heating vapor cells
%     \item Software
%     \item Next steps---other types of magnetometers
% \end{itemize}}
% \bigskip

% \todo{
% \noindent Outline:
% \begin{itemize}
%     \item Step-by-step process --- scan resonance, locking, optical pumping, FID
%     \item \textbf{Concrete example of $M_z$ magnetometer based on Cs}
%     \item Cross-reference to other articles
%     \item Hardware and software
%     \item Next steps --- other types of magnetometers
% \end{itemize}}
% \bigskip

% \todo{Useful reference: \cite{Budker1999}}

Further technical details of setting up an atomic magnetometer in the lab may vary greatly depending on the operating mode (Sec.~\ref{sec:4-OperatingModes}) --- see, e.g. \cite{Budker1999}. 
%for a didactic example, we refer the reader to \cite{Budker1999}.
Regardless of the specific design, however, one usually needs to go through the following general steps:
%, after all components are in place, the following operation steps can be identified:
\begin{enumerate}
    \item Align the vapor cell and all optomechanics with respect to the laser beam(s).
    \item Lock each laser to the correct frequency.
    \item Prepare the atomic spins via optical pumping, and allow them to evolve in the presence of the magnetic field to be measured.
    \item Scan the magnetic resonance to obtain the magnetometer signal in frequency space (Fig.~\ref{fig:lorentzian}).
    \item Tuning to resonance, for example via phase-locking techniques, enables tracking changes in the magnetic field as a function of time.
\end{enumerate}

As mentioned previously, the linewidth of the magnetic resonance in step (iv) is proportional to the transverse coherence rate $1/T_2$, so in principle one may be able to extract the $T_2$ value directly from the magnetometer signal. However, the strength of optical fields affects this linewidth, so the most precise $T_2$ measurements are usually conducted using  a pulsed pump/probe regime whereby a free-induction-decay (FID) signal can be obtained.
%--- similar to what is done in NMR experiments (which utilize nuclear, rather than electron, spins). 
Another common approach to finding the $T_2$ time is to measure power-broadened resonance linewidths and extrapolate to zero laser power \cite{Scholtes2014}. Whether or not such an investigation is necessary or desirable is context-dependent; for many applications it suffices to perform characterization of magnetometer time-series data as discussed in Sec.~\ref{sec:6-Characterization}. However, linewidth measurements can be an important tool for noise characterization (Sec.~\ref{subsec:3.1-Sensitivity}), as various noise sources --- including magnetic-field gradients and anything else that couples to the atomic spins --- manifest themselves as broadening mechanisms.

In terms of data acquisition, there is great freedom of choice as regards both hardware and software. Commercial data-acquisition (DAQ) systems are available with a corresponding price tag; lower-budget options may involve open-source software and inexpensive hardware such as a computer sound card \cite{Groeger2006}. When sampling an oscillating signal, care should be taken to ensure that the sampling rate is fast enough to avoid aliasing effects. 
% \AMF{Would you guys say that there is some optimal sampling rate for? I occasionally argue with a few colleagues who claim that they gain in SNR from oversampling their OPM output at as high a rate as possible...}

% \subsubsection{Heating vapor cells}

% In atomic magnetometry and related fields, we are always interested in enhancing the interaction between light and atoms. This can be accomplished either by using a large number of atoms (large collective dipole operator) or by using strong light (large electric-field amplitude)---but not so strong as to cause decoherence. In order to increase the density of atoms without increasing the overall cell size, one can heat the vapor cells. Different techniques are possible, including hot-air heating and coil-based heating systems using high-resistance wire \todo{[REFS]}. Care must be taken to avoid heating gradients which can damage the cells, particularly if antirelaxation coating is used---another reason why buffer-gas cells are preferred for use in SERF magnetometers. 
% cell performance

%\pagebreak
\section{Characterize and optimize your magnetometer}
\label{sec:6-Characterization} 
% \textbf{Anne's section?}
% \bigskip

% \todo{\noindent Original suggested outline:
% \begin{itemize}
% %    \item Cell performance---practical measurement of T1,T2
%     \item Noise characterization---measuring different noise sources
%     \item Sensitivity and bandwidth---how to measure in real life
%     \item Broadening
%     \item \textbf{Mathematica notebook as supplement}
%     \item How to make a good magnetic-field reference (Or, the impossibility of actually characterizing your sensor)
% \end{itemize}}

% \todo{Introduce Georg's time-series data and Mathematica notebook. Explain how to produce the frequency spectrum and extract the sensitivity and bandwidth --- include plots. Mention noise characterization.}

% https://opg.optica.org/oe/fulltext.cfm?uri=oe-26-13-17350&id=392701

% \AMF{Previously we discussed including broadening, heading error, and Allan deviation in the analysis here, as well as how to make a good magnetic-field reference. Do we still want to add these elements or is it too much?}

To illustrate how to analyze actual data, in \ref{sec:Sup-DataAnalysis} we provide magnetometer time-series data from \cite{Bison2018}, along with an accompanying analysis notebook. Although the data were obtained with a Hanle magnetometer, the illustrated analysis techniques are broadly applicable. The analysis is performed in Mathematica, but any other analysis program with similar capabilities can be used instead. 

Measurement of magnetic signals in the time domain may be interesting and useful in itself, especially in the case of transient and/or triggered signals such as those encountered in biological applications. However, to characterize magnetometer performance as per the figures of merit introduced in Sec.~\ref{sec:3-Requirements}, it is generally more instructive to work in the frequency domain. This is usually done by applying a fast Fourier transform (FFT) to the time-series data, in order to produce a frequency spectrum. 

\begin{figure}[h]
  \centering
  \includegraphics[width=0.9\columnwidth]{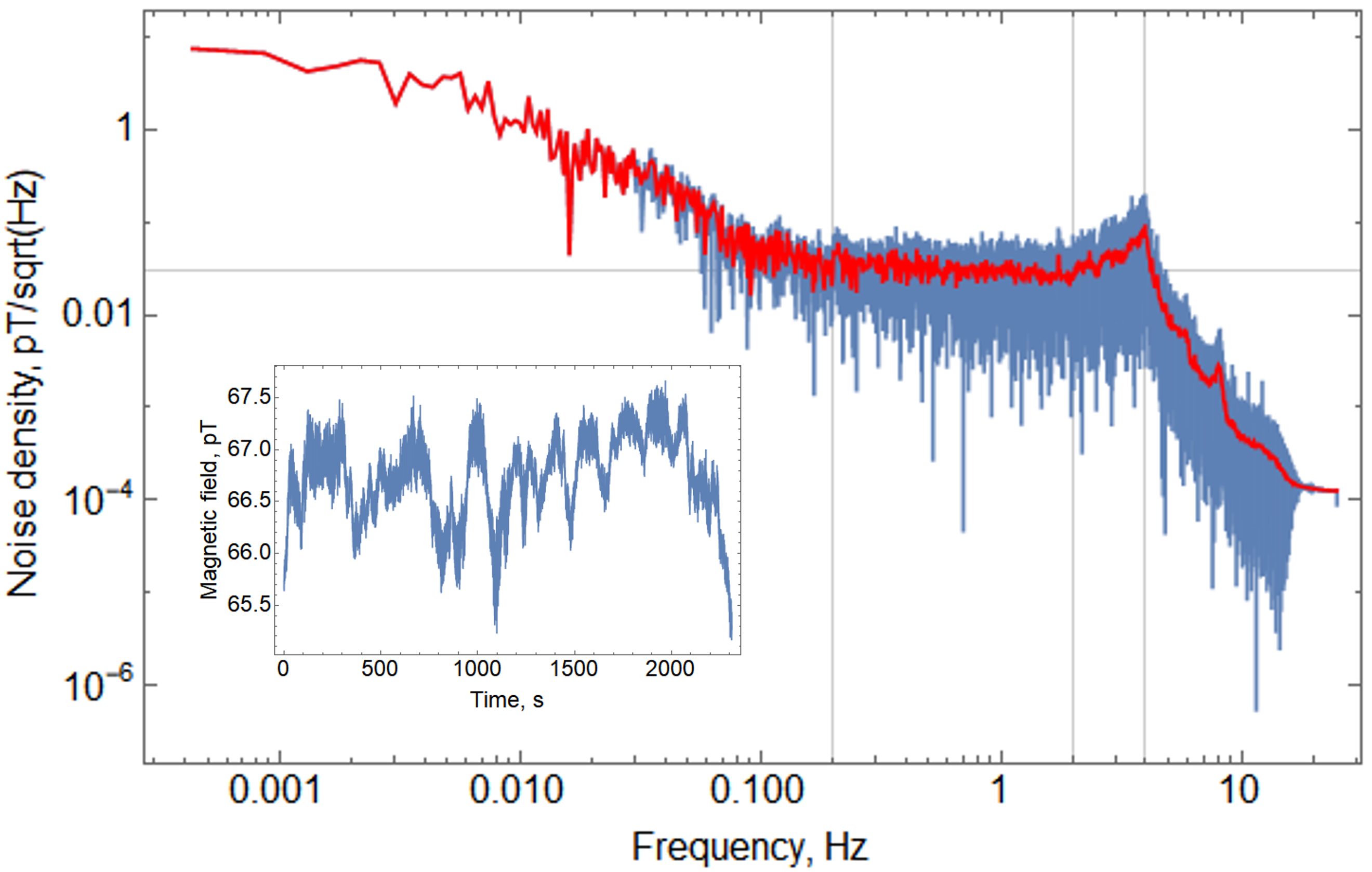}
  \caption{Noise-density frequency spectrum obtained by applying a fast Fourier transform (FFT) to raw magnetometer time-series data (inset) from \cite{Bison2018}. The red and blue curves show the results with and without averaging, respectively. Vertical lines separate frequency regions dominated by different noise behavior, as explained in the text; the horizontal line indicates magnetometer sensitivity in the bandwidth of flat frequency response.}
\label{fig:spectrum}
\end{figure}

Fig.~\ref{fig:spectrum} illustrates this analysis process for the case where no magnetic signal is present, such that the intrinsic magnetometer noise can be characterized. Looking at the noise-density spectrum, we see that the noise behavior is highly frequency-dependent and can be characterized as follows:
\begin{itemize}
    \item The noise is highest at lower frequencies, dominated by $1/f$ (so-called ``pink'' or ``flicker'') noise from background magnetic fields. %\AMF{What can we say is the physical origin? Is it to do with the cell heater or some other slowly varying parameter?}
    \item Between 0.2 and 2 Hz, we have the flat or ``white''-noise region of the spectrum. This region defines the useful bandwidth of the magnetometer and corresponds to a sensitivity of $30~\text{fT}/\sqrt{\text{Hz}}$.
    \item The noise increases between 2 and 4 Hz, in this case due to an acoustic resonance of the magnetically shielded room in which measurements were conducted.
    \item Above 4 Hz, the magnetometer response drops off rapidly, here because of the low-pass filter of the lock-in amplifier used for data acquisition.
\end{itemize}

For purposes of characterizing magnetometer response, data may also be obtained by measuring known magnetic-field reference signals produced using a calibrated coil system --- standard procedure for device optimization before any measurement of unknown fields is attempted. Peaks should then appear in the frequency spectrum corresponding to the frequencies of the applied fields; any peaks already appearing in the absence of applied fields are likely due to frequency-dependent noise sources.

% \AMF{Should we add some concluding paragraph here or in a separate section?}

% \subsection{Noise}
% \label{subsection:6.1-Noise}

% For measurement of relatively small magnetic signals, We would like the sensitivity of our magnetometer to be as good as possible---in the ideal case, it is only limited by fundamental quantum mechanics. It is important to identify the various contributions to noise in our measurements, and to understand how they behave.
% \IBN{Haven't you already discussed some noises in part 2??}

% \subsubsection{Electronic noise}

% \subsubsection{Photon shot noise}

% \subsubsection{Atomic noise}

% \paragraph{Spin-projection noise}

% \paragraph{Back-action noise}

% \subsubsection{Technical noise}

%\pagebreak
\section{Typical applications of atomic magnetometers}
\label{sec:7-Applications} 
% \textbf{Optional section to be written later}
% \bigskip

% Real-world examples with biological/nonbiological systems, applied/fundamental physics

% \todo{Here I would suggest just making a table of different applications with references, rather than writing a text---move this to Section 4.}

% \subsection{Typical applications of atomic magnetometers}
% \label{subsec:4.3-Applications}

In Table~\ref{tab:magn_applications}, we summarize the main current applications of atomic magnetometers. Not surprisingly, many of these diverse applications value high sensitivity and high accuracy at frequencies below tens of Hz. Non-invasive detection of biological magnetic signals is probably the most sought-after example, since such measurements can be carried out in a magnetically shielded environment to take maximum advantage of SERF magnetometers' exceptional sensitivity. Another important feature of atomic sensors is their negligible internal magnetic signature, making them desirable both for biomedical and surveillance applications. In summary, alkali-metal atomic magnetometers are admittedly still in their early steps toward widespread commercial success, and we may optimistically expect the emergence of new applications as fuller capabilities of atomic systems are explored in research laboratories. 

\begin{table}[h]
\small
\centering
\begin{tabular}{|p{0.25\textwidth}|p{0.7\textwidth}|} 
\hline
Biomedical          & %\textit{in vitro}:
magnetomyography (MMG)~\cite{Jensen2016,Broser2018},  %\\ 
%
%                   & \textit{in vivo}: 
magnetoencephalography (MEG)~\cite{Xia2006,Tierney2019,Boto2018,Sheng2017,Petrenko2021,Limes2020,Zhang2020,Kowalczyk2021,Borna2020,Gialopsou2021}, (fetal) magnetocardiography (MCG and fMCG)~\cite{Sander2020,Batie2018,Bison2003,Belfi2007,Limes2020,Sulai2019}, magnetic-field imaging (MFI)~\cite{Lembke2014}\\ 

                    & magnetic biomarkers~\cite{Johnson2012,Bougas2018,Bi2021,Colombo2016,Jaufenthaler2021}  \\ 

                    & biomagnetism of plants~\cite{Corsini2011,Fabricant2021} and livestock~\cite{Sutter2020}  \\ 
\hline
Zero-/low-field NMR  &   \cite{Blanchard2021,Savukov2005,Bodenstedt2021,Ledbetter2008,Xu2006,Jiang2019,Tayler2017,Begus2017,Bevilacqua2019APL,Barskiy2019,Put2021}  \\ 
\hline
Geophysics         &   \cite{Dang2010,Zhang2021PRL} \\ 
\hline
Aerospace           &    \cite{Korth2016,Bennett2021} \\ 
\hline
Laser guide stars           &    \cite{PedrerosBustos2018} \\ 
\hline
Defense and industry            &   underwater surveillance~\cite{Deans2018}, electromagnetic induction imaging~\cite{Marmugi2020,Jensen2019,Bevilacqua2021,Maddox2022} \\ 
\hline
Fundamental science & search for new physics beyond particle standard model~\cite{Afach2018,JacksonKimball2017,Klassen2020,Afach2021,Su2021,Rosner2022}, with comagnetometry \cite{Limes2018,Mora2019,Yang2021,Terrano2022,Padniuk2022} \\ 
& precision measurements ~\cite{Mora2019,Klassen2020} \\
                    & \textit{quantum applications}: squeezing-enhanced magnetometers~\cite{Kuzmich1997,Auzinsh2004,Sewell2012,Wolfgramm2010,Troullinou2021,Li2022}, quantum non-demolition measurements of atomic spins~\cite{Kuzmich1999,Shah2010}, quantum information~\cite{Siddons2009,Wang2014}, entanglement~\cite{Wasilewski2010} \\
\hline
\end{tabular}
\caption{Common applications of atomic magnetometers. \label{tab:magn_applications}}
\end{table}
\clearpage

%\pagebreak
\section*{Acknowledgments}

We thank our many colleagues from the Workshop on Optically Pumped Magnetometers (WOPM) for their generous support and assistance throughout the writing process, in particular Theo Scholtes.
A.F. thanks Danila Barskiy, Dmitry Budker, Pavel Fadeev, Till Lenz, and Hendrik Bekker for helpful discussion and editing.
Some portions of the text were adapted from \cite{Fabricant2014}.

\printbibliography

@article{Limes2018,
    title = {{3He-129Xe comagnetometery using 87Rb detection and decoupling}},
    year = {2018},
    journal = {Physical Review Letters},
    author = {Limes, M. E. and Sheng, D. and Romalis, M. V.},
    number = {3},
    month = {1},
    pages = {033401},
    volume = {120},
    publisher = {American Physical Society},
    url = {https://journals.aps.org/prl/abstract/10.1103/PhysRevLett.120.033401},
    doi = {10.1103/PhysRevLett.120.033401},
    issn = {10797114},
    pmid = {29400512}
}

@article{Borna2017,
    title = {{A 20-channel magnetoencephalography system based on optically pumped magnetometers}},
    year = {2017},
    journal = {Physics in Medicine and Biology},
    author = {Borna, Amir and Carter, Tony R. and Goldberg, Josh D. and Colombo, Anthony P. and Jau, Yuan Yu and Berry, Christopher and McKay, Jim and Stephen, Julia and Weisend, Michael and Schwindt, Peter D.D.},
    number = {23},
    month = {11},
    pages = {8909--8923},
    volume = {62},
    publisher = {IOP Publishing},
    url = {https://iopscience.iop.org/article/10.1088/1361-6560/aa93d1},
    doi = {10.1088/1361-6560/aa93d1},
    issn = {13616560},
    pmid = {29035875}
}

@article{Bertrand2021,
    title = {{A 4He vector zero-field optically pumped magnetometer operated in the Earth-field}},
    year = {2021},
    journal = {Review of Scientific Instruments},
    author = {Bertrand, F. and Jager, T. and Boness, A. and Fourcault, W. and Le Gal, G. and Palacios-Laloy, A. and Paulet, J. and L{\'{e}}ger, J. M.},
    number = {10},
    month = {10},
    pages = {105005},
    volume = {92},
    publisher = {AIP Publishing LLCAIP Publishing},
    url = {https://aip.scitation.org/doi/abs/10.1063/5.0062791},
    doi = {10.1063/5.0062791},
    issn = {10897623},
    pmid = {34717435}
}

@article{Cohen2019,
    title = {{A cold atom radio-frequency magnetometer}},
    year = {2019},
    journal = {Applied Physics Letters},
    author = {Cohen, Yuval and Jadeja, Krishna and Sula, Sindi and Venturelli, Michela and Deans, Cameron and Marmugi, Luca and Renzoni, Ferruccio},
    number = {7},
    month = {2},
    pages = {073505},
    volume = {114},
    publisher = {AIP Publishing LLC AIP Publishing},
    url = {https://aip.scitation.org/doi/abs/10.1063/1.5084004},
    doi = {10.1063/1.5084004},
    issn = {00036951},
    arxivId = {1902.08258}
}

@article{Nardelli2020,
    title = {{A conformal array of microfabricated optically-pumped first-order gradiometers for magnetoencephalography}},
    year = {2020},
    journal = {EPJ Quantum Technology},
    author = {Nardelli, N. V. and Perry, A. R. and Krzyzewski, S. P. and Knappe, S. A.},
    number = {1},
    month = {9},
    pages = {1--11},
    volume = {7},
    publisher = {SpringerOpen},
    url = {https://epjquantumtechnology.springeropen.com/articles/10.1140/epjqt/s40507-020-00086-4},
    doi = {10.1140/epjqt/s40507-020-00086-4},
    issn = {21960763}
}

@article{Ingleby2022,
    title = {{A digital alkali spin maser}},
    year = {2022},
    journal = {Scientific Reports 2022 12:1},
    author = {Ingleby, Stuart and Griffin, Paul and Dyer, Terry and Mrozowski, Marcin and Riis, Erling},
    number = {1},
    month = {7},
    pages = {1--7},
    volume = {12},
    publisher = {Nature Publishing Group},
    url = {https://www.nature.com/articles/s41598-022-16910-z},
    isbn = {0123456789},
    doi = {10.1038/s41598-022-16910-z},
    issn = {2045-2322},
    arxivId = {2207.11009}
}

@article{Siddons2009,
    title = {{A gigahertz-bandwidth atomic probe based on the slow-light Faraday effect}},
    year = {2009},
    journal = {Nature Photonics},
    author = {Siddons, Paul and Bell, Nia C. and Cai, Yifei and Adams, Charles S. and Hughes, Ifan G.},
    number = {4},
    month = {3},
    pages = {225--229},
    volume = {3},
    publisher = {Nature Publishing Group},
    url = {https://www.nature.com/articles/nphoton.2009.27},
    doi = {10.1038/nphoton.2009.27},
    issn = {17494885}
}

@article{Owston1967,
    title = {{A Hall effect magnetometer for small magnetic fields}},
    year = {1967},
    journal = {Journal of Scientific Instruments},
    author = {Owston, C. N.},
    number = {9},
    month = {9},
    pages = {798--800},
    volume = {44},
    publisher = {IOP Publishing},
    url = {https://iopscience.iop.org/article/10.1088/0950-7671/44/9/441},
    doi = {10.1088/0950-7671/44/9/441},
    issn = {09507671}
}

@article{Groeger2006EPJD,
    title = {{A high-sensitivity laser-pumped Mx magnetometer}},
    year = {2006},
    journal = {European Physical Journal D},
    author = {Groeger, S. and Bison, G. and Schenker, J. L. and Wynands, R. and Weis, A.},
    number = {2},
    month = {2},
    pages = {239--247},
    volume = {38},
    publisher = {Springer},
    url = {https://link.springer.com/article/10.1140/epjd/e2006-00037-y},
    doi = {10.1140/epjd/e2006-00037-y},
    issn = {14346060},
    arxivId = {physics/0406105}
}

@article{Rosner2022,
    title = {{A highly drift-stable atomic magnetometer for fundamental physics experiments}},
    year = {2022},
    journal = {Applied Physics Letters},
    author = {Rosner, M. and Beck, D. and Fierlinger, P. and Filter, H. and Klau, C. and Kuchler, F. and R{\"{o}}{\ss}ner, P. and Sturm, M. and Wurm, D. and Sun, Z.},
    number = {16},
    month = {4},
    pages = {161102},
    volume = {120},
    publisher = {AIP Publishing LLCAIP Publishing},
    url = {https://aip.scitation.org/doi/abs/10.1063/5.0083854},
    doi = {10.1063/5.0083854},
    issn = {0003-6951},
    arxivId = {2201.06936}
}

@article{Castagna2009,
    title = {{A large sample study of spin relaxation and magnetometric sensitivity of paraffin-coated Cs vapor cells}},
    year = {2009},
    journal = {Applied Physics B: Lasers and Optics},
    author = {Castagna, N. and Bison, G. and Di Domenico, G. and Hofer, A. and Knowles, P. and MacChione, C. and Saudan, H. and Weis, A.},
    number = {4},
    month = {4},
    pages = {763--772},
    volume = {96},
    publisher = {Springer},
    url = {https://link.springer.com/article/10.1007/s00340-009-3464-5},
    isbn = {0034000934645},
    doi = {10.1007/s00340-009-3464-5},
    issn = {09462171},
    arxivId = {0812.4425}
}

@article{Brossel1952,
    title = {{A New "Double Resonance" Method for Investigating Atomic Energy Levels. Application to Hg 3P1}},
    year = {1952},
    journal = {Physical Review},
    author = {Brossel, Jean and Bitter, Francis},
    number = {3},
    month = {5},
    pages = {308},
    volume = {86},
    publisher = {American Physical Society},
    url = {https://journals.aps.org/pr/abstract/10.1103/PhysRev.86.308},
    doi = {10.1103/PhysRev.86.308},
    issn = {0031899X}
}

@article{Pulz1999,
    title = {{A new optically pumped tandem magnetometer: Principles and experiences}},
    year = {1999},
    journal = {Measurement Science and Technology},
    author = {Pulz, E. and J{\"{a}}ckel, K. H. and Linthe, H. J.},
    number = {11},
    month = {11},
    pages = {1025--1031},
    volume = {10},
    publisher = {IOP Publishing},
    url = {https://iopscience.iop.org/article/10.1088/0957-0233/10/11/309},
    doi = {10.1088/0957-0233/10/11/309},
    issn = {09570233}
}

@article{Waters1958,
    title = {{A nuclear magnetometer}},
    year = {1958},
    journal = {Journal of Scientific Instruments},
    author = {Waters, G. S. and Francis, P. D.},
    number = {3},
    month = {3},
    pages = {88--93},
    volume = {35},
    publisher = {IOP Publishing},
    url = {https://iopscience.iop.org/article/10.1088/0950-7671/35/3/302},
    doi = {10.1088/0950-7671/35/3/302},
    issn = {09507671}
}

@article{Grujic2015,
    title = {{A sensitive and accurate atomic magnetometer based on free spin precession}},
    year = {2015},
    journal = {European Physical Journal D},
    author = {Gruji{\'{c}}, Zoran D. and Koss, Peter A. and Bison, Georg and Weis, Antoine},
    number = {5},
    month = {5},
    pages = {1--10},
    volume = {69},
    publisher = {Springer},
    url = {https://link.springer.com/article/10.1140/epjd/e2015-50875-3},
    doi = {10.1140/epjd/e2015-50875-3},
    issn = {14346079}
}

@article{Groeger2006,
    title = {{A sound card based multi-channel frequency measurement system}},
    year = {2006},
    journal = {The European Physical Journal - Applied Physics},
    author = {Groeger, S. and Bison, G. and Knowles, P. E. and Weis, A.},
    number = {3},
    month = {3},
    pages = {221--224},
    volume = {33},
    publisher = {EDP Sciences},
    url = {https://www.cambridge.org/core/journals/the-european-physical-journal-applied-physics/article/abs/sound-card-based-multichannel-frequency-measurement-system/3D3452C22E03D5CCEDF25CB6CA87EC3D},
    doi = {10.1051/EPJAP:2006020},
    issn = {1286-0042},
    arxivId = {physics/0506068}
}

@article{Kominis2008,
    title = {{A subfemtotesla multichannel atomic magnetometer}},
    year = {2003},
    journal = {Nature 2003 422:6932},
    author = {Kominis, I. K. and Kornack, T. W. and Allred, J. C. and Romalis, M. V.},
    number = {6932},
    month = {4},
    pages = {596--599},
    volume = {422},
    publisher = {Nature Publishing Group},
    url = {https://www.nature.com/articles/nature01484},
    doi = {10.1038/nature01484},
    issn = {1476-4687},
    pmid = {12686995}
}

@article{Fabricant2021,
    title = {{Action potentials induce biomagnetic fields in carnivorous Venus flytrap plants}},
    year = {2021},
    journal = {Scientific Reports},
    author = {Fabricant, Anne and Iwata, Geoffrey Z. and Scherzer, Sönke and Bougas, Lykourgos and Rolfs, Katharina and Jodko-W{\l}adzi{\'{n}}ska, Anna and Voigt, Jens and Hedrich, Rainer and Budker, Dmitry},
    number = {1},
    month = {1},
    pages = {1--7},
    volume = {11},
    publisher = {Nature Publishing Group},
    url = {https://www.nature.com/articles/s41598-021-81114-w},
    isbn = {0123456789},
    doi = {10.1038/s41598-021-81114-w},
    issn = {20452322},
    pmid = {33446898}
}

@article{Deans2018,
    title = {{Active underwater detection with an array of atomic magnetometers}},
    year = {2018},
    journal = {Applied Optics},
    author = {Deans, Cameron and Marmugi, Luca and Renzoni, Ferruccio},
    number = {10},
    month = {4},
    pages = {2346},
    volume = {57},
    publisher = {Optica Publishing Group},
    url = {https://opg.optica.org/ao/abstract.cfm?uri=ao-57-10-2346},
    doi = {10.1364/ao.57.002346},
    issn = {1559-128X},
    pmid = {29714214},
    arxivId = {1803.07846}
}

@misc{Steck2021,
    title = {{Alkali D Line Data}},
    year = {2021},
    author = {Steck, Daniel A.},
    url = {https://steck.us/alkalidata/}
}

@misc{Klassen2020,
    title = {{All-Optical Cs Magnetometry System for a Neutron Electric Dipole Moment Experiment - Master thesis, University of Manitoba}},
    year = {2020},
    author = {Klassen, Wolfgang},
    month = {3},
    publisher = {University of Manitoba},
    url = {https://mspace.lib.umanitoba.ca/handle/1993/34596},
    address = {Winnipeg, Canada}
}

@article{Yang2021,
    title = {{All-optical single-species cesium atomic comagnetometer with optical free induction decay detection}},
    year = {2021},
    journal = {Applied Physics B: Lasers and Optics},
    author = {Yang, Yucheng and Wu, Teng and Chen, Jingbiao and Peng, Xiang and Guo, Hong},
    number = {3},
    month = {3},
    pages = {1--11},
    volume = {127},
    publisher = {Springer Science and Business Media Deutschland GmbH},
    url = {https://link.springer.com/article/10.1007/s00340-021-07594-w},
    doi = {10.1007/s00340-021-07594-w},
    issn = {09462171},
    arxivId = {2007.15343}
}

@article{Patton2014,
    title = {{All-optical vector atomic magnetometer}},
    year = {2014},
    journal = {Physical Review Letters},
    author = {Patton, B. and Zhivun, E. and Hovde, D. C. and Budker, D.},
    number = {1},
    month = {7},
    pages = {013001},
    volume = {113},
    publisher = {American Physical Society},
    url = {https://journals.aps.org/prl/abstract/10.1103/PhysRevLett.113.013001},
    doi = {10.1103/PhysRevLett.113.013001},
    issn = {10797114},
    pmid = {25032923},
    arxivId = {1403.7545}
}

@misc{Allan2012,
    title = {{Allan Variance -- Overview by David W. Allan}},
    author = {Allan, David W.},
    url = {http://www.allanstime.com/AllanVariance/index.html}
}

@article{Black2001,
    title = {{An introduction to Pound–Drever–Hall laser frequency stabilization}},
    year = {2001},
    journal = {American Journal of Physics},
    author = {Black, Eric D.},
    number = {1},
    month = {12},
    pages = {79--87},
    volume = {69},
    publisher = {American Association of Physics TeachersAAPT},
    url = {https://aapt.scitation.org/doi/abs/10.1119/1.1286663},
    doi = {10.1119/1.1286663},
    issn = {0002-9505}
}

@article{Schultze2017,
    title = {{An optically pumped magnetometer working in the light-shift dispersed Mz mode}},
    year = {2017},
    journal = {Sensors (Switzerland)},
    author = {Schultze, Volkmar and Schillig, Bastian and Ijsselsteijn, Rob and Scholtes, Theo and Woetzel, Stefan and Stolz, Ronny},
    number = {3},
    month = {3},
    pages = {561},
    volume = {17},
    publisher = {Multidisciplinary Digital Publishing Institute},
    url = {https://www.mdpi.com/1424-8220/17/3/561},
    doi = {10.3390/s17030561},
    issn = {14248220},
    pmid = {28287414}
}

@article{Gould2008,
    title = {{Animal navigation: The evolution of magnetic orientation}},
    year = {2008},
    journal = {Current Biology},
    author = {Gould, James L.},
    number = {11},
    month = {6},
    pages = {R482-R484},
    volume = {18},
    publisher = {Cell Press},
    url = {https://www.cell.com/current-biology/fulltext/S0960-9822(08)00394-1},
    doi = {10.1016/j.cub.2008.03.052},
    issn = {09609822},
    pmid = {18522823}
}

@book{Corney2006,
    title = {{Atomic and Laser Spectroscopy}},
    year = {2006},
    author = {Corney, Alan},
    publisher = {Oxford University Press},
    address = {}
}

@book{Budker2008,
    title = {{Atomic Physics: An Exploration Through Problems and Solutions}},
    year = {2008},
    booktitle = {Physics Today},
    author = {Budker, Dmitry and Kimball, Derek F. and Demille, David P.},
    edition = {2},
    number = {3},
    month = {7},
    pages = {},
    volume = {},
    publisher = {Oxford University Press},
    url = {https://global.oup.com/academic/product/atomic-physics-9780199532414},
    isbn = {9780199532414}
}

@article{Auzinsh2004,
    title = {{Can a quantum nondemolition measurement improve the sensitivity of an atomic magnetometer?}},
    year = {2004},
    journal = {Physical Review Letters},
    author = {Auzinsh, M. and Budker, D. and Kimball, D. F. and Rochester, S. M. and Stalnaker, J. E. and Sushkov, A. O. and Yashchuk, V. V.},
    number = {17},
    month = {10},
    pages = {173002},
    volume = {93},
    publisher = {American Physical Society},
    url = {https://journals.aps.org/prl/abstract/10.1103/PhysRevLett.93.173002},
    doi = {10.1103/PhysRevLett.93.173002},
    issn = {00319007},
    arxivId = {physics/0403097}
}

@article{Vasilakis2014,
    title = {{Cavity enhanced quantum limited magnetometry}},
    year = {2014},
    journal = {Research in Optical Sciences (2014), paper QTu3B.6},
    author = {Vasilakis, G. and Shen, H. and Jensen, K. and Salart, D. and Balabas, A. Fabricant and Polzik, E. S.},
    month = {3},
    pages = {QTu3B.6},
    publisher = {Optica Publishing Group},
    url = {https://opg.optica.org/abstract.cfm?uri=QIM-2014-QTu3B.6},
    isbn = {9781557529954},
    doi = {10.1364/QIM.2014.QTU3B.6},
    issn = {21622701}
}

@article{Belfi2007,
    title = {{Cesium coherent population trapping magnetometer for cardiosignal detection in an unshielded environment}},
    year = {2007},
    journal = {Journal of the Optical Society of America B},
    author = {Belfi, J. and Bevilacqua, G. and Biancalana, V. and Cartaleva, S. and Dancheva, Y. and Moi, L.},
    number = {9},
    month = {9},
    pages = {2357},
    volume = {24},
    publisher = {Optical Society of America},
    url = {https://www.osapublishing.org/josab/abstract.cfm?uri=josab-24-9-2357},
    doi = {10.1364/josab.24.002357},
    issn = {0740-3224}
}

@article{Schultze2012,
    title = {{Characteristics and performance of an intensity-modulated optically pumped magnetometer in comparison to the classical Mx magnetometer}},
    year = {2012},
    journal = {Optics Express},
    author = {Schultze, Volkmar and IJsselsteijn, Rob and Scholtes, Theo and Woetzel, Stefan and Meyer, Hans-Georg},
    number = {13},
    month = {6},
    pages = {14201},
    volume = {20},
    publisher = {Optical Society of America},
    url = {https://opg.optica.org/oe/abstract.cfm?uri=oe-20-13-14201&id=238230},
    doi = {10.1364/oe.20.014201},
    issn = {1094-4087},
    pmid = {22714483}
}

@article{Li2017,
    title = {{Characterization of high-temperature performance of cesium vapor cells with anti-relaxation coating}},
    year = {2017},
    journal = {Journal of Applied Physics},
    author = {Li, Wenhao and Balabas, Mikhail and Peng, Xiang and Pustelny, Szymon and Wickenbrock, Arne and Guo, Hong and Budker, Dmitry},
    number = {6},
    month = {2},
    pages = {063104},
    volume = {121},
    publisher = {AIP Publishing LLCAIP Publishing},
    url = {https://aip.scitation.org/doi/abs/10.1063/1.4976017},
    doi = {10.1063/1.4976017},
    issn = {10897550},
    arxivId = {1609.04867}
}

@article{Afach2018,
    title = {{Characterization of the global network of optical magnetometers to search for exotic physics (GNOME)}},
    year = {2018},
    journal = {Physics of the Dark Universe},
    author = {Afach, S. and Budker, D. and DeCamp, G. and Dumont, V. and Gruji{\'{c}}, Z. D. and Guo, H. and Jackson Kimball, D. F. and Kornack, T. W. and Lebedev, V. and Li, W. and Masia-Roig, H. and Nix, S. and Padniuk, M. and Palm, C. A. and Pankow, C. and Penaflor, A. and Peng, X. and Pustelny, S. and Scholtes, T. and Smiga, J. A. and Stalnaker, J. E. and Weis, A. and Wickenbrock, A. and Wurm, D.},
    month = {12},
    pages = {162--180},
    volume = {22},
    publisher = {Elsevier},
    url = {https://www.sciencedirect.com/science/article/pii/S2212686418301031},
    doi = {10.1016/j.dark.2018.10.002},
    issn = {22126864},
    arxivId = {1807.09391}
}

@article{Sulai2019,
    title = {{Characterizing atomic magnetic gradiometers for fetal magnetocardiography}},
    year = {2019},
    journal = {Review of Scientific Instruments},
    author = {Sulai, I. A. and Deland, Z. J. and Bulatowicz, M. D. and Wahl, C. P. and Wakai, R. T. and Walker, T. G.},
    number = {8},
    month = {8},
    pages = {085003},
    volume = {90},
    publisher = {AIP Publishing LLCAIP Publishing},
    url = {https://aip.scitation.org/doi/abs/10.1063/1.5091007},
    doi = {10.1063/1.5091007},
    issn = {10897623},
    pmid = {31472627},
    arxivId = {1906.03227}
}

@article{Schwindt2004,
    title = {{Chip-scale atomic magnetometer}},
    year = {2004},
    journal = {Applied Physics Letters},
    author = {Schwindt, Peter D.D. and Knappe, Svenja and Shah, Vishal and Hollberg, Leo and Kitching, John and Liew, Li Anne and Moreland, John},
    number = {26},
    month = {12},
    pages = {6409--6411},
    volume = {85},
    publisher = {American Institute of PhysicsAIP},
    url = {https://aip.scitation.org/doi/abs/10.1063/1.1839274},
    doi = {10.1063/1.1839274},
    issn = {00036951}
}

@article{Schwindt2007,
    title = {{Chip-scale atomic magnetometer with improved sensitivity by use of the Mx technique}},
    year = {2007},
    journal = {Applied Physics Letters},
    author = {Schwindt, Peter D.D. and Lindseth, Brad and Knappe, Svenja and Shah, Vishal and Kitching, John and Liew, Li Anne},
    number = {8},
    month = {2},
    pages = {081102},
    volume = {90},
    publisher = {American Institute of PhysicsAIP},
    url = {https://aip.scitation.org/doi/abs/10.1063/1.2709532},
    doi = {10.1063/1.2709532},
    issn = {00036951}
}

@article{Hong2021,
    title = {{Chip-scale ultra-low field atomic magnetometer based on coherent population trapping}},
    year = {2021},
    journal = {Sensors},
    author = {Hong, Hyun Gue and Park, Sang Eon and Lee, Sang Bum and Heo, Myoung Sun and Park, Jongcheol and Kim, Tae Hyun and Kim, Hee Yeon and Kwon, Taeg Yong},
    number = {4},
    month = {2},
    pages = {1--10},
    volume = {21},
    publisher = {Multidisciplinary Digital Publishing Institute},
    url = {https://www.mdpi.com/1424-8220/21/4/1517},
    doi = {10.3390/s21041517},
    issn = {14248220},
    pmid = {33671625}
}

@article{Terrano2022,
    title = {{Comagnetometer probes of dark matter and new physics}},
    year = {2022},
    journal = {Quantum Science and Technology},
    author = {Terrano, W. A. and Romalis, M. V.},
    number = {1},
    month = {11},
    pages = {014001},
    volume = {7},
    publisher = {IOP Publishing},
    url = {https://iopscience.iop.org/article/10.1088/2058-9565/ac1ae0},
    doi = {10.1088/2058-9565/ac1ae0},
    issn = {20589565},
    arxivId = {2106.09210}
}

@article{JacksonKimball2017,
    title = {{Constraints on long-range spin-gravity and monopole-dipole couplings of the proton}},
    year = {2017},
    journal = {Physical Review D},
    author = {Jackson Kimball, Derek F. and Dudley, Jordan and Li, Yan and Patel, Dilan and Valdez, Julian},
    number = {7},
    month = {10},
    pages = {075004},
    volume = {96},
    publisher = {American Physical Society},
    url = {https://journals.aps.org/prd/abstract/10.1103/PhysRevD.96.075004},
    doi = {10.1103/PhysRevD.96.075004},
    issn = {24700029},
    arxivId = {1707.00745}
}

@misc{Park2019,
    title = {{Construction of a Single Beam SERF Magnetometer using Potassium Atoms for GNOME - Bachelor thesis, Oberlin College}},
    year = {2019},
    author = {Park, Sun Yool},
    month = {4},
    publisher = {Department of Physics and Astronomy of Oberlin College},
    institution = {Oberlin College},
    address = {Oberlin, OH, USA}
}

@article{Li2020,
    title = {{Continuous high-sensitivity and high-bandwidth atomic magnetometer}},
    year = {2020},
    journal = {Physical Review Applied},
    author = {Li, Rujie and Baynes, Fred N. and Luiten, André N. and Perrella, Christopher},
    number = {6},
    month = {12},
    pages = {064067},
    volume = {14},
    publisher = {American Physical Society},
    url = {https://journals.aps.org/prapplied/abstract/10.1103/PhysRevApplied.14.064067},
    doi = {https://doi.org/10.1103/PhysRevApplied.14.064067},
    issn = {23317019}
}

@article{Sebbag2021,
    title = {{Demonstration of an integrated nanophotonic chip-scale alkali vapor magnetometer using inverse design}},
    year = {2021},
    journal = {Light: Science and Applications},
    author = {Sebbag, Yoel and Talker, Eliran and Naiman, Alex and Barash, Yefim and Levy, Uriel},
    number = {1},
    month = {3},
    pages = {1--8},
    volume = {10},
    publisher = {Nature Publishing Group},
    url = {https://www.nature.com/articles/s41377-021-00499-5},
    doi = {10.1038/s41377-021-00499-5},
    issn = {20477538}
}

@article{Vicarini2018,
    title = {{Demonstration of the mass-producible feature of a Cs vapor microcell technology for miniature atomic clocks}},
    year = {2018},
    journal = {Sensors and Actuators, A: Physical},
    author = {Vicarini, R. and Maurice, V. and Abdel Hafiz, M. and Rutkowski, J. and Gorecki, C. and Passilly, N. and Ribetto, L. and Gaff, V. and Volant, V. and Galliou, S. and Boudot, R.},
    month = {9},
    pages = {99--106},
    volume = {280},
    publisher = {Elsevier},
    url = {https://www.sciencedirect.com/science/article/abs/pii/S0924424718305387},
    doi = {10.1016/j.sna.2018.07.032},
    issn = {09244247}
}

@book{Blum2012,
    title = {{Density Matrix Theory and Applications}},
    year = {2012},
    author = {Blum, Karl},
    series = {Springer Series on Atomic, Optical, and Plasma Physics},
    volume = {64},
    publisher = {Springer Berlin Heidelberg},
    url = {http://link.springer.com/10.1007/978-3-642-20561-3},
    address = {Berlin, Heidelberg},
    isbn = {978-3-642-20560-6},
    doi = {10.1007/978-3-642-20561-3}
}

@article{Batie2018,
    title = {{Detection of fetal arrhythmia by using optically pumped magnetometers}},
    year = {2018},
    journal = {JACC: Clinical Electrophysiology},
    author = {Batie, Margo and Bitant, Sarah and Strasburger, Janette F. and Shah, Vishal and Alem, Orang and Wakai, Ronald T.},
    number = {2},
    month = {2},
    pages = {284--287},
    volume = {4},
    publisher = {NIH Public Access},
    url = {https://www.ncbi.nlm.nih.gov/pmc/articles/PMC5841248},
    doi = {10.1016/j.jacep.2017.08.009},
    issn = {2405500X},
    pmid = {29527577}
}

@article{Kowalczyk2021,
    title = {{Detection of human auditory evoked brain signals with a resilient nonlinear optically pumped magnetometer}},
    year = {2021},
    journal = {NeuroImage},
    author = {Kowalczyk, Anna U. and Bezsudnova, Yulia and Jensen, Ole and Barontini, Giovanni},
    month = {2},
    pages = {117497},
    volume = {226},
    publisher = {Academic Press},
    url = {https://www.sciencedirect.com/science/article/pii/S1053811920309824},
    doi = {10.1016/j.neuroimage.2020.117497},
    issn = {10959572},
    pmid = {33132074}
}

@article{Jensen2019,
    title = {{Detection of low-conductivity objects using eddy current measurements with an optical magnetometer}},
    year = {2019},
    journal = {Physical Review Research},
    author = {Jensen, Kasper and Zugenmaier, Michael and Arnbak, Jens and St{\ae}rkind, Hans and Balabas, Mikhail V. and Polzik, Eugene S.},
    number = {3},
    month = {11},
    pages = {033087},
    volume = {1},
    publisher = {American Physical Society},
    url = {https://journals.aps.org/prresearch/abstract/10.1103/PhysRevResearch.1.033087},
    doi = {10.1103/PhysRevResearch.1.033087},
    issn = {26431564},
    arxivId = {1905.01661}
}

@article{Cohen-Tannoudji1969,
    title = {{Detection of the static magnetic field produced by the oriented nuclei of optically pumped 3He gas}},
    year = {1969},
    journal = {Physical Review Letters},
    author = {Cohen-Tannoudji, C. and Dupont-Roc, J. and Haroche, S. and Lalo{\"{e}}, F.},
    number = {15},
    month = {4},
    pages = {758},
    volume = {22},
    publisher = {American Physical Society},
    url = {https://journals.aps.org/prl/abstract/10.1103/PhysRevLett.22.758},
    doi = {10.1103/PhysRevLett.22.758},
    issn = {00319007}
}

@article{Preston1998,
    title = {{Doppler‐free saturated absorption: Laser spectroscopy}},
    year = {1998},
    journal = {American Journal of Physics},
    author = {Preston, Daryl W.},
    number = {11},
    month = {7},
    pages = {1432},
    volume = {64},
    publisher = {American Association of Physics TeachersAAPT},
    url = {https://aapt.scitation.org/doi/abs/10.1119/1.18457},
    doi = {10.1119/1.18457},
    issn = {0002-9505}
}

@article{Belfi2009,
    title = {{Dual channel self-oscillating optical magnetometer}},
    year = {2009},
    journal = {Journal of the Optical Society of America B},
    author = {Belfi, J. and Bevilacqua, G. and Biancalana, V. and Cartaleva, S. and Dancheva, Y. and Khanbekyan, K. and Moi, L.},
    number = {5},
    month = {5},
    pages = {910},
    volume = {26},
    publisher = {Optical Society of America},
    url = {https://www.osapublishing.org/josab/abstract.cfm?uri=josab-26-5-910},
    doi = {10.1364/josab.26.000910},
    issn = {0740-3224},
    arxivId = {0812.1160}
}

@article{Bison2003,
    title = {{Dynamical mapping of the human cardiomagnetic field with a room-temperature, laser-optical sensor}},
    year = {2003},
    journal = {Optics Express},
    author = {Bison, G. and Wynands, R. and Weis, A.},
    number = {8},
    month = {4},
    pages = {904},
    volume = {11},
    publisher = {Optical Society of America},
    url = {https://www.osapublishing.org/oe/abstract.cfm?uri=oe-11-8-904},
    doi = {10.1364/oe.11.000904},
    issn = {10944087}
}

@article{Aschenbrenner1936,
    title = {{Eine Anordnung zur Registrierung rascher Magnetischer St{\"{o}}rungen}},
    year = {1936},
    journal = {Hochfreq. Electroakustik},
    author = {Aschenbrenner, Hans and Goubau, Georg},
    number = {6},
    pages = {177--181},
    volume = {47}
}

@article{Marmugi2020,
    title = {{Electromagnetic induction imaging with atomic magnetometers: Progress and perspectives}},
    year = {2020},
    journal = {Applied Sciences (Switzerland)},
    author = {Marmugi, Luca and Renzoni, Ferruccio},
    number = {18},
    month = {9},
    pages = {6370},
    volume = {10},
    publisher = {Multidisciplinary Digital Publishing Institute},
    url = {https://www.mdpi.com/2076-3417/10/18/6370/htm https://www.mdpi.com/2076-3417/10/18/6370},
    doi = {10.3390/APP10186370},
    issn = {20763417}
}

@article{Bevilacqua2021,
    title = {{Electromagnetic induction imaging: signal detection based on tuned-dressed optical magnetometry}},
    year = {2021},
    journal = {Optics Express, Vol. 29, Issue 23, pp. 37081-37090},
    author = {Bevilacqua, Giuseppe and Biancalana, Valerio and Dancheva, Yordanka and Dancheva, Yordanka and Fregosi, Alessandro and Fregosi, Alessandro and Napoli, Gaetano and Vigilante, Antonio},
    number = {23},
    month = {11},
    pages = {37081--37090},
    volume = {29},
    publisher = {Optica Publishing Group},
    url = {https://opg.optica.org/oe/abstract.cfm?uri=oe-29-23-37081},
    doi = {10.1364/OE.437930},
    issn = {1094-4087},
    pmid = {34808787},
    arxivId = {2107.06819}
}

@phdthesis{Julsgaard2003,
    title = {{Entanglement and Quantum Interactions with Macroscopic Gas Samples}},
    year = {2003},
    author = {Julsgaard, Brian},
    month = {10},
    school = {Center for Quantum Optics (QUANTOP)},
    address = {Aarhus, Denmark}
}

@article{Koshev2021,
    title = {{Evolution of MEG: A first MEG-feasible fluxgate magnetometer}},
    year = {2021},
    journal = {Human Brain Mapping},
    author = {Koshev, Nikolay and Butorina, Anna and Skidchenko, Ekaterina and Kuzmichev, Alexey and Ossadtchi, Alexei and Ostras, Maxim and Fedorov, Maxim and Vetoshko, Petr},
    number = {15},
    month = {10},
    pages = {4844--4856},
    volume = {42},
    publisher = {John Wiley {\&} Sons, Ltd},
    url = {https://onlinelibrary.wiley.com/doi/10.1002/hbm.25582},
    doi = {10.1002/hbm.25582},
    issn = {10970193},
    pmid = {34327772}
}

@article{Nagel1998,
    title = {{Experimental realization of coherent dark-state magnetometers}},
    year = {1998},
    journal = {Europhysics Letters},
    author = {Nagel, A. and Graf, L. and Naumov, A. and Mariotti, E. and Biancalana, V. and Meschede, D. and Wynands, R.},
    number = {1},
    month = {10},
    pages = {31--36},
    volume = {44},
    publisher = {EDP Sciences},
    url = {https://epljournal.edpsciences.org/articles/epl/abs/1998/19/44106/44106.html},
    doi = {10.1209/epl/i1998-00430-0},
    issn = {02955075}
}

@article{Auzinsh2008,
    title = {{F -resolved magneto-optical resonances in the D1 excitation of cesium: Experiment and theory}},
    year = {2008},
    journal = {Physical Review A - Atomic, Molecular, and Optical Physics},
    author = {Auzinsh, M. and Ferber, R. and Gahbauer, F. and Jarmola, A. and Kalvans, L.},
    number = {1},
    month = {7},
    pages = {013417},
    volume = {78},
    publisher = {American Physical Society},
    url = {https://journals.aps.org/pra/abstract/10.1103/PhysRevA.78.013417},
    doi = {10.1103/PhysRevA.78.013417},
    issn = {10502947},
    arxivId = {0803.0201}
}

@article{Lebedev2020,
    title = {{Fast and robust optically pumped cesium magnetometer}},
    year = {2020},
    journal = {Advanced Optical Technologies},
    author = {Lebedev, Victor and Hartwig, Stefan and Middelmann, Thomas},
    number = {5},
    month = {10},
    pages = {275--286},
    volume = {9},
    publisher = {De Gruyter Open Ltd},
    url = {https://www.degruyter.com/document/doi/10.1515/aot-2020-0024/html},
    doi = {10.1515/aot-2020-0024},
    issn = {21928584}
}

@article{Bodenstedt2021,
    title = {{Fast-field-cycling ultralow-field nuclear magnetic relaxation dispersion}},
    year = {2021},
    journal = {Nature Communications},
    author = {Bodenstedt, Sven and Mitchell, Morgan W. and Tayler, Michael C.D.},
    number = {1},
    month = {6},
    pages = {1--8},
    volume = {12},
    publisher = {Nature Publishing Group},
    url = {https://www.nature.com/articles/s41467-021-24248-9},
    doi = {10.1038/s41467-021-24248-9},
    issn = {20411723},
    pmid = {34193862},
    arxivId = {2012.05546}
}

@article{Lucivero2021,
    title = {{Femtotesla direct magnetic gradiometer using a single multipass cell}},
    year = {2021},
    journal = {Physical Review Applied},
    author = {Lucivero, V. G. and Lee, W. and Dural, N. and Romalis, M. V.},
    number = {1},
    month = {1},
    pages = {014004},
    volume = {15},
    publisher = {American Physical Society},
    url = {https://journals.aps.org/prapplied/abstract/10.1103/PhysRevApplied.15.014004},
    doi = {10.1103/PhysRevApplied.15.014004},
    issn = {23317019},
    arxivId = {2009.13212}
}

@article{Florez2021,
    title = {{Floquet description of optically pumped magnetometers}},
    year = {2021},
    journal = {Physical Review A},
    author = {Florez, Hans Marin and Pyragius, Tadas},
    number = {3},
    month = {3},
    pages = {033113},
    volume = {103},
    publisher = {American Physical Society},
    url = {https://link.aps.org/doi/10.1103/PhysRevA.103.033113},
    doi = {10.1103/PhysRevA.103.033113},
    issn = {2469-9926}
}

@article{Jiang2021,
    title = {{Floquet maser}},
    year = {2021},
    journal = {Science Advances},
    author = {Jiang, Min and Su, Haowen and Wu, Ze and Peng, Xinhua and Budker, Dmitry},
    number = {8},
    month = {2},
    pages = {719--736},
    volume = {7},
    publisher = {American Association for the Advancement of Science},
    url = {https://www.science.org/doi/abs/10.1126/sciadv.abe0719},
    doi = {10.1126/sciadv.abe0719},
    issn = {23752548},
    pmid = {33597242}
}

@article{Shah2018,
    title = {{Fully integrated standalone zero field optically pumped magnetometer for biomagnetism}},
    year = {2018},
    journal = {Proc. SPIE 10548, Steep Dispersion Engineering and Opto-Atomic Precision Metrology XI},
    author = {Shah, Vishal and Osborne, James and Orton, Jeff and Alem, Orang},
    month = {2},
    pages = {105481G},
    volume = {10548},
    publisher = {SPIE},
    url = {https://www.spiedigitallibrary.org/conference-proceedings-of-spie/10548/105481G/Fully-integrated-standalone-zero-field-optically-pumped-magnetometer-for-biomagnetism/10.1117/12.2299197.short},
    isbn = {9781510615816},
    doi = {10.1117/12.2299197},
    issn = {1996756X}
}

@misc{GEM2022,
    title = {{GEM GSMP Potassium Magnetometer for High Precision and Accuracy - GEM Systems}},
    url = {https://www.gemsys.ca/ultra-high-sensitivity-potassium}
}

@phdthesis{Krauter2011,
    title = {{Generation and application of entanglement of room temperature ensembles of atoms}},
    year = {2011},
    author = {Krauter, Hanna},
    month = {2},
    url = {https://nbi.ku.dk/english/theses/phd-theses/hanna-krauter/hannah_thesis.pdf},
    school = {University of Copenhagen},
    address = {Copenhagen}
}

@article{Lee2021,
    title = {{Heading errors in all-optical alkali-metal-vapor magnetometers in geomagnetic fields}},
    year = {2021},
    journal = {Physical Review A},
    author = {Lee, W. and Lucivero, V. G. and Romalis, M. V. and Limes, M. E. and Foley, E. L. and Kornack, T. W.},
    number = {6},
    month = {6},
    pages = {063103},
    volume = {103},
    publisher = {American Physical Society},
    url = {https://journals.aps.org/pra/abstract/10.1103/PhysRevA.103.063103},
    doi = {10.1103/PhysRevA.103.063103},
    issn = {24699934},
    arxivId = {2103.01358}
}

@incollection{Heil2017,
    title = {{Helium magnetometers}},
    year = {2017},
    booktitle = {High Sensitvity Magnetometers. Smart Sensors, Measurement and Instrumentation},
    author = {Heil, Werner},
    editor = {Grosz, A. and Haji-Sheikh, M. and Mukhopadhyay, S.},
    pages = {493--521},
    volume = {19},
    publisher = {Springer},
    url = {https://link.springer.com/chapter/10.1007/978-3-319-34070-8_16},
    address = {Cham},
    isbn = {978-3-319-34068-5},
    doi = {10.1007/978-3-319-34070-8}
}

@article{Fourcault2021,
    title = {{Helium-4 magnetometers for room-temperature biomedical imaging: toward collective operation and photon-noise limited sensitivity}},
    year = {2021},
    journal = {Optics Express},
    author = {Fourcault, William and Romain, Rudy and Le Gal, Gwenael and Bertrand, François and Josselin, Vincent and Le Prado, Matthieu and Labyt, Etienne and Palacios-Laloy, Agustin},
    number = {10},
    month = {5},
    pages = {14467},
    volume = {29},
    publisher = {Optical Society of America},
    url = {https://opg.optica.org/oe/fulltext.cfm?uri=oe-29-10-14467&id=450547},
    doi = {10.1364/oe.420031},
    issn = {10944087},
    pmid = {33985169}
}

@article{Cai2020,
    title = {{Herriott-cavity-assisted all-optical atomic vector magnetometer}},
    year = {2020},
    journal = {Physical Review A},
    author = {Cai, B. and Hao, C. P. and Qiu, Z. R. and Yu, Q. Q. and Xiao, W. and Sheng, D.},
    number = {5},
    month = {5},
    pages = {053436},
    volume = {101},
    publisher = {American Physical Society},
    url = {https://journals.aps.org/pra/abstract/10.1103/PhysRevA.101.053436},
    doi = {10.1103/PhysRevA.101.053436},
    issn = {24699934},
    arxivId = {2005.11196}
}

@article{Shah2010,
    title = {{High bandwidth atomic magnetometery with continuous quantum nondemolition measurements}},
    year = {2010},
    journal = {Physical Review Letters},
    author = {Shah, V. and Vasilakis, G. and Romalis, M. V.},
    number = {1},
    month = {1},
    pages = {013601},
    volume = {104},
    publisher = {American Physical Society},
    url = {https://journals.aps.org/prl/abstract/10.1103/PhysRevLett.104.013601},
    doi = {10.1103/PhysRevLett.104.013601},
    issn = {00319007},
    arxivId = {0907.2241}
}

@article{Yang2020,
    title = {{High bandwidth three-axis magnetometer based on optically polarized 85Rb under unshielded environment}},
    year = {2020},
    journal = {Journal of Physics D: Applied Physics},
    author = {Yang, Hongying and Zhang, Ke and Wang, Yanhua and Zhao, Nan},
    number = {6},
    month = {12},
    pages = {065002},
    volume = {53},
    publisher = {IOP Publishing},
    url = {https://iopscience.iop.org/article/10.1088/1361-6463/ab541a},
    doi = {10.1088/1361-6463/ab541a},
    issn = {13616463}
}

@book{Grosz2017,
    title = {{High Sensitivity Magnetometers}},
    year = {2017},
    booktitle = {Smart Sensors, Measurement and Instrumentation},
    editor = {Grosz, Asaf and Haji-Sheikh, Michael J. and Mukhopadhyay, Subhas C.},
    pages = {},
    series = {Smart Sensors, Measurement and Instrumentation},
    volume = {19},
    publisher = {Springer International Publishing},
    url = {http://link.springer.com/10.1007/978-3-319-34070-8},
    address = {Cham, Switzerland},
    isbn = {9783319340685},
    doi = {10.1007/978-3-319-34070-8},
    issn = {21948410}
}

@article{Tiporlini2013,
    title = {{High sensitivity optically pumped quantum magnetometer}},
    year = {2013},
    journal = {The Scientific World Journal},
    author = {Tiporlini, Valentina and Alameh, Kamal},
    volume = {2013},
    url = {https://www.hindawi.com/journals/tswj/2013/858379},
    doi = {10.1155/2013/858379},
    issn = {1537744X},
    pmid = {23766716}
}

@article{Jimenez-Martinez2012,
    title = {{High-bandwidth optical magnetometer}},
    year = {2012},
    journal = {Journal of the Optical Society of America B},
    author = {Jim{\'{e}}nez-Mart{\'{i}}nez, Ricardo and Griffith, W. Clark and Knappe, Svenja and Kitching, John and Prouty, Mark},
    number = {12},
    month = {12},
    pages = {3398},
    volume = {29},
    publisher = {Optical Society of America},
    url = {https://www.osapublishing.org/josab/abstract.cfm?uri=josab-29-12-3398},
    doi = {10.1364/josab.29.003398},
    issn = {0740-3224}
}

@article{Allred2002,
    title = {{High-sensitivity atomic magnetometer unaffected by spin-exchange relaxation}},
    year = {2002},
    journal = {Physical Review Letters},
    author = {Allred, J. C. and Lyman, R. N. and Kornack, T. W. and Romalis, M. V.},
    number = {13},
    month = {9},
    pages = {1308011--1308014},
    volume = {89},
    publisher = {American Physical Society},
    url = {https://journals.aps.org/prl/abstract/10.1103/PhysRevLett.89.130801},
    doi = {10.1103/physrevlett.89.130801},
    issn = {00319007},
    pmid = {12225013}
}

@article{Ishikawa2021,
    title = {{High-temperature lithium atomic magnetometry by symmetric hyperfine coherent population trapping resonances}},
    year = {2021},
    journal = {Journal of the Optical Society of America B},
    author = {Ishikawa, Kiyoshi},
    number = {7},
    month = {7},
    pages = {2155},
    volume = {38},
    publisher = {Optical Society of America},
    url = {https://www.osapublishing.org/josab/abstract.cfm?uri=josab-38-7-2155},
    doi = {10.1364/josab.423749},
    issn = {0740-3224}
}

@article{Gialopsou2021,
    title = {{Improved spatio-temporal measurements of visually evoked fields using optically-pumped magnetometers}},
    year = {2021},
    journal = {Scientific Reports 2021 11:1},
    author = {Gialopsou, Aikaterini and Abel, Christopher and James, T. M. and Coussens, Thomas and Bason, Mark G. and Puddy, Reuben and Di Lorenzo, Francesco and Rolfs, Katharina and Voigt, Jens and Sander, Tilmann and Cercignani, Mara and Kr{\"{u}}ger, Peter},
    number = {1},
    month = {11},
    pages = {1--11},
    volume = {11},
    publisher = {Nature Publishing Group},
    url = {https://www.nature.com/articles/s41598-021-01854-7},
    isbn = {0123456789},
    doi = {10.1038/s41598-021-01854-7},
    issn = {2045-2322},
    pmid = {34789806}
}

@article{Li2022,
    title = {{Improving sensitivity of an amplitude-modulated magneto-optical atomic magnetometer using squeezed light}},
    year = {2022},
    journal = {arXiv},
    author = {Li, Jiahui and Novikova, Irina},
    month = {7},
    url = {https://arxiv.org/abs/2207.12962v1},
    doi = {10.48550/arxiv.2207.12962},
    arxivId = {2207.12962}
}

@article{DeVore2016,
    title = {{Improving student understanding of lock-in amplifiers}},
    year = {2016},
    journal = {American Journal of Physics},
    author = {DeVore, Seth and Gauthier, Alexandre and Levy, Jeremy and Singh, Chandralekha},
    number = {1},
    month = {12},
    pages = {52--56},
    volume = {84},
    publisher = {American Association of Physics TeachersAAPT},
    url = {https://aapt.scitation.org/doi/abs/10.1119/1.4934957},
    doi = {10.1119/1.4934957},
    issn = {0002-9505}
}

@article{Leger2015,
    title = {{In-flight performance of the Absolute Scalar Magnetometer vector mode on board the Swarm satellites}},
    year = {2015},
    journal = {Earth, Planets and Space},
    author = {L{\'{e}}ger, Jean Michel and Jager, Thomas and Bertrand, François and Hulot, Gauthier and Brocco, Laura and Vigneron, Pierre and Lalanne, Xavier and Chulliat, Arnaud and Fratter, Isabelle},
    number = {1},
    month = {4},
    pages = {1--12},
    volume = {67},
    publisher = {Springer Berlin},
    url = {https://earth-planets-space.springeropen.com/articles/10.1186/s40623-015-0231-1},
    doi = {10.1186/s40623-015-0231-1},
    issn = {18805981}
}

@article{Novikova2005,
    title = {{Influence of a buffer gas on nonlinear magneto-optical polarization rotation}},
    year = {2005},
    journal = {Journal of the Optical Society of America B},
    author = {Novikova, Irina and Matsko, Andrey B. and Welch, George R.},
    number = {1},
    month = {1},
    pages = {44},
    volume = {22},
    publisher = {Optica Publishing Group},
    url = {https://opg.optica.org/josab/abstract.cfm?uri=josab-22-1-44},
    doi = {10.1364/josab.22.000044},
    issn = {0740-3224}
}

@article{Knappe-Grueneberg2008,
    title = {{Influence of demagnetization coil configuration on residual field in an extremely magnetically shielded room: Model and measurements}},
    year = {2008},
    journal = {Journal of Applied Physics},
    author = {Knappe-Grueneberg, Silvia and Schnabel, Allard and Wuebbeler, Gerd and Burghoff, Martin},
    number = {7},
    month = {3},
    pages = {07E925},
    volume = {103},
    publisher = {American Institute of PhysicsAIP},
    url = {https://aip.scitation.org/doi/abs/10.1063/1.2837876},
    doi = {10.1063/1.2837876},
    issn = {0021-8979}
}

@article{Oelsner2022,
    title = {{Integrated optically pumped magnetometer for measurements within Earth’s magnetic field}},
    year = {2022},
    journal = {Physical Review Applied},
    author = {Oelsner, G. and IJsselsteijn, R. and Scholtes, T. and Kr{\"{u}}ger, A. and Schultze, V. and Seyffert, G. and Werner, G. and J{\"{a}}ger, M. and Chwala, A. and Stolz, R.},
    number = {2},
    month = {2},
    pages = {024034},
    volume = {17},
    publisher = {American Physical Society (APS)},
    url = {https://journals.aps.org/prapplied/abstract/10.1103/PhysRevApplied.17.024034},
    doi = {10.1103/physrevapplied.17.024034},
    issn = {23317019},
    arxivId = {2008.01570}
}

@article{Campbell2021,
    title = {{Intrinsic pulsed magnetic gradiometer in Earth's field}},
    year = {2021},
    journal = {arXiv},
    author = {Campbell, Kaleb and Wang, Ying-Ju and Savukov, Igor and Schwindt, Peter and Jau, Yuan-Yu and Shah, Vishal},
    month = {11},
    url = {https://arxiv.org/abs/2111.12310v1},
    doi = {10.48550/arxiv.2111.12310},
    arxivId = {2111.12310}
}

@article{Scholtes2014,
    title = {{Intrinsic relaxation rates of polarized Cs vapor in miniaturized cells}},
    year = {2014},
    journal = {Applied Physics B: Lasers and Optics},
    author = {Scholtes, Theo and Woetzel, Stefan and IJsselsteijn, Rob and Schultze, Volkmar and Meyer, Hans Georg},
    number = {1},
    month = {10},
    pages = {211--218},
    volume = {117},
    publisher = {Springer Verlag},
    url = {https://link.springer.com/article/10.1007/s00340-014-5824-z},
    doi = {10.1007/S00340-014-5824-z},
    issn = {09462171}
}

@article{Seltzer2010,
    title = {{Investigation of antirelaxation coatings for alkali-metal vapor cells using surface science techniques}},
    year = {2010},
    journal = {Journal of Chemical Physics},
    author = {Seltzer, S. J. and Michalak, D. J. and Donaldson, M. H. and Balabas, M. V. and Barber, S. K. and Bernasek, S. L. and Bouchiat, M. A. and Hexemer, A. and Hibberd, A. M. and Kimball, D. F.Jackson and Jaye, C. and Karaulanov, T. and Narducci, F. A. and Rangwala, S. A. and Robinson, H. G. and Shmakov, A. K. and Voronov, D. L. and Yashchuk, V. V. and Pines, A. and Budker, D.},
    number = {14},
    month = {10},
    pages = {144703},
    volume = {133},
    publisher = {American Institute of PhysicsAIP},
    url = {https://aip.scitation.org/doi/abs/10.1063/1.3489922},
    doi = {10.1063/1.3489922},
    issn = {00219606},
    arxivId = {1002.4417}
}

@article{Pouliot2018,
    title = {{Investigations of optical pumping for magnetometry using an auto-locking laser system}},
    year = {2018},
    journal = {Proc. SPIE 10637, Laser Technology for Defense and Security XIV},
    author = {Pouliot, Alexander and Beica, Hermina C. and Carew, Adam and Barrett, Brynle and Vorozcovs, Andrew and Carlse, Gehrig and Kumarakrishnan, Anantharaman},
    month = {5},
    pages = {40--47},
    volume = {10637},
    publisher = {SPIE},
    url = {https://www.spiedigitallibrary.org/conference-proceedings-of-spie/10637/106370A/Investigations-of-optical-pumping-for-magnetometry-using-an-auto-locking/10.1117/12.2304598.full},
    isbn = {9781510617858},
    doi = {10.1117/12.2304598},
    issn = {1996756X},
    arxivId = {1805.07391}
}

@article{Tayler2017,
    title = {{Invited Review Article: Instrumentation for nuclear magnetic resonance in zero and ultralow magnetic field}},
    year = {2017},
    journal = {Review of Scientific Instruments},
    author = {Tayler, Michael C.D. and Theis, Thomas and Sjolander, Tobias F. and Blanchard, John W. and Kentner, Arne and Pustelny, Szymon and Pines, Alexander and Budker, Dmitry},
    number = {9},
    month = {9},
    pages = {091101},
    volume = {88},
    publisher = {AIP Publishing LLC AIP Publishing},
    url = {https://aip.scitation.org/doi/abs/10.1063/1.5003347},
    doi = {10.1063/1.5003347},
    issn = {10897623}
}

@article{Han2017,
    title = {{Is light narrowing possible with dense-vapor paraffin coated cells for atomic magnetometers?}},
    year = {2017},
    journal = {AIP Advances},
    author = {Han, Runqi and Balabas, Mikhail and Hovde, Chris and Li, Wenhao and Roig, Hector Masia and Wang, Tao and Wickenbrock, Arne and Zhivun, Elena and You, Zheng and Budker, Dmitry},
    number = {12},
    month = {12},
    pages = {125224},
    volume = {7},
    publisher = {AIP Publishing LLCAIP Publishing},
    url = {https://aip.scitation.org/doi/abs/10.1063/1.4997691},
    doi = {10.1063/1.4997691},
    issn = {21583226}
}

@inproceedings{Pratt2021,
    title = {{Kernel Flux: a whole-head 432-magnetometer optically-pumped magnetoencephalography (OP-MEG) system for brain activity imaging during natural human experiences}},
    year = {2021},
    booktitle = {Proc. SPIE 11700, Optical and Quantum Sensing and Precision Metrology},
    author = {Pratt, Ethan J and Ledbetter, Micah and Jim{\'{e}}nez-Mart{\'{i}}nez, Ricardo and Shapiro, Benjamin and Solon, Amelia and Iwata, Geoffrey Z and Garber, Steve and Gormley, Jeff and Decker, Dakota and Delgadillo, David and Dellis, Argyrios T and Phillips, Jake and Sundar, Guhan and Leung, Jerry and Coyne, Jim and McKinley, Mike and Lopez, Gilbert and Homan, Scott and Marsh, Lucas and Zhang, Mary and Maurice, Vincent and Siepser, Benjamin and Giovannoli, Teresa and Leverett, Brandon and Lerner, Gabriel and Seidman, Scott and DeLuna, Vicente and Wright-Freeman, Kayla and Kates-Harbeck, Julian and Lasser, Teague and Mohseni, Hooman and Sharp, TJ and Zorzos, Anthony and Lara, Antonio H and Kouhzadi, Ali and Ojeda, Alejandro and Chopra, Pronoy and Bednarke, Zachary and Henninger, Michael and Alford, Jamu K},
    number = {5},
    month = {3},
    pages = {1170032},
    volume = {11700},
    publisher = {SPIE},
    url = {https://www.spiedigitallibrary.org/conference-proceedings-of-spie/11700/1170032/Kernel-Flux--a-whole-head-432-magnetometer-optically-pumped/10.1117/12.2581794.full},
    isbn = {9781510642355},
    doi = {10.1117/12.2581794},
    issn = {1996756X}
}

@book{Milonni2010,
    title = {{Laser Physics}},
    year = {2010},
    author = {Milonni, Peter W. and Eberly, Joseph H.},
    month = {3},
    publisher = {John Wiley {\&} Sons, Inc.},
    url = {https://onlinelibrary.wiley.com/doi/book/10.1002/9780470409718},
    address = {Hoboken, NJ, USA},
    isbn = {9780470387719},
    doi = {10.1002/9780470409718}
}

@article{Aleksandrov1995,
    title = {{Laser pumping in the scheme of an Mx-magnetometer}},
    year = {1995},
    journal = {Optics and Spectroscopy (English translation of Optika i Spektroskopiya)},
    author = {Aleksandrov, E. B. and Balabas, M. V. and Vershovskii, A. K. and Ivanov, A. E. and Yakobson, N. N. and Velichanskii, V. L. and Senkov, N. V.},
    number = {2},
    pages = {292--298},
    volume = {78},
    url = {https://ui.adsabs.harvard.edu/abs/1995OptSp..78..292A/abstract},
    issn = {0030400X}
}

@article{Groeger2006SAA,
    title = {{Laser-pumped cesium magnetometers for high-resolution medical and fundamental research}},
    year = {2006},
    journal = {Sensors and Actuators, A: Physical},
    author = {Groeger, S. and Bison, G. and Knowles, P. E. and Wynands, R. and Weis, A.},
    number = {1-2 SPEC. ISS.},
    month = {5},
    pages = {1--5},
    volume = {129},
    publisher = {Elsevier},
    doi = {10.1016/j.sna.2005.09.036},
    issn = {09244247}
}

@article{Brazhnikov2022,
    title = {{Level-crossing resonances on open atomic transitions in a buffered Cs vapor cell: Linewidth narrowing, high contrast, and atomic magnetometry applications}},
    year = {2022},
    journal = {Physical Review A},
    author = {Brazhnikov, D. V. and Vishnyakov, V. I. and Goncharov, A. N. and Alipieva, E. and Andreeva, C. and Taskova, E.},
    number = {1},
    month = {7},
    pages = {013113},
    volume = {106},
    publisher = {American Physical Society},
    url = {https://journals.aps.org/pra/abstract/10.1103/PhysRevA.106.013113},
    doi = {10.1103/PhysRevA.106.013113},
    issn = {24699934}
}

@article{Karlen2017,
    title = {{Lifetime assessment of RbN{\_}3-filled MEMS atomic vapor cells with Al{\_}2O{\_}3 coating}},
    year = {2017},
    journal = {Optics Express},
    author = {Karlen, Sylvain and Gobet, Jean and Overstolz, Thomas and Haesler, Jacques and Lecomte, Steve},
    number = {3},
    month = {2},
    pages = {2187},
    volume = {25},
    publisher = {Optica Publishing Group},
    url = {https://opg.optica.org/oe/abstract.cfm?uri=oe-25-3-2187},
    doi = {10.1364/oe.25.002187},
    issn = {10944087},
    pmid = {29519066}
}

@article{Blanchard2021,
    title = {{Lower than low: Perspectives on zero- to ultralow-field nuclear magnetic resonance}},
    year = {2021},
    journal = {Journal of Magnetic Resonance},
    author = {Blanchard, John W. and Budker, Dmitry and Trabesinger, Andreas},
    month = {2},
    pages = {106886},
    volume = {323},
    publisher = {Academic Press},
    url = {https://www.sciencedirect.com/science/article/abs/pii/S1090780720302044},
    doi = {10.1016/j.jmr.2020.106886},
    issn = {10907807},
    pmid = {33518173}
}

@article{Jiang2019,
    title = {{Magnetic gradiometer for the detection of zero- to ultralow-field nuclear magnetic resonance}},
    year = {2019},
    journal = {Physical Review Applied},
    author = {Jiang, Min and Frutos, Román Picazo and Wu, Teng and Blanchard, John W. and Peng, Xinhua and Budker, Dmitry},
    number = {2},
    month = {2},
    pages = {024005},
    volume = {11},
    publisher = {American Physical Society},
    url = {https://journals.aps.org/prapplied/abstract/10.1103/PhysRevApplied.11.024005},
    doi = {10.1103/PhysRevApplied.11.024005},
    issn = {23317019},
    arxivId = {1808.02743}
}

@article{Johnson2012,
    title = {{Magnetic relaxometry with an atomic magnetometer and SQUID sensors on targeted cancer cells}},
    year = {2012},
    journal = {Journal of Magnetism and Magnetic Materials},
    author = {Johnson, Cort and Adolphi, Natalie L. and Butler, Kimberly L. and Lovato, Debbie M. and Larson, Richard and Schwindt, Peter D.D. and Flynn, Edward R.},
    number = {17},
    month = {8},
    pages = {2613--2619},
    volume = {324},
    publisher = {North-Holland},
    url = {https://www.sciencedirect.com/science/article/pii/S0304885312002582},
    doi = {10.1016/j.jmmm.2012.03.015},
    issn = {03048853}
}

@incollection{Weis2017,
    title = {{Magnetic resonance based atomic magnetometers}},
    year = {2017},
    booktitle = {High Sensitivity Magnetometers. Smart Sensors, Measurement and Instrumentation},
    author = {Weis, Antoine and Bison, Georg and Gruji{\'{c}}, Zoran D.},
    editor = {Grosz, A. and Haji-Sheikh, M. and Mukhopadhyay, S.},
    chapter = {},
    pages = {361--424},
    volume = {19},
    publisher = {Springer},
    url = {https://link.springer.com/chapter/10.1007/978-3-319-34070-8_13},
    address = {Cham},
    isbn = {978-3-319-34068-5},
    doi = {10.1007/978-3-319-34070-8}
}

@article{Xu2006,
    title = {{Magnetic resonance imaging with an optical atomic magnetometer}},
    year = {2006},
    journal = {Proceedings of the National Academy of Sciences of the United States of America},
    author = {Xu, Shoujun and Yashchuk, Valeriy V. and Donaldson, Marcus H. and Rochester, Simon M. and Budker, Dmitry and Pines, Alexander},
    number = {34},
    month = {8},
    pages = {12668--12671},
    volume = {103},
    publisher = {National Academy of Sciences},
    url = {https://www.pnas.org/content/103/34/12668 https://www.pnas.org/content/103/34/12668.abstract},
    doi = {10.1073/pnas.0605396103},
    issn = {00278424},
    pmid = {16885210}
}

@article{Sewell2012,
    title = {{Magnetic sensitivity beyond the projection noise limit by spin squeezing}},
    year = {2012},
    journal = {Physical Review Letters},
    author = {Sewell, R. J. and Koschorreck, M. and Napolitano, M. and Dubost, B. and Behbood, N. and Mitchell, M. W.},
    number = {25},
    month = {12},
    pages = {253605},
    volume = {109},
    publisher = {American Physical Society},
    url = {https://journals.aps.org/prl/abstract/10.1103/PhysRevLett.109.253605},
    doi = {10.1103/PhysRevLett.109.253605},
    issn = {00319007},
    arxivId = {1111.6969}
}

@article{Xu2021,
    title = {{Magnetic sensitivity of cryptochrome 4 from a migratory songbird}},
    year = {2021},
    journal = {Nature},
    author = {Xu, Jingjing and Jarocha, Lauren E. and Zollitsch, Tilo and Konowalczyk, Marcin and Henbest, Kevin B. and Richert, Sabine and Golesworthy, Matthew J. and Schmidt, Jessica and D{\'{e}}jean, Victoire and Sowood, Daniel J.C. and Bassetto, Marco and Luo, Jiate and Walton, Jessica R. and Fleming, Jessica and Wei, Yujing and Pitcher, Tommy L. and Moise, Gabriel and Herrmann, Maike and Yin, Hang and Wu, Haijia and Bart{\"{o}}lke, Rabea and K{\"{a}}sehagen, Stefanie J. and Horst, Simon and Dautaj, Glen and Murton, Patrick D.F. and Gehrckens, Angela S. and Chelliah, Yogarany and Takahashi, Joseph S. and Koch, Karl Wilhelm and Weber, Stefan and Solov’yov, Ilia A. and Xie, Can and Mackenzie, Stuart R. and Timmel, Christiane R. and Mouritsen, Henrik and Hore, P. J.},
    number = {7864},
    month = {6},
    pages = {535--540},
    volume = {594},
    publisher = {Nature Publishing Group},
    url = {https://www.nature.com/articles/s41586-021-03618-9},
    doi = {10.1038/s41586-021-03618-9},
    issn = {1476-4687},
    pmid = {34163056}
}

@article{Papoyan2016,
    title = {{Magnetic-field-compensation optical vector magnetometer}},
    year = {2016},
    journal = {Applied Optics},
    author = {Papoyan, Aram and Shmavonyan, Svetlana and Khanbekyan, Alen and Khanbekyan, Karen and Marinelli, Carmela and Mariotti, Emilio},
    number = {4},
    month = {2},
    pages = {892},
    volume = {55},
    publisher = {Optica Publishing Group},
    url = {https://opg.optica.org/ao/abstract.cfm?uri=ao-55-4-892},
    doi = {10.1364/ao.55.000892},
    issn = {0003-6935}
}

@article{Zhang2021IEEE,
    title = {{Magnetocardiography measurements by microfabricated atomic magnetometer with a 3-D spherical alkali vapor cell}},
    year = {2021},
    journal = {IEEE Transactions on Instrumentation and Measurement},
    author = {Zhang, Jin and Liu, Kangni and Zhang, Jianfeng and Wang, Ziji and Shang, Jintang},
    volume = {70},
    publisher = {Institute of Electrical and Electronics Engineers Inc.},
    url = {https://ieeexplore.ieee.org/document/9585396},
    doi = {10.1109/TIM.2021.3120375},
    issn = {15579662}
}

@article{Jensen2018,
    title = {{Magnetocardiography on an isolated animal heart with a room-temperature optically pumped magnetometer}},
    year = {2018},
    journal = {Scientific Reports},
    author = {Jensen, Kasper and Skarsfeldt, Mark Alexander and St{\ae}rkind, Hans and Arnbak, Jens and Balabas, Mikhail V. and Olesen, Søren Peter and Bentzen, Bo Hjorth and Polzik, Eugene S.},
    number = {1},
    month = {11},
    pages = {1--9},
    volume = {8},
    publisher = {Nature Publishing Group},
    url = {https://www.nature.com/articles/s41598-018-34535-z},
    doi = {10.1038/s41598-018-34535-z},
    issn = {20452322},
    pmid = {30385784},
    arxivId = {1806.10954}
}

@article{Sheng2017,
    title = {{Magnetoencephalography with a Cs-based high-sensitivity compact atomic magnetometer}},
    year = {2017},
    journal = {Review of Scientific Instruments},
    author = {Sheng, Jingwei and Wan, Shuangai and Sun, Yifan and Dou, Rongshe and Guo, Yuhao and Wei, Kequan and He, Kaiyan and Qin, Jie and Gao, Jia Hong},
    number = {9},
    month = {9},
    volume = {88},
    publisher = {Rev Sci Instrum},
    url = {https://pubmed.ncbi.nlm.nih.gov/28964239},
    doi = {10.1063/1.5001730},
    issn = {10897623},
    pmid = {28964239}
}

@article{Xia2006,
    title = {{Magnetoencephalography with an atomic magnetometer}},
    year = {2006},
    journal = {Applied Physics Letters},
    author = {Xia, H. and Ben-Amar Baranga, A. and Hoffman, D. and Romalis, M. V.},
    number = {21},
    month = {11},
    pages = {211104},
    volume = {89},
    publisher = {American Institute of PhysicsAIP},
    url = {https://aip.scitation.org/doi/abs/10.1063/1.2392722},
    doi = {10.1063/1.2392722},
    issn = {00036951}
}

@article{Matsko2005,
    title = {{Magnetometer based on the opto-electronic microwave oscillator}},
    year = {2005},
    journal = {Optics Communications},
    author = {Matsko, Andrey B. and Strekalov, Dmitry and Maleki, Lute},
    number = {1-3},
    month = {3},
    pages = {141--148},
    volume = {247},
    publisher = {North-Holland},
    url = {https://www.sciencedirect.com/science/article/abs/pii/S0030401804011757},
    doi = {10.1016/j.optcom.2004.11.047},
    issn = {00304018}
}

@incollection{Gawlik2013,
    title = {{Magnetometry with cold atoms}},
    year = {2013},
    booktitle = {Optical Magnetometry},
    author = {Gawlik, W. and Higbie, J. M.},
    editor = {Budker, Dmitry and Jackson Kimball, Derek F.},
    chapter = {9},
    month = {5},
    pages = {167--189},
    volume = {},
    publisher = {Cambridge University Press},
    url = {https://www.cambridge.org/core/books/optical-magnetometry/magnetometry-with-cold-atoms/AE8C14285BFF2E77508641AFCABC47A9},
    isbn = {9780511846380},
    doi = {10.1017/CBO9780511846380.010}
}

@incollection{Jensen2017,
    title = {{Magnetometry with nitrogen-vacancy centers in diamond}},
    year = {2017},
    booktitle = {High Sensitivity Magnetometers. Smart Sensors, Measurement and Instrumentation},
    author = {Jensen, Kasper and Kehayias, Pauli and Budker, Dmitry},
    editor = {Grosz, A. and Haji-Sheikh, M. and Mukhopadhyay, S.},
    pages = {553--576},
    volume = {19},
    publisher = {Springer},
    url = {https://link.springer.com/chapter/10.1007/978-3-319-34070-8_18},
    address = {Cham},
    isbn = {978-3-319-34068-5},
    doi = {10.1007/978-3-319-34070-8}
}

@article{Hoffmann1982,
    title = {{Magnetoresistance and non-Ohmic conductivity of thin platinum films at low temperatures}},
    year = {1982},
    journal = {Physical Review B},
    author = {Hoffmann, H. and Hofmann, F. and Schoepe, W.},
    number = {8},
    month = {4},
    pages = {5563--5565},
    volume = {25},
    publisher = {American Physical Society},
    url = {https://journals.aps.org/prb/abstract/10.1103/PhysRevB.25.5563},
    doi = {10.1103/PhysRevB.25.5563},
    issn = {01631829}
}

@article{Mora2019,
    title = {{Measurement of the ratio between g-factors of the ground states of 87Rb and 85Rb}},
    year = {2019},
    journal = {Annalen der Physik},
    author = {Mora, Jason and Cobos, Aracely and Fuentes, Dominic and Jackson Kimball, Derek F.},
    number = {5},
    month = {5},
    pages = {1800281},
    volume = {531},
    publisher = {John Wiley {\&} Sons, Ltd},
    url = {https://onlinelibrary.wiley.com/doi/10.1002/andp.201800281},
    doi = {10.1002/andp.201800281},
    issn = {15213889},
    arxivId = {1809.04053}
}

@article{Woetzel2011,
    title = {{Microfabricated atomic vapor cell arrays for magnetic field measurements}},
    year = {2011},
    journal = {Review of Scientific Instruments},
    author = {Woetzel, S. and Schultze, V. and Ijsselsteijn, R. and Schulz, T. and Anders, S. and Stolz, R. and Meyer, H. G.},
    number = {3},
    month = {3},
    pages = {033111},
    volume = {82},
    publisher = {American Institute of PhysicsAIP},
    url = {https://aip.scitation.org/doi/abs/10.1063/1.3559304},
    doi = {10.1063/1.3559304},
    issn = {00346748},
    pmid = {21456722}
}

@article{Korth2016,
    title = {{Miniature atomic scalar magnetometer for space based on the rubidium isotope 87Rb}},
    year = {2016},
    journal = {Journal of Geophysical Research: Space Physics},
    author = {Korth, Haje and Strohbehn, Kim and Tejada, Francisco and Andreou, Andreas G. and Kitching, John and Knappe, Svenja and Lehtonen, S. John and London, Shaughn M. and Kafel, Matiwos},
    number = {8},
    month = {8},
    pages = {7870--7880},
    volume = {121},
    publisher = {John Wiley {\&} Sons, Ltd},
    url = {https://agupubs.onlinelibrary.wiley.com/doi/10.1002/2016JA022389},
    isbn = {10.1002/2016},
    doi = {10.1002/2016JA022389},
    issn = {21699402}
}

@article{Aleksandrov2009,
    title = {{Modern radio-optical methods in quantum magnetometry}},
    year = {2009},
    journal = {Physics-Uspekhi},
    author = {Aleksandrov, Evgenii B. and Vershovskii, Anton K.},
    number = {6},
    month = {6},
    pages = {573--601},
    volume = {52},
    publisher = {Uspekhi Fizicheskikh Nauk (UFN) Journal},
    url = {https://iopscience.iop.org/article/10.3367/UFNe.0179.200906f.0605},
    issn = {1063-7869}
}

@article{Boto2018,
    title = {{Moving magnetoencephalography towards real-world applications with a wearable system}},
    year = {2018},
    journal = {Nature},
    author = {Boto, Elena and Holmes, Niall and Leggett, James and Roberts, Gillian and Shah, Vishal and Meyer, Sofie S. and Mu{\~{n}}oz, Leonardo Duque and Mullinger, Karen J. and Tierney, Tim M. and Bestmann, Sven and Barnes, Gareth R. and Bowtell, Richard and Brookes, Matthew J.},
    number = {7698},
    month = {3},
    pages = {657--661},
    volume = {555},
    publisher = {Nature Publishing Group},
    url = {https://www.nature.com/articles/nature26147},
    doi = {10.1038/nature26147},
    issn = {14764687},
    pmid = {29562238}
}

@article{Allen1972,
    title = {{Narrow line rubidium magnetometer for high accuracy field measurements}},
    year = {1972},
    journal = {Journal of geomagnetism and geoelectricity},
    author = {Allen, J. H. and Bender, P. L.},
    number = {1},
    pages = {105--125},
    volume = {24},
    publisher = {Society of Geomagnetism and Earth, Planetary and Space Sciences},
    url = {https://www.jstage.jst.go.jp/article/jgg1949/24/1/24_1_105/_article},
    doi = {10.5636/jgg.24.105},
    issn = {00221392}
}

@article{Holmes2021,
    title = {{Naturalistic hyperscanning with wearable magnetoencephalography}},
    year = {2021},
    journal = {bioRxiv},
    author = {Holmes, Niall and Rea, Molly and Hill, Ryan M. and Boto, Elena and Stuart, Andrew and Leggett, James and Edwards, Lucy J. and Rhodes, Natalie and Shah, Vishal and Osborne, James and Fromhold, T. Mark and Glover, Paul and Montague, P. Read and Brookes, Matthew J. and Bowtell, Richard},
    month = {9},
    pages = {2021.09.07.459124},
    publisher = {Cold Spring Harbor Laboratory},
    url = {https://www.biorxiv.org/content/10.1101/2021.09.07.459124v1},
    doi = {10.1101/2021.09.07.459124}
}

@article{Savukov2005,
    title = {{NMR detection with an atomic magnetometer}},
    year = {2005},
    journal = {Physical Review Letters},
    author = {Savukov, I. M. and Romalis, M. V.},
    number = {12},
    month = {4},
    pages = {123001},
    volume = {94},
    publisher = {American Physical Society},
    url = {https://journals.aps.org/prl/abstract/10.1103/PhysRevLett.94.123001},
    doi = {10.1103/PhysRevLett.94.123001},
    issn = {00319007},
    arxivId = {physics/0411163}
}

@article{Jensen2016,
    title = {{Non-invasive detection of animal nerve impulses with an atomic magnetometer operating near quantum limited sensitivity}},
    year = {2016},
    journal = {Scientific Reports},
    author = {Jensen, Kasper and Budvytyte, Rima and Thomas, Rodrigo A. and Wang, Tian and Fuchs, Annette M. and Balabas, Mikhail V. and Vasilakis, Georgios and Mosgaard, Lars D. and St{\ae}rkind, Hans C. and Muller, Jorg H. and Heimburg, Thomas and Olesen, Søren Peter and Polzik, Eugene S.},
    number = {1},
    month = {7},
    pages = {1--7},
    volume = {6},
    publisher = {Nature Publishing Group},
    url = {https://www.nature.com/articles/srep29638},
    doi = {10.1038/srep29638},
    issn = {20452322},
    pmid = {27417378},
    arxivId = {1601.03273}
}

@article{Borna2020,
    title = {{Non-invasive functional-brain-imaging with an OPM-based magnetoencephalography system}},
    year = {2020},
    journal = {PLOS ONE},
    author = {Borna, Amir and Carter, Tony R. and Colombo, Anthony P. and Jau, Yuan Yu and McKay, Jim and Weisend, Michael and Taulu, Samu and Stephen, Julia M. and Schwindt, Peter D.D.},
    number = {1},
    month = {1},
    pages = {e0227684},
    volume = {15},
    publisher = {Public Library of Science},
    url = {https://journals.plos.org/plosone/article?id=10.1371/journal.pone.0227684},
    isbn = {1111111111},
    doi = {10.1371/JOURNAL.PONE.0227684},
    issn = {1932-6203},
    pmid = {31978102}
}

@article{Bougas2018,
    title = {{Nondestructive in-line sub-picomolar detection of magnetic nanoparticles in flowing complex fluids}},
    year = {2018},
    journal = {Scientific Reports},
    author = {Bougas, Lykourgos and Langenegger, Lukas D. and Mora, Carlos A. and Zeltner, Martin and Stark, Wendelin J. and Wickenbrock, Arne and Blanchard, John W. and Budker, Dmitry},
    number = {1},
    month = {2},
    pages = {1--8},
    volume = {8},
    publisher = {Nature Publishing Group},
    url = {https://www.nature.com/articles/s41598-018-21802-2},
    doi = {10.1038/s41598-018-21802-2},
    issn = {20452322},
    pmid = {29472727},
    arxivId = {1801.05665}
}

@article{Budker1999,
    title = {{Nonlinear laser spectroscopy and magneto-optics}},
    year = {1999},
    journal = {American Journal of Physics},
    author = {Budker, Dmitry and Orlando, Donald J. and Yashchuk, Valeriy},
    number = {7},
    month = {7},
    pages = {584--592},
    volume = {67},
    publisher = {American Association of Physics TeachersAAPT},
    url = {https://aapt.scitation.org/doi/abs/10.1119/1.19328},
    doi = {10.1119/1.19328},
    issn = {0002-9505}
}

@article{Gawlik2006,
    title = {{Nonlinear magneto-optical rotation with amplitude modulated light}},
    year = {2006},
    journal = {Applied Physics Letters},
    author = {Gawlik, W. and Krzemie{\'{n}}, L. and Pustelny, S. and Sangla, D. and Zachorowski, J. and Graf, M. and Sushkov, A. O. and Budker, D.},
    number = {13},
    month = {3},
    pages = {131108},
    volume = {88},
    publisher = {American Institute of PhysicsAIP},
    url = {https://aip.scitation.org/doi/abs/10.1063/1.2190457},
    doi = {10.1063/1.2190457},
    issn = {0003-6951},
    arxivId = {physics/0510207}
}

@article{Budker2002PRA,
    title = {{Nonlinear magneto-optical rotation with frequency-modulated light}},
    year = {2002},
    journal = {Physical Review A - Atomic, Molecular, and Optical Physics},
    author = {Budker, D. and Kimball, D. F. and Yashchuk, V. V. and Zolotorev, M.},
    number = {5},
    month = {5},
    pages = {4},
    volume = {65},
    publisher = {American Physical Society},
    url = {https://journals.aps.org/pra/abstract/10.1103/PhysRevA.65.055403},
    doi = {10.1103/PhysRevA.65.055403},
    issn = {10941622},
    arxivId = {physics/0201025}
}

@article{Maul2018,
    title = {{Nuclear hyperpolarization of He 3 by magnetized plasmas}},
    year = {2018},
    journal = {Physical Review A},
    author = {Maul, A. and Bl{\"{u}}mler, P. and Nacher, P. J. and Otten, E. and Tastevin, G. and Heil, W.},
    number = {6},
    month = {12},
    pages = {063405},
    volume = {98},
    publisher = {American Physical Society},
    url = {https://journals.aps.org/pra/abstract/10.1103/PhysRevA.98.063405},
    doi = {10.1103/PhysRevA.98.063405},
    issn = {24699934}
}

@article{Wang2014,
    title = {{Optical continuous-variable quadratic phase gate via Faraday interaction}},
    year = {2014},
    journal = {Optics Express},
    author = {Wang, Ming-Feng and Jiang, Nian-Quan and Zheng, Yi-Zhuang},
    number = {8},
    month = {4},
    pages = {9182},
    volume = {22},
    publisher = {Optica Publishing Group},
    url = {https://opg.optica.org/oe/abstract.cfm?uri=oe-22-8-9182},
    doi = {10.1364/oe.22.009182},
    issn = {10944087}
}

@article{Begus2017,
    title = {{Optical detection of low frequency NQR signals: A step forward from conventional NQR}},
    year = {2017},
    journal = {Journal of Physics D: Applied Physics},
    author = {Begus, S. and Pirnat, J. and Jazbinsek, V. and Trontelj, Z.},
    number = {9},
    month = {2},
    pages = {095601},
    volume = {50},
    publisher = {IOP Publishing},
    url = {https://iopscience.iop.org/article/10.1088/1361-6463/aa4f23},
    doi = {10.1088/1361-6463/aa4f23},
    issn = {13616463}
}

@article{Budker2007,
    title = {{Optical magnetometry}},
    year = {2007},
    journal = {Nature Physics},
    author = {Budker, Dmitry and Romalis, Michael},
    number = {4},
    month = {4},
    pages = {227--234},
    volume = {3},
    publisher = {Nature Publishing Group},
    url = {https://www.nature.com/articles/nphys566},
    doi = {10.1038/nphys566},
    issn = {17452481},
    arxivId = {physics/0611246}
}

@book{Budker2013,
    title = {{Optical Magnetometry}},
    year = {2013},
    editor = {Budker, Dmitry and Jackson Kimball, Derek F.},
    month = {5},
    publisher = {Cambridge University Press},
    url = {http://ebooks.cambridge.org/ref/id/CBO9780511846380},
    address = {Cambridge},
    isbn = {9780511846380},
    doi = {10.1017/CBO9780511846380}
}

@article{Lembke2014,
    title = {{Optical multichannel room temperature magnetic field imaging system for clinical application}},
    year = {2014},
    journal = {Biomedical Optics Express},
    author = {Lembke, G and Ern{\'{e}}, S N and Nowak, H and Menhorn, B and Pasquarelli, A and Bison, G},
    number = {3},
    month = {3},
    pages = {876},
    volume = {5},
    publisher = {Optica Publishing Group},
    url = {https://opg.optica.org/boe/abstract.cfm?uri=boe-5-3-876},
    doi = {10.1364/boe.5.000876},
    issn = {2156-7085}
}

@article{Happer1972,
    title = {{Optical pumping}},
    year = {1972},
    journal = {Reviews of Modern Physics},
    author = {Happer, William},
    number = {2},
    month = {4},
    pages = {169--249},
    volume = {44},
    publisher = {American Physical Society},
    url = {https://journals.aps.org/rmp/abstract/10.1103/RevModPhys.44.169},
    doi = {10.1103/RevModPhys.44.169},
    issn = {00346861}
}

@article{Bell1961,
    title = {{Optically driven spin precession}},
    year = {1961},
    journal = {Physical Review Letters},
    author = {Bell, William E. and Bloom, Arnold L.},
    number = {6},
    month = {3},
    pages = {280--281},
    volume = {6},
    publisher = {American Physical Society},
    url = {https://journals.aps.org/prl/abstract/10.1103/PhysRevLett.6.280},
    doi = {10.1103/PhysRevLett.6.280},
    issn = {00319007}
}

@book{Auzinsh2010,
    title = {{Optically Polarized Atoms: Understanding Light-Atom Interactions}},
    year = {2010},
    booktitle = {Choice Reviews Online},
    author = {Auzinsh, Marcis and Budker, Dmitry and Rochester, Simon},
    number = {10},
    month = {7},
    pages = {},
    volume = {},
    publisher = {Oxford University Press},
    url = {https://global.oup.com/academic/product/optically-polarized-atoms-9780199565122},
    isbn = {9780199565122}
}

@article{Alexandrov1992,
    title = {{Optically pumped atomic magnetometers after three decades}},
    year = {1992},
    journal = {Optical Engineering},
    author = {Alexandrov, Evgeny B.},
    number = {4},
    month = {4},
    pages = {711},
    volume = {31},
    publisher = {SPIE},
    url = {https://www.spiedigitallibrary.org/journals/optical-engineering/volume-31/issue-4/0000/Optically-pumped-atomic-magnetometers-after-three-decades/10.1117/12.56132.short},
    doi = {10.1117/12.56132},
    issn = {00913286}
}

@article{Sander2020,
    title = {{Optically pumped magnetometers enable a new level of biomagnetic measurements}},
    year = {2020},
    journal = {Advanced Optical Technologies},
    author = {Sander, Tilmann and Jodko-W{\l}adzi{\'{n}}ska, Anna and Hartwig, Stefan and Br{\"{u}}hl, Rüdiger and Middelmann, Thomas},
    number = {5},
    month = {10},
    pages = {247--251},
    volume = {9},
    publisher = {De Gruyter Open Ltd},
    url = {https://www.degruyter.com/document/doi/10.1515/aot-2020-0027/html},
    doi = {10.1515/aot-2020-0027},
    issn = {21928584}
}

@incollection{Romalis2022,
    title = {{Optically pumped magnetometers for biomagnetic measurements}},
    year = {2022},
    booktitle = {Flexible High Performance Magnetic Field Sensors},
    author = {Romalis, Michael V.},
    editor = {Labyt, Etienne and Sander, Tilmann and Wakai, Ronald},
    chapter = {},
    edition = {},
    pages = {3--15},
    publisher = {Springer},
    url = {https://link.springer.com/chapter/10.1007/978-3-031-05363-4_1},
    address = {Cham},
    isbn = {978-3-031-05362-7},
    doi = {10.1007/978-3-031-05363-4}
}

@article{Broser2018,
    title = {{Optically pumped magnetometers for magneto-myography to study the innervation of the hand}},
    year = {2018},
    journal = {IEEE Transactions on Neural Systems and Rehabilitation Engineering},
    author = {Broser, Philip J. and Knappe, Svenja and Kajal, Diljit Singh and Noury, Nima and Alem, Orang and Shah, Vishal and Braun, Christoph},
    number = {11},
    month = {11},
    pages = {2226--2230},
    volume = {26},
    publisher = {Institute of Electrical and Electronics Engineers Inc.},
    doi = {10.1109/TNSRE.2018.2871947},
    issn = {15344320},
    pmid = {30273154}
}

@article{Tierney2019,
    title = {{Optically pumped magnetometers: From quantum origins to multi-channel magnetoencephalography}},
    year = {2019},
    journal = {NeuroImage},
    author = {Tierney, Tim M. and Holmes, Niall and Mellor, Stephanie and L{\'{o}}pez, José David and Roberts, Gillian and Hill, Ryan M. and Boto, Elena and Leggett, James and Shah, Vishal and Brookes, Matthew J. and Bowtell, Richard and Barnes, Gareth R.},
    month = {10},
    pages = {598--608},
    volume = {199},
    publisher = {Academic Press},
    url = {https://www.sciencedirect.com/science/article/pii/S1053811919304550},
    doi = {10.1016/j.neuroimage.2019.05.063},
    issn = {10959572},
    pmid = {31141737}
}

@article{Rochester2012,
    title = {{Orientation-to-alignment conversion and spin squeezing}},
    year = {2012},
    journal = {Physical Review A - Atomic, Molecular, and Optical Physics},
    author = {Rochester, S. M. and Ledbetter, M. P. and Zigdon, T. and Wilson-Gordon, A. D. and Budker, D.},
    number = {2},
    month = {2},
    pages = {022125},
    volume = {85},
    publisher = {American Physical Society},
    url = {https://journals.aps.org/pra/abstract/10.1103/PhysRevA.85.022125},
    doi = {10.1103/PhysRevA.85.022125},
    issn = {10502947},
    arxivId = {1106.3538}
}

@incollection{Janosek2017,
    title = {{Parallel fluxgate magnetometers}},
    year = {2017},
    booktitle = {High Sensitivity Magnetometers. Smart Sensors, Measurement and Instrumentation},
    author = {Janosek, Michal},
    editor = {Grosz, A. and Haji-Sheikh, M. and Mukhopadhyay, S.},
    pages = {41--61},
    volume = {19},
    publisher = {Springer},
    url = {https://link.springer.com/chapter/10.1007/978-3-319-34070-8_2},
    address = {Cham},
    isbn = {978-3-319-34068-5},
    doi = {10.1007/978-3-319-34070-8}
}

@article{LeGal2021,
    title = {{Parametric resonance magnetometer based on elliptically polarized light yielding three-axis measurement with isotropic sensitivity}},
    year = {2021},
    journal = {Applied Physics Letters},
    author = {Le Gal, Gwenael and Rouve, Laure Line and Palacios-Laloy, Agustin},
    number = {25},
    month = {6},
    pages = {254001},
    volume = {118},
    publisher = {AIP Publishing LLC AIP Publishing},
    url = {https://aip.scitation.org/doi/abs/10.1063/5.0047124},
    doi = {10.1063/5.0047124},
    issn = {00036951}
}

@article{Staehler2001,
    title = {{Picotesla magnetometry with coherent dark states}},
    year = {2001},
    journal = {Europhysics Letters},
    author = {St{\"{a}}hler, M. and Knappe, S. and Affolderbach, C. and Kemp, W. and Wynands, R.},
    number = {3},
    month = {5},
    pages = {323--328},
    volume = {54},
    publisher = {EDP Sciences},
    url = {https://epljournal.edpsciences.org/articles/epl/abs/2001/09/6509/6509.html},
    doi = {10.1209/epl/i2001-00245-y},
    issn = {02955075}
}

@article{PedrerosBustos2018,
    title = {{Polarization-driven spin precession of mesospheric sodium atoms}},
    year = {2018},
    journal = {Optics Letters},
    author = {Pedreros Bustos, Felipe and Calia, Domenico Bonaccini and Budker, Dmitry and Centrone, Mauro and Hellemeier, Joschua and Hickson, Paul and Holzl{\"{o}}hner, Ronald and Rochester, Simon},
    number = {23},
    month = {12},
    pages = {5825},
    volume = {43},
    publisher = {Optica Publishing Group},
    url = {https://opg.optica.org/ol/abstract.cfm?uri=ol-43-23-5825},
    doi = {10.1364/ol.43.005825},
    issn = {0146-9592},
    pmid = {30499952}
}

@article{Balabas2010,
    title = {{Polarized alkali-metal vapor with minute-long transverse spin-relaxation time}},
    year = {2010},
    journal = {Physical Review Letters},
    author = {Balabas, M. V. and Karaulanov, T. and Ledbetter, M. P. and Budker, D.},
    number = {7},
    month = {8},
    pages = {070801},
    volume = {105},
    publisher = {American Physical Society},
    url = {https://journals.aps.org/prl/abstract/10.1103/PhysRevLett.105.070801},
    doi = {10.1103/PhysRevLett.105.070801},
    issn = {00319007},
    arxivId = {1005.1617}
}

@article{Limes2020,
    title = {{Portable magnetometry for detection of biomagnetism in ambient environments}},
    year = {2020},
    journal = {Physical Review Applied},
    author = {Limes, M. E. and Foley, E. L. and Kornack, T. W. and Caliga, S. and McBride, S. and Braun, A. and Lee, W. and Lucivero, V. G. and Romalis, M. V.},
    number = {1},
    month = {7},
    pages = {011002},
    volume = {14},
    publisher = {American Physical Society},
    url = {https://journals.aps.org/prapplied/abstract/10.1103/PhysRevApplied.14.011002},
    doi = {10.1103/PhysRevApplied.14.011002},
    issn = {23317019},
    arxivId = {2001.03534}
}

@article{Bennett2021,
    title = {{Precision magnetometers for aerospace applications: A review}},
    year = {2021},
    journal = {Sensors},
    author = {Bennett, James S. and Vyhnalek, Brian E. and Greenall, Hamish and Bridge, Elizabeth M. and Gotardo, Fernando and Forstner, Stefan and Harris, Glen I. and Miranda, Félix A. and Bowen, Warwick P.},
    number = {16},
    month = {8},
    pages = {5568},
    volume = {21},
    publisher = {Multidisciplinary Digital Publishing Institute},
    url = {https://www.mdpi.com/1424-8220/21/16/5568/htm https://www.mdpi.com/1424-8220/21/16/5568},
    doi = {10.3390/s21165568},
    issn = {14248220},
    pmid = {34451010}
}

@article{Bloom1962,
    title = {{Principles of operation of the rubidium vapor magnetometer}},
    year = {1962},
    journal = {Applied Optics},
    author = {Bloom, Arnold L.},
    number = {1},
    month = {1},
    pages = {61},
    volume = {1},
    publisher = {Optical Society of America},
    url = {https://www.osapublishing.org/ao/abstract.cfm?uri=ao-1-1-61},
    doi = {10.1364/ao.1.000061},
    issn = {0003-6935}
}

@article{Gerginov2017,
    title = {{Pulsed operation of a miniature scalar optically pumped magnetometer}},
    year = {2017},
    journal = {Journal of the Optical Society of America B},
    author = {Gerginov, Vladislav and Krzyzewski, Sean and Knappe, Svenja},
    number = {7},
    month = {7},
    pages = {1429},
    volume = {34},
    publisher = {Optical Society of America},
    url = {https://www.osapublishing.org/josab/abstract.cfm?uri=josab-34-7-1429},
    doi = {10.1364/josab.34.001429},
    issn = {0740-3224}
}

@article{Jaufenthaler2021,
    title = {{Pulsed optically pumped magnetometers: Addressing dead time and bandwidth for the unshielded magnetorelaxometry of magnetic nanoparticles}},
    year = {2021},
    journal = {Sensors (Switzerland)},
    author = {Jaufenthaler, Aaron and Kornack, Thomas and Lebedev, Victor and Limes, Mark E. and K{\"{o}}rber, Rainer and Liebl, Maik and Baumgarten, Daniel},
    number = {4},
    month = {2},
    pages = {1--19},
    volume = {21},
    publisher = {Multidisciplinary Digital Publishing Institute},
    url = {https://www.mdpi.com/1424-8220/21/4/1212/htm https://www.mdpi.com/1424-8220/21/4/1212},
    doi = {10.3390/s21041212},
    issn = {14248220},
    pmid = {33572285}
}

@article{Bi2021,
    title = {{Quantitative analysis of magnetic cobalt particles with an optically pumped atomic magnetometer}},
    year = {2021},
    journal = {Applied Physics Letters},
    author = {Bi, Xin and Ruan, Limeng and Liu, Zehua and Li, Kan and Ruan, Yi and Zheng, Wenqiang and Lin, Qiang},
    number = {8},
    month = {2},
    pages = {084101},
    volume = {118},
    publisher = {AIP Publishing LLC AIP Publishing},
    url = {https://aip.scitation.org/doi/abs/10.1063/5.0039565},
    doi = {10.1063/5.0039565},
    issn = {00036951}
}

@article{Jaklevic1964,
    title = {{Quantum interference effects in Josephson tunneling}},
    year = {1964},
    journal = {Physical Review Letters},
    author = {Jaklevic, R. C. and Lambe, John and Silver, A. H. and Mercereau, J. E.},
    number = {7},
    month = {2},
    pages = {159--160},
    volume = {12},
    publisher = {American Physical Society},
    url = {https://journals.aps.org/prl/abstract/10.1103/PhysRevLett.12.159},
    doi = {10.1103/PhysRevLett.12.159},
    issn = {00319007}
}

@article{Wasilewski2010,
    title = {{Quantum noise limited and entanglement-assisted magnetometry}},
    year = {2010},
    journal = {Physical Review Letters},
    author = {Wasilewski, W. and Jensen, K. and Krauter, H. and Renema, J. J. and Balabas, M. V. and Polzik, E. S.},
    number = {13},
    month = {3},
    pages = {133601},
    volume = {104},
    publisher = {American Physical Society},
    url = {https://journals.aps.org/prl/abstract/10.1103/PhysRevLett.104.133601},
    doi = {10.1103/PhysRevLett.104.133601},
    issn = {00319007},
    arxivId = {0907.2453}
}

@article{Kuzmich1999,
    title = {{Quantum nondemolition measurements of collective atomic spin}},
    year = {1999},
    journal = {Physical Review A - Atomic, Molecular, and Optical Physics},
    author = {Kuzmich, A. and Mandel, L. and Janis, J. and Young, Y. E. and Ejnisman, R. and Bigelow, N. P.},
    number = {3},
    month = {9},
    pages = {2346--2350},
    volume = {60},
    publisher = {American Physical Society},
    url = {https://journals.aps.org/pra/abstract/10.1103/PhysRevA.60.2346},
    doi = {10.1103/PhysRevA.60.2346},
    issn = {10941622}
}

@book{Fox2006,
    title = {{Quantum Optics: An Introduction}},
    year = {2006},
    author = {Fox, Mark},
    month = {4},
    pages = {},
    publisher = {Oxford University Press},
    url = {https://global.oup.com/academic/product/quantum-optics-9780198566724},
    isbn = {9780198566724}
}

@article{Horrom2012,
    title = {{Quantum-enhanced magnetometer with low-frequency squeezing}},
    year = {2012},
    journal = {Physical Review A - Atomic, Molecular, and Optical Physics},
    author = {Horrom, Travis and Singh, Robinjeet and Dowling, Jonathan P. and Mikhailov, Eugeniy E.},
    number = {2},
    month = {8},
    pages = {023803},
    volume = {86},
    publisher = {American Physical Society},
    url = {https://journals.aps.org/pra/abstract/10.1103/PhysRevA.86.023803},
    doi = {10.1103/PhysRevA.86.023803},
    issn = {10502947},
    arxivId = {1202.3831}
}

@misc{Fabricant2014,
    title = {{Quantum-Limited Optical Magnetometry with Cesium Microcells - Master thesis, Center for Quantop Optics (QUANTOP), Niels Bohr Institute}},
    year = {2014},
    author = {Fabricant, Anne},
    publisher = {QUANTOP, Niels Bohr Institute, University of Copenhagen},
    institution = {Center for Quantop Optics (QUANTOP), Niels Bohr Institute},
    address = {Copenhagen, Denmark}
}

@article{Matsko2001,
    title = {{Radiation Trapping in Coherent Media}},
    year = {2001},
    journal = {Physical Review Letters},
    author = {Matsko, A. B. and Novikova, I. and Scully, M. O. and Welch, G. R.},
    number = {13},
    month = {9},
    pages = {133601},
    volume = {87},
    doi = {10.1103/PhysRevLett.87.133601},
    issn = {0031-9007}
}

@article{Zhang2020,
    title = {{Recording brain activities in unshielded Earth’s field with optically pumped atomic magnetometers}},
    year = {2020},
    journal = {Science Advances},
    author = {Zhang, Rui and Xiao, Wei and Ding, Yudong and Feng, Yulong and Peng, Xiang and Shen, Liang and Sun, Chenxi and Wu, Teng and Wu, Yulong and Yang, Yucheng and Zheng, Zhaoyu and Zhang, Xiangzhi and Chen, Jingbiao and Guo, Hong},
    number = {24},
    month = {6},
    pages = {8792--8804},
    volume = {6},
    publisher = {American Association for the Advancement of Science},
    url = {https://www.science.org/doi/full/10.1126/sciadv.aba8792},
    doi = {10.1126/sciadv.aba8792},
    issn = {23752548},
    pmid = {32582858}
}

@article{Sutter2020,
    title = {{Recording the heart beat of cattle using a gradiometer system of optically pumped magnetometers}},
    year = {2020},
    journal = {Computers and Electronics in Agriculture},
    author = {Sutter, Jens U. and Lewis, Oliver and Robinson, Clive and McMahon, Anthony and Boyce, Robert and Bragg, Rachel and Macrae, Alastair and Orton, Jeffrey and Shah, Vishal and Ingleby, Stuart J. and Griffin, Paul F. and Riis, Erling},
    month = {10},
    pages = {105651},
    volume = {177},
    publisher = {Elsevier},
    url = {https://www.sciencedirect.com/science/article/pii/S0168169920311480},
    doi = {10.1016/J.COMPAG.2020.105651},
    issn = {0168-1699}
}

@article{Graf2005,
    title = {{Relaxation of atomic polarization in paraffin-coated cesium vapor cells}},
    year = {2005},
    journal = {Physical Review A - Atomic, Molecular, and Optical Physics},
    author = {Graf, M. T. and Kimball, D. F. and Rochester, S. M. and Kerner, K. and Wong, C. and Budker, D. and Alexandrov, E. B. and Balabas, M. V. and Yashchuk, V. V.},
    number = {2},
    month = {8},
    pages = {023401},
    volume = {72},
    publisher = {American Physical Society},
    url = {https://journals.aps.org/pra/abstract/10.1103/PhysRevA.72.023401},
    doi = {10.1103/PhysRevA.72.023401},
    issn = {10502947},
    arxivId = {physics/0503202}
}

@article{Bouchiat1966,
    title = {{Relaxation of optically pumped rb atoms on paraffin-coated walls}},
    year = {1966},
    journal = {Physical Review},
    author = {Bouchiat, M. A. and Brossel, J.},
    number = {1},
    month = {7},
    pages = {41--54},
    volume = {147},
    publisher = {American Physical Society},
    url = {https://journals.aps.org/pr/abstract/10.1103/PhysRev.147.41},
    doi = {10.1103/PhysRev.147.41},
    issn = {0031899X}
}

@article{Budker2002RMP,
    title = {{Resonant nonlinear magneto-optical effects in atoms}},
    year = {2002},
    journal = {Reviews of Modern Physics},
    author = {Budker, D. and Gawlik, W. and Kimball, D. F. and Rochester, S. M. and Yashchuk, V. V. and Weis, A.},
    number = {4},
    month = {11},
    pages = {1153--1201},
    volume = {74},
    publisher = {American Physical Society},
    url = {https://journals.aps.org/rmp/abstract/10.1103/RevModPhys.74.1153},
    doi = {10.1103/RevModPhys.74.1153},
    issn = {00346861},
    arxivId = {physics/0203077}
}

@article{Ding2018,
    title = {{Response of a Bell-Bloom magnetometer to a magnetic field of arbitrary direction}},
    year = {2018},
    journal = {Sensors (Switzerland)},
    author = {Ding, Zhichao and Yuan, Jie and Long, Xingwu},
    number = {5},
    month = {5},
    pages = {1401},
    volume = {18},
    publisher = {Multidisciplinary Digital Publishing Institute},
    url = {https://www.mdpi.com/1424-8220/18/5/1401/htm https://www.mdpi.com/1424-8220/18/5/1401},
    doi = {10.3390/s18051401},
    issn = {14248220},
    pmid = {29724059}
}

@article{Padniuk2022,
    title = {{Response of atomic spin-based sensors to magnetic and nonmagnetic perturbations}},
    year = {2022},
    journal = {Scientific Reports 2022 12:1},
    author = {Padniuk, Mikhail and Kopciuch, Marek and Cipolletti, Riccardo and Wickenbrock, Arne and Budker, Dmitry and Pustelny, Szymon},
    number = {1},
    month = {1},
    pages = {1--9},
    volume = {12},
    publisher = {Nature Publishing Group},
    url = {https://www.nature.com/articles/s41598-021-03609-w},
    isbn = {0123456789},
    doi = {10.1038/s41598-021-03609-w},
    issn = {2045-2322},
    pmid = {35013346},
    arxivId = {2107.05501}
}

@article{Scott1962,
    title = {{Review of gyromagnetic ratio experiments}},
    year = {1962},
    journal = {Reviews of Modern Physics},
    author = {Scott, G. G.},
    number = {1},
    month = {1},
    pages = {102},
    volume = {34},
    publisher = {American Physical Society},
    url = {https://journals.aps.org/rmp/abstract/10.1103/RevModPhys.34.102},
    doi = {10.1103/RevModPhys.34.102},
    issn = {00346861}
}

@article{Chalupczak2012,
    title = {{Room temperature femtotesla radio-frequency atomic magnetometer}},
    year = {2012},
    journal = {Applied Physics Letters},
    author = {Chalupczak, W. and Godun, R. M. and Pustelny, S. and Gawlik, W.},
    number = {24},
    month = {6},
    pages = {242401},
    volume = {100},
    publisher = {American Institute of PhysicsAIP},
    url = {https://aip.scitation.org/doi/abs/10.1063/1.4729016},
    doi = {10.1063/1.4729016},
    issn = {00036951}
}

@article{Gerginov2020,
    title = {{Scalar magnetometry below 100 fT/sqrt(Hz) in a microfabricated cell}},
    year = {2020},
    journal = {IEEE Sensors Journal},
    author = {Gerginov, Vladislav and Pomponio, Marco and Knappe, Svenja},
    number = {21},
    month = {11},
    pages = {12684--12690},
    volume = {20},
    publisher = {Institute of Electrical and Electronics Engineers Inc.},
    url = {https://ieeexplore.ieee.org/document/9117011},
    doi = {10.1109/JSEN.2020.3002193},
    issn = {15581748}
}

@article{Su2021,
    title = {{Search for exotic spin-dependent interactions with a spin-based amplifier}},
    year = {2021},
    journal = {Science Advances},
    author = {Su, Haowen and Wang, Yuanhong and Jiang, Min and Ji, Wei and Fadeev, Pavel and Hu, Dongdong and Peng, Xinhua and Budker, Dmitry},
    number = {47},
    month = {11},
    pages = {9535},
    volume = {7},
    publisher = {American Association for the Advancement of Science},
    url = {https://www.science.org/doi/full/10.1126/sciadv.abi9535},
    doi = {10.1126/sciadv.abi9535},
    issn = {23752548},
    pmid = {34788098},
    arxivId = {2103.15282}
}

@inproceedings{Corsini2011,
    title = {{Search for plant biomagnetism with a sensitive atomic magnetometer}},
    year = {2011},
    booktitle = {Journal of Applied Physics},
    author = {Corsini, Eric and Acosta, Victor and Baddour, Nicolas and Higbie, James and Lester, Brian and Licht, Paul and Patton, Brian and Prouty, Mark and Budker, Dmitry},
    number = {7},
    month = {4},
    pages = {074701},
    volume = {109},
    publisher = {American Institute of PhysicsAIP},
    url = {https://aip.scitation.org/doi/abs/10.1063/1.3560920},
    doi = {10.1063/1.3560920},
    issn = {00218979},
    arxivId = {1006.3578}
}

@article{Afach2021,
    title = {{Search for topological defect dark matter with a global network of optical magnetometers}},
    year = {2021},
    journal = {Nature Physics},
    author = {Afach, Samer and Buchler, Ben C. and Budker, Dmitry and Dailey, Conner and Derevianko, Andrei and Dumont, Vincent and Figueroa, Nataniel L. and Gerhardt, Ilja and Gruji{\'{c}}, Zoran D. and Guo, Hong and Hao, Chuanpeng and Hamilton, Paul S. and Hedges, Morgan and Jackson Kimball, Derek F. and Kim, Dongok and Khamis, Sami and Kornack, Thomas and Lebedev, Victor and Lu, Zheng Tian and Masia-Roig, Hector and Monroy, Madeline and Padniuk, Mikhail and Palm, Christopher A. and Park, Sun Yool and Paul, Karun V. and Penaflor, Alexander and Peng, Xiang and Pospelov, Maxim and Preston, Rayshaun and Pustelny, Szymon and Scholtes, Theo and Segura, Perrin C. and Semertzidis, Yannis K. and Sheng, Dong and Shin, Yun Chang and Smiga, Joseph A. and Stalnaker, Jason E. and Sulai, Ibrahim and Tandon, Dhruv and Wang, Tao and Weis, Antoine and Wickenbrock, Arne and Wilson, Tatum and Wu, Teng and Wurm, David and Xiao, Wei and Yang, Yucheng and Yu, Dongrui and Zhang, Jianwei},
    number = {12},
    month = {12},
    pages = {1396--1401},
    volume = {17},
    publisher = {Nature Publishing Group},
    url = {https://www.nature.com/articles/s41567-021-01393-y},
    doi = {10.1038/s41567-021-01393-y},
    issn = {17452481},
    arxivId = {2102.13379}
}

@article{Sargsyan2017,
    title = {{Selective reflection from an Rb layer with a thickness below {$\lambda$}/12 and applications}},
    year = {2017},
    journal = {Optics Letters},
    author = {Sargsyan, Armen and Papoyan, Aram and Hughes, Ifan G. and Adams, Charles S. and Sarkisyan, David},
    number = {8},
    month = {4},
    pages = {1476},
    volume = {42},
    publisher = {Optica Publishing Group},
    url = {https://opg.optica.org/ol/abstract.cfm?uri=ol-42-8-1476},
    doi = {10.1364/ol.42.001476},
    issn = {0146-9592},
    pmid = {28409777}
}

@article{Bevilacqua2019PRA,
    title = {{Self-adaptive loop for external-disturbance reduction in a differential measurement setup}},
    year = {2019},
    journal = {Physical Review Applied},
    author = {Bevilacqua, Giuseppe and Biancalana, Valerio and Dancheva, Yordanka and Vigilante, Antonio},
    number = {1},
    month = {1},
    pages = {014029},
    volume = {11},
    publisher = {American Physical Society},
    url = {https://journals.aps.org/prapplied/abstract/10.1103/PhysRevApplied.11.014029},
    doi = {10.1103/PhysRevApplied.11.014029},
    issn = {23317019}
}

@article{Bison2018,
    title = {{Sensitive and stable vector magnetometer for operation in zero and finite fields}},
    year = {2018},
    journal = {Optics Express},
    author = {Bison, G. and Bondar, V. and Schmidt-Wellenburg, P. and Schnabel, A. and Voigt, J.},
    number = {13},
    month = {6},
    pages = {17350},
    volume = {26},
    publisher = {Optica Publishing Group},
    url = {https://opg.optica.org/oe/abstract.cfm?uri=oe-26-13-17350},
    doi = {10.1364/oe.26.017350},
    issn = {10944087},
    pmid = {30119547}
}

@article{Budker2000,
    title = {{Sensitive magnetometry based on nonlinear magneto-optical rotation}},
    year = {2000},
    journal = {Physical Review A - Atomic, Molecular, and Optical Physics},
    author = {Budker, D. and Kimball, D. F. and Rochester, S. M. and Yashchuk, V. V. and Zolotorev, M.},
    number = {4},
    month = {9},
    pages = {7},
    volume = {62},
    publisher = {American Physical Society},
    url = {https://journals.aps.org/pra/abstract/10.1103/PhysRevA.62.043403},
    doi = {10.1103/PhysRevA.62.043403},
    issn = {10941622}
}

@article{Fu2020,
    title = {{Sensitive magnetometry in challenging environments}},
    year = {2020},
    journal = {AVS Quantum Science},
    author = {Fu, Kai Mei C. and Iwata, Geoffrey Z. and Wickenbrock, Arne and Budker, Dmitry},
    number = {4},
    month = {12},
    pages = {044702},
    volume = {2},
    publisher = { American Vacuum Society AVS },
    url = {https://avs.scitation.org/doi/abs/10.1116/5.0025186},
    doi = {10.1116/5.0025186},
    issn = {26390213},
    arxivId = {2008.00082}
}

@article{Jimenez-Martinez2010,
    title = {{Sensitivity comparison of Mx and frequency-modulated bell-bloom Cs magnetometers in a microfabricated cell}},
    year = {2010},
    journal = {IEEE Transactions on Instrumentation and Measurement},
    author = {Jim{\'{e}}nez-Mart{\'{i}}nez, Ricardo and Griffith, W. Clark and Wang, Ying Ju and Knappe, Svenja and Kitching, John and Smith, Ken and Prouty, Mark D.},
    number = {2},
    month = {2},
    pages = {372--378},
    volume = {59},
    url = {https://ieeexplore.ieee.org/document/5247087},
    doi = {10.1109/TIM.2009.2023829},
    issn = {00189456}
}

@article{Li2018,
    title = {{SERF atomic magnetometer - Recent advances and applications: A review}},
    year = {2018},
    journal = {IEEE Sensors Journal},
    author = {Li, Jundi and Quan, Wei and Zhou, Binquan and Wang, Zhuo and Lu, Jixi and Hu, Zhaohui and Liu, Gang and Fang, Jiancheng},
    number = {20},
    month = {10},
    pages = {8198--8207},
    volume = {18},
    publisher = {Institute of Electrical and Electronics Engineers Inc.},
    url = {https://ieeexplore.ieee.org/document/8425967},
    doi = {10.1109/JSEN.2018.2863707},
    issn = {1530437X}
}

@article{Lucivero2014,
    title = {{Shot-noise-limited magnetometer with sub-picotesla sensitivity at room temperature}},
    year = {2014},
    journal = {Review of Scientific Instruments},
    author = {Lucivero, Vito Giovanni and Anielski, Pawel and Gawlik, Wojciech and Mitchell, Morgan W.},
    number = {11},
    month = {11},
    pages = {113108},
    volume = {85},
    publisher = {American Institute of PhysicsAIP},
    url = {https://aip.scitation.org/doi/abs/10.1063/1.4901588},
    doi = {10.1063/1.4901588},
    issn = {10897623},
    pmid = {25430099},
    arxivId = {1403.7796}
}

@article{Jimenez-Martinez2018,
    title = {{Signal tracking beyond the time resolution of an atomic sensor by Kalman filtering}},
    year = {2018},
    journal = {Physical Review Letters},
    author = {Jim{\'{e}}nez-Mart{\'{i}}nez, Ricardo and Ko{\l}ody{\'{n}}ski, Jan and Troullinou, Charikleia and Lucivero, Vito Giovanni and Kong, Jia and Mitchell, Morgan W.},
    number = {4},
    month = {1},
    pages = {040503},
    volume = {120},
    publisher = {American Physical Society},
    url = {https://journals.aps.org/prl/abstract/10.1103/PhysRevLett.120.040503},
    doi = {10.1103/PhysRevLett.120.040503},
    issn = {10797114},
    pmid = {29437429},
    arxivId = {1707.08131}
}

@article{Oelsner2019,
    title = {{Sources of heading errors in optically pumped magnetometers operated in the Earth's magnetic field}},
    year = {2019},
    journal = {Physical Review A},
    author = {Oelsner, G. and Schultze, V. and Ijsselsteijn, R. and Wittk{\"{a}}mper, F. and Stolz, R.},
    number = {1},
    month = {1},
    pages = {013420},
    volume = {99},
    publisher = {American Physical Society},
    url = {https://journals.aps.org/pra/abstract/10.1103/PhysRevA.99.013420},
    doi = {10.1103/PhysRevA.99.013420},
    issn = {24699934},
    arxivId = {1809.00627}
}

@article{Kuzmich1997,
    title = {{Spin squeezing in an ensemble of atoms illuminated with squeezed light}},
    year = {1997},
    journal = {Physical Review Letters},
    author = {Kuzmich, A. and M{\o}lmer, Klaus and Polzik, E. S.},
    number = {24},
    month = {12},
    pages = {4782--4785},
    volume = {79},
    publisher = {American Physical Society},
    url = {https://journals.aps.org/prl/abstract/10.1103/PhysRevLett.79.4782},
    doi = {10.1103/PhysRevLett.79.4782},
    issn = {10797114}
}

@article{Troullinou2021,
    title = {{Squeezed-light enhancement and backaction evasion in a high sensitivity optically pumped magnetometer}},
    year = {2021},
    journal = {Physical Review Letters},
    author = {Troullinou, C. and Jim{\'{e}}nez-Mart{\'{i}}nez, R. and Kong, J. and Lucivero, V. G. and Mitchell, M. W.},
    number = {19},
    month = {11},
    pages = {193601},
    volume = {127},
    publisher = {American Physical Society},
    url = {https://journals.aps.org/prl/abstract/10.1103/PhysRevLett.127.193601},
    doi = {10.1103/PhysRevLett.127.193601},
    issn = {10797114},
    pmid = {34797131},
    arxivId = {2108.01519}
}

@article{Wolfgramm2010,
    title = {{Squeezed-light optical magnetometry}},
    year = {2010},
    journal = {Physical Review Letters},
    author = {Wolfgramm, Florian and Cer{\`{e}}, Alessandro and Beduini, Federica A. and Predojevi{\'{c}}, Ana and Koschorreck, Marco and Mitchell, Morgan W.},
    number = {5},
    month = {7},
    pages = {053601},
    volume = {105},
    publisher = {American Physical Society},
    url = {https://journals.aps.org/prl/abstract/10.1103/PhysRevLett.105.053601},
    doi = {10.1103/PhysRevLett.105.053601},
    issn = {00319007},
    arxivId = {1008.1721}
}

@article{Zhang2021PRL,
    title = {{Stand-off magnetometry with directional emission from sodium vapors}},
    year = {2021},
    journal = {Physical Review Letters},
    author = {Zhang, Rui and Klinger, Emmanuel and Pedreros Bustos, Felipe and Akulshin, Alexander and Guo, Hong and Wickenbrock, Arne and Budker, Dmitry},
    number = {17},
    month = {10},
    pages = {173605},
    volume = {127},
    publisher = {American Physical Society},
    url = {https://journals.aps.org/prl/abstract/10.1103/PhysRevLett.127.173605},
    doi = {10.1103/PhysRevLett.127.173605},
    issn = {10797114},
    pmid = {34739270},
    arxivId = {2103.07358}
}

@article{Vasilakis2011,
    title = {{Stroboscopic backaction evasion in a dense alkali-metal vapor}},
    year = {2011},
    journal = {Physical Review Letters},
    author = {Vasilakis, G. and Shah, V. and Romalis, M. V.},
    number = {14},
    month = {4},
    pages = {143601},
    volume = {106},
    publisher = {American Physical Society},
    url = {https://journals.aps.org/prl/abstract/10.1103/PhysRevLett.106.143601},
    doi = {10.1103/PhysRevLett.106.143601},
    issn = {00319007},
    arxivId = {1011.2682}
}

@article{Bevilacqua2019APL,
    title = {{Sub-millimetric ultra-low-field MRI detected in situ by a dressed atomic magnetometer}},
    year = {2019},
    journal = {Applied Physics Letters},
    author = {Bevilacqua, Giuseppe and Biancalana, Valerio and Dancheva, Yordanka and Vigilante, Antonio},
    number = {17},
    month = {10},
    pages = {174102},
    volume = {115},
    publisher = {AIP Publishing LLC AIP Publishing},
    url = {https://aip.scitation.org/doi/abs/10.1063/1.5123653},
    doi = {10.1063/1.5123653},
    issn = {00036951},
    arxivId = {1908.01283}
}

@article{Sheng2013,
    title = {{Subfemtotesla scalar atomic magnetometry using multipass cells}},
    year = {2013},
    journal = {Physical Review Letters},
    author = {Sheng, D. and Li, S. and Dural, N. and Romalis, M. V.},
    number = {16},
    month = {4},
    pages = {160802},
    volume = {110},
    publisher = {American Physical Society},
    url = {https://journals.aps.org/prl/abstract/10.1103/PhysRevLett.110.160802},
    doi = {https://doi.org/10.1103/PhysRevLett.110.160802},
    issn = {00319007},
    arxivId = {1208.1099}
}

@article{Shah2007,
    title = {{Subpicotesla atomic magnetometry with a microfabricated vapour cell}},
    year = {2007},
    journal = {Nature Photonics},
    author = {Shah, Vishal and Knappe, Svenja and Schwindt, Peter D.D. and Kitching, John},
    number = {11},
    month = {11},
    pages = {649--652},
    volume = {1},
    publisher = {Nature Publishing Group},
    url = {https://www.nature.com/articles/nphoton.2007.201},
    doi = {10.1038/nphoton.2007.201},
    issn = {17494885}
}

@article{Bao2018,
    title = {{Suppression of the nonlinear Zeeman effect and heading error in Earth-field-range alkali-vapor magnetometers}},
    year = {2018},
    journal = {Physical Review Letters},
    author = {Bao, Guzhi and Wickenbrock, Arne and Rochester, Simon and Zhang, Weiping and Budker, Dmitry},
    number = {3},
    month = {1},
    pages = {033202},
    volume = {120},
    publisher = {American Physical Society},
    url = {https://journals.aps.org/prl/abstract/10.1103/PhysRevLett.120.033202},
    doi = {10.1103/PhysRevLett.120.033202},
    issn = {10797114},
    pmid = {29400503},
    arxivId = {1708.05262}
}

@article{Wiltschko2021,
    title = {{The magnetic compass of birds: The role of cryptochrome}},
    year = {2021},
    journal = {Frontiers in Physiology},
    author = {Wiltschko, Roswitha and Nie{\ss}ner, Christine and Wiltschko, Wolfgang},
    month = {5},
    pages = {667000},
    volume = {12},
    publisher = {Frontiers Media S.A.},
    url = {https://www.frontiersin.org/articles/10.3389/fphys.2021.667000/full},
    doi = {10.3389/fphys.2021.667000},
    issn = {1664042X}
}

@article{Huang2015,
    title = {{Three-axis atomic magnetometer based on spin precession modulation}},
    year = {2015},
    journal = {Applied Physics Letters},
    author = {Huang, H. C. and Dong, H. F. and Hu, X. Y. and Chen, L. and Gao, Y.},
    number = {18},
    month = {11},
    pages = {182403},
    volume = {107},
    publisher = {AIP Publishing LLCAIP Publishing},
    url = {https://aip.scitation.org/doi/abs/10.1063/1.4935096},
    doi = {10.1063/1.4935096},
    issn = {00036951}
}

@article{Maddox2022,
    title = {{Through-skin pilot-hole detection and localization with a mechanically translatable atomic magnetometer}},
    year = {2022},
    journal = {Applied Physics Letters},
    author = {Maddox, Benjamin and Cohen, Yuval and Renzoni, Ferruccio},
    number = {1},
    month = {1},
    pages = {014002},
    volume = {120},
    publisher = {AIP Publishing LLCAIP Publishing},
    url = {https://aip.scitation.org/doi/abs/10.1063/5.0081274},
    doi = {10.1063/5.0081274},
    issn = {0003-6951}
}

@article{Colombo2016,
    title = {{Towards a mechanical MPI scanner based on atomic magnetometry}},
    year = {2016},
    journal = {International Journal on Magnetic Particle Imaging},
    author = {Colombo, Simone and Lebedev, Victor and Tonyushkin, Alexey and Grujic, Zoran Dragan and Dolgovskiy, Vladimir and Weis, Antoine},
    number = {1},
    volume = {3},
    url = {https://journal.iwmpi.org/index.php/iwmpi/article/view/77},
    arxivId = {1611.09047}
}

@article{Petrenko2021,
    title = {{Towards the non-zero field cesium magnetic sensor array for magnetoencephalography}},
    year = {2021},
    journal = {IEEE Sensors Journal},
    author = {Petrenko, Mikhail V. and Dmitriev, Sergei P. and Pazgalev, Anatoly S. and Ossadtchi, Alex E. and Vershovskii, Anton K.},
    number = {17},
    month = {9},
    pages = {18626--18632},
    volume = {21},
    publisher = {Institute of Electrical and Electronics Engineers Inc.},
    url = {https://ieeexplore.ieee.org/document/9455403},
    doi = {10.1109/JSEN.2021.3089455},
    issn = {15581748}
}

@incollection{Palacios-Laloy2022,
    title = {{Tri-axial helium-4 optically pumped magnetometers for MEG}},
    year = {2022},
    booktitle = {Flexible High Performance Magnetic Field Sensors},
    author = {Palacios-Laloy, A. and Le Prado, M. and Labyt, E.},
    pages = {79--110},
    publisher = {Springer, Cham},
    url = {https://link.springer.com/chapter/10.1007/978-3-031-05363-4_6},
    doi = {10.1007/978-3-031-05363-4}
}

@article{Dang2010,
    title = {{Ultrahigh sensitivity magnetic field and magnetization measurements with an atomic magnetometer}},
    year = {2010},
    journal = {Applied Physics Letters},
    author = {Dang, H. B. and Maloof, A. C. and Romalis, M. V.},
    number = {15},
    month = {10},
    pages = {151110},
    volume = {97},
    publisher = {American Institute of PhysicsAIP},
    url = {https://aip.scitation.org/doi/abs/10.1063/1.3491215},
    doi = {10.1063/1.3491215},
    issn = {00036951},
    arxivId = {0910.2206}
}

@article{Nikiel2014,
    title = {{Ultrasensitive 3He magnetometer for measurements of high magnetic fields}},
    year = {2014},
    journal = {European Physical Journal D},
    author = {Nikiel, Anna and Bl{\"{u}}mler, Peter and Heil, Werner and Hehn, Manfred and Karpuk, Sergej and Maul, Andreas and Otten, Ernst and Schreiber, Laura M. and Terekhov, Maxim},
    number = {11},
    month = {11},
    pages = {1--12},
    volume = {68},
    publisher = {Springer},
    url = {https://link.springer.com/article/10.1140/epjd/e2014-50401-3},
    doi = {10.1140/epjd/e2014-50401-3},
    issn = {14346079}
}

@article{Rushton2022,
    title = {{Unshielded portable optically pumped magnetometer for the remote detection of conductive objects using eddy current measurements}},
    year = {2022},
    journal = {arXiv},
    author = {Rushton, L M and Pyragius, T and Meraki, A and Elson, L and Jensen, K},
    month = {6},
    url = {https://arxiv.org/abs/2206.04631v1},
    doi = {10.48550/arxiv.2206.04631},
    arxivId = {2206.04631}
}

@article{Seltzer2004,
    title = {{Unshielded three-axis vector operation of a spin-exchange-relaxation-free atomic magnetometer}},
    year = {2004},
    journal = {Applied Physics Letters},
    author = {Seltzer, S. J. and Romalis, M. V.},
    number = {20},
    month = {11},
    pages = {4804--4806},
    volume = {85},
    publisher = {American Institute of PhysicsAIP},
    url = {https://aip.scitation.org/doi/abs/10.1063/1.1814434},
    doi = {10.1063/1.1814434},
    issn = {00036951}
}

@article{Yudin2010,
    title = {{Vector magnetometry based on electromagnetically induced transparency in linearly polarized light}},
    year = {2010},
    journal = {Physical Review A - Atomic, Molecular, and Optical Physics},
    author = {Yudin, V. I. and Taichenachev, A. V. and Dudin, Y. O. and Velichansky, V. L. and Zibrov, A. S. and Zibrov, S. A.},
    number = {3},
    month = {9},
    pages = {033807},
    volume = {82},
    publisher = {American Physical Society},
    url = {https://journals.aps.org/pra/abstract/10.1103/PhysRevA.82.033807},
    doi = {10.1103/PhysRevA.82.033807},
    issn = {10502947}
}

@article{Pyragius2019,
    title = {{Voigt-effect-based three-dimensional vector magnetometer}},
    year = {2019},
    journal = {Physical Review A},
    author = {Pyragius, Tadas and Florez, Hans Marin and Fernholz, Thomas},
    number = {2},
    month = {8},
    pages = {023416},
    volume = {100},
    publisher = {American Physical Society},
    url = {https://journals.aps.org/pra/abstract/10.1103/PhysRevA.100.023416},
    doi = {10.1103/PhysRevA.100.023416},
    issn = {24699934}
}

@article{Put2021,
    title = {{Zero- to Ultralow-Field NMR Spectroscopy of Small Biomolecules}},
    year = {2021},
    journal = {Analytical Chemistry},
    author = {Put, Piotr and Pustelny, Szymon and Budker, Dmitry and Druga, Emanuel and Sjolander, Tobias F. and Pines, Alexander and Barskiy, Danila A.},
    number = {6},
    month = {2},
    pages = {3226--3232},
    volume = {93},
    publisher = {American Chemical Society},
    url = {https://pubs.acs.org/doi/10.1021/acs.analchem.0c04738},
    doi = {10.1021/acs.analchem.0c04738},
    issn = {0003-2700}
}

@article{Barskiy2019,
    title = {{Zero-field nuclear magnetic resonance of chemically exchanging systems}},
    year = {2019},
    journal = {Nature Communications},
    author = {Barskiy, Danila A. and Tayler, Michael C. D. and Marco-Rius, Irene and Kurhanewicz, John and Vigneron, Daniel B. and Cikrikci, Sevil and Aydogdu, Ayca and Reh, Moritz and Pravdivtsev, Andrey N. and H{\"{o}}vener, Jan-Bernd and Blanchard, John W. and Wu, Teng and Budker, Dmitry and Pines, Alexander},
    number = {1},
    month = {7},
    pages = {3002},
    volume = {10},
    publisher = {Nature Publishing Group},
    url = {https://www.nature.com/articles/s41467-019-10787-9},
    doi = {10.1038/s41467-019-10787-9},
    issn = {2041-1723},
    pmid = {31278303}
}

@article{Ledbetter2008,
    title = {{Zero-field remote detection of NMR with a microfabricated atomic magnetometer}},
    year = {2008},
    journal = {Proceedings of the National Academy of Sciences of the United States of America},
    author = {Ledbetter, M. P. and Savukov, I. M. and Budker, D. and Shah, V. and Knappe, S. and Kitching, J. and Michalak, D. J. and Xu, S. and Pines, A.},
    number = {7},
    month = {2},
    pages = {2286--2290},
    volume = {105},
    publisher = {National Academy of Sciences},
    url = {https://www.pnas.org/content/105/7/2286 https://www.pnas.org/content/105/7/2286.abstract},
    doi = {10.1073/pnas.0711505105},
    issn = {00278424},
    pmid = {18287080}
}

\pagebreak
\appendix

\section{Optical pumping in cesium}
\label{sec:Sup-OpticalPumping}
\begin{figure}[h]
    \centering
    \includegraphics[width=0.29\columnwidth]{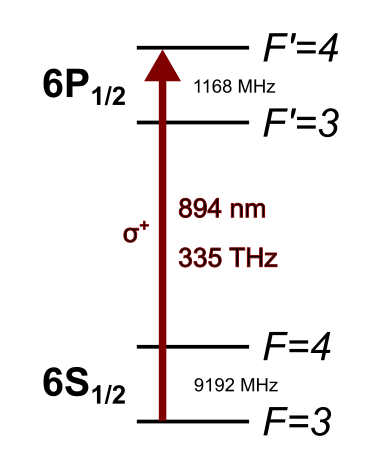} %\quad
    \includegraphics[width=0.7\columnwidth]{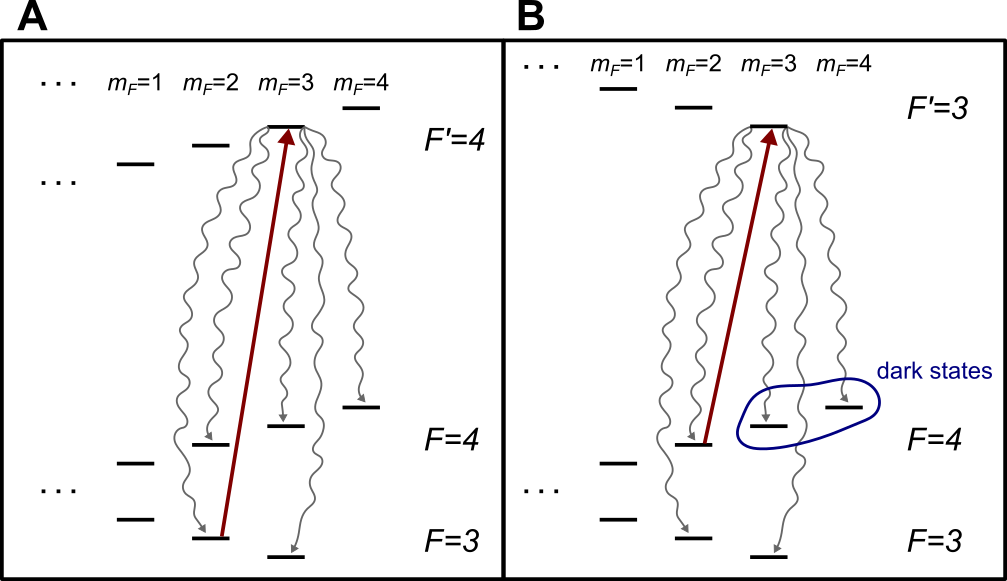}
    \caption{Example of optical pumping on the cesium D1 line (see Sec.~\ref{sec:2-BlochEq} for notation and further details), illustrating one of the excitations and associated decay paths. Here we do not consider the possible presence of buffer gas, which may complicate the pumping process in an atomic vapor cell. A) Right-circularly polarized light tuned to the $F=3 \longrightarrow F^\prime=4$ transition tends to pump the atom into the extreme hyperfine substate $F=3,\,m_F=3$. This particular transition is ``leaky'' because an excited atom may also decay to the $F=4$ ground state. B) Pumping on the $F=4 \longrightarrow F^\prime=3$ transition is preferred since it allows us to make use of the ``dark states'' $F=4,\,m_F=3$ and $m_F=4$.}
    \label{fig:magnetometer}
\end{figure}

% \pagebreak
\section{Sensitivity of an atomic magnetometer}
\label{sec:Sup-Sensitivity}
% \todo{\textbf{Update to reflect our $M_z$ geometry and notation. Currently this derivation is done for a SERF/RF configuration.}}
% \bigskip

\noindent \textit{How does the sensitivity of the magnetometer scale with the number of atoms $N$ and the measurement time $\tau$?}
\bigskip

\begin{figure}[h]
    \centering
   \includegraphics[width=0.8\columnwidth]{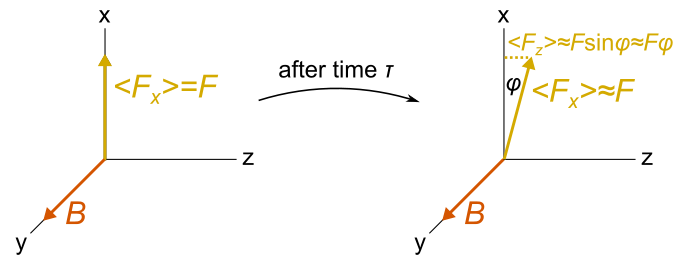}
    \caption{Spin evolution in an atomic magnetometer.}
    \label{fig:magnetometer1}
\end{figure}

\noindent Here we choose $\hat{x}$ as atomic-spin polarization (pump) axis, $\hat{y}$ as magnetic-field direction, and $\hat{z}$ as probe direction, as shown in Fig.~\ref{fig:magnetometer1}. This is a typical geometry for SERF magnetometers (Table~\ref{tab:magn_overview}), although the same derivation is possible for other operating modes.

After polarizing an atom along $\hat{x}$, we want to measure the angular-momentum projection $m_F=F,F-1,...,-F$ along $\hat{z}$ for some measurement duration $\tau$. Meanwhile, the atomic spin is evolving under the influence of the external magnetic field $\vec{B}$ according to the Bloch equation (Sec.~\ref{subsec:2.2-SpinRelaxation}):
\begin{equation}
    \frac{\text{d}\vec{F}}{\text{d}t} = \gamma\,\vec{B}\times\vec{F} \,.
\end{equation}
The gyromagnetic ratio $\gamma$ of the atomic system can also be expressed as $\gamma = g\,\mu_B / \hbar$, where $g$ is the ground-state Land\'e factor and $\mu_B$ is the Bohr magneton. For simplicity, we initially assume a static magnetic field.

Treating the expectation value of the angular momentum as a classical vector, we find for the evolution of $\left\langle F_z \right\rangle$:
\begin{equation}
    \frac{\text{d}\left\langle F_z \right\rangle}{\text{d}t} = -\gamma\,B \left\langle F_x \right\rangle \,.
\end{equation}
Starting from $\left\langle F_z \right\rangle = 0$ because the atom is polarized along $\hat{x}$, and integrating over a small measurement time $\tau$ during which $\left\langle F_x \right\rangle$ can be considered constant (practically speaking, a time much less than the longitudinal spin-relaxation time $T_1$), we obtain
\begin{equation}
    \left\langle F_z \left( \tau \right) \right\rangle = -\gamma\,B \left\langle F_x \right\rangle \tau \,.
\end{equation}
Looking at Fig.~\ref{fig:magnetometer1}, which shows the spin precession angle $\phi$, we see that
\begin{equation}
    \sin\phi \approx \left\langle F_z \right\rangle / \left\langle F_x \right\rangle \approx \phi \,.
    \label{eq:phi}
\end{equation}
Setting $F \approx \left\langle F_x \right\rangle \sim 1$ for an order-of-magnitude  (for our example of cesium, we typically have $F=4)$, we end up with the following expression:
\begin{equation}
    \phi \approx \gamma\,B\,\tau \,.
    \label{eq:phi1}
\end{equation}
This spin precession may be detected via optical rotation or some other technique; in any case we assume that our magnetometer has 100\% detection efficiency. (Note that for polarization-based detection, the angle of optical rotation of the probe light is proportional to $\phi$.) Hence the uncertainty in $B$ depends on the uncertainty in $\phi$:
\begin{equation}
    \delta B \approx \frac{\delta \phi}{\gamma\,\tau} \,.
    \label{eq:uncertaintyB}
\end{equation}

The same result can also be derived for a magnetic field $\vec{B}=B\cos\left(\Omega t\right)\hat{y}$ oscillating at the Larmor frequency $\Omega$. Then we find
\begin{equation}
     \frac{\text{d}\left\langle F_z \right\rangle}{\text{d}t} = -\gamma B \left\langle F_x \right\rangle \frac{\sin\left(\Omega \tau\right)}{\Omega} \,.
\end{equation}
Taking $F \sim 1$ as before and noting that typically $\Omega t \ll 1$, we arrive again at Eq.~\ref{eq:phi1}.

If we are limited only by the quantum noise of the atom (spin-projection noise, Sec.~\ref{subsec:3.1-Sensitivity}), $\delta \phi$ is fundamentally set by the Heisenberg uncertainty relation:
\begin{equation}
    \Delta F_y \, \Delta F_z \geq \frac{\left<F_x\right>}{2} \,,
\end{equation}
or, assuming minimum uncertainty $\Delta F_y = \Delta F_z$,
\begin{equation}
    \Delta F_z = \sqrt{\frac{\left<F_x\right>}{2}} \approx \sqrt{\frac{F}{2}} \sim 1 \,.
\end{equation}
We also have 
\begin{equation}
    \Delta F_z \approx F\,\delta\phi \sim \delta\phi \,,
\end{equation}
such that we find an uncertainty (standard deviation) in the precession angle of order 1\,rad, i.e. $\delta \phi \sim 1$. So Eq.~\ref{eq:uncertaintyB} becomes
\begin{equation}
    \delta B \approx \frac{1}{\gamma\,\tau} \,.
\end{equation}

Basic statistics tells us that if we perform the same measurement on $N$ independent atoms, the total measurement uncertainty (standard error) improves by $\sqrt{N}$. Similarly, if we repeatedly perform measurements of duration $\tau$ for a total time $t$, we gain another factor of $\sqrt{t/\tau}$. Thus we arrive at the total sensitivity
\begin{equation}
    \delta B_{\text{PN}} = \delta B \frac{1}{\sqrt{N}} \frac{1}{\sqrt{t/\tau}} \approx \frac{1}{\gamma} \frac{1}{\sqrt{N\,\tau\,t}} = \frac{\hbar}{g\,\mu_B} \frac{1}{\sqrt{N\,\tau\,t}} \,.
\end{equation}
Although we would like $\tau$ to be as long as possible, in reality this time is limited by the coherence time $T_2$ of the atomic ensemble, and therefore
\begin{equation}
    \delta B_{\text{PN}} \approx \frac{\hbar}{g\,\mu_B} \frac{1}{\sqrt{N\,T_2\,t}} \,.
\end{equation}
% In the idealized case where spin relaxation is absent and one could continuously interrogate the spins for a time $t=T_2$, the sensitivity becomes
% \begin{equation}
%     \delta B_{\text{PN}} \approx \frac{\hbar}{g\,\mu_B} \frac{1}{t\sqrt{N}} \,.
% \end{equation}

% As an example, we can calculate the expected sensitivity limit of a magnetometer based on $10^6$ lithium-6 atoms. This atom has nuclear spin $I=1$, such that the ground state $2\text{S}_{1/2}$ splits into two hyperfine sublevels with $F=1/2$ and $F=3/2$; the Land\'e factor is $\left|g\right|=2/3$. For $T \sim 1 \,\text{min}$, we find
% \begin{equation}
%     \delta B_{\text{tot}} \sim 0.1 \,\text{fT} = 1 \,\text{pG} \,,
% \end{equation}
% In magnetometry, sensitivity is usually given in magnetic-field resolution per square root of the measurement bandwidth. In the simplest case, the bandwidth is limited by the duration of the measurement $\delta\nu \approx 1/t$, where we ignore factors of order unity.
% which in standard sensitivity units corresponds to $\sim\,1 \,\text{fT}/\sqrt{\text{Hz}}\,=\,10 \,\text{pG}/\sqrt{\text{Hz}}$. Note that we have not considered fundamental noise introduced by the detection process, e.g. photon shot noise. In the case where there is spin relaxation such that $\tau \sim 1 \,\text{s}$, the sensitivity becomes an order of magnitude worse.

% \pagebreak
% \section{Cell characterization}
% \label{sec:Sup-CellCharacterization}
% \input{./Supplementary/CellCharacterization}

% \pagebreak
\section{Data-analysis example}
\label{sec:Sup-DataAnalysis}
See the provided data file ``MagData.bin'' and accompanying Mathematica analysis notebook ``MagAnalysis.nb'', used to produce the results in Sec.~\ref{sec:6-Characterization}.

% \pagebreak
% \printbibliography
%\bibliographystyle{unsrt}
%\bibliography{references}

\end{document}